\documentclass[fleqn,usenatbib]{mnras}

\usepackage[T1]{fontenc}
\usepackage{ae,aecompl}
 
\usepackage{graphicx}	
\usepackage{amsmath}	

\usepackage{amssymb}	
\usepackage{float}
\usepackage{caption}
\usepackage{subcaption}

\title[Velocity structure in the Centaurus cluster]{The velocity structure of the Intracluster Medium of the Centaurus cluster.} 

\author[Gatuzz et al.]{
Efrain Gatuzz$^{1}$\thanks{E-mail: egatuzz@mpe.mpg.de},
J. S. Sanders$^{1}$,
R. Canning$^{2}$,
K. Dennerl$^{1}$,
A. C. Fabian$^{3}$,
C. Pinto$^{4}$,\newauthor
H. Russell$^{5}$,
T. Tamura$^{6}$,
S. A. Walker$^{7,8}$
and J. ZuHone$^{9}$
\\
$^{1}$ Max-Planck-Institut f\"ur extraterrestrische Physik, Gie{\ss}enbachstra{\ss}e 1, 85748 Garching, Germany\\
$^{2}$ Kavli Institute for Particle Astrophysics and Cosmology, Stanford University, 452 Lomita Mall, Stanford, CA 94305-4085, USA\\ 
$^{3}$ Institute of Astronomy, Madingley Road, Cambridge CB3 0HA, UK\\ 
$^{4}$ INAF - IASF Palermo, Via U. La Malfa 153, I-90146 Palermo, Italy \\
$^{5}$ School of Physics \& Astronomy, University of Nottingham, University Park, Nottingham NG7 2RD, UK\\ 
$^{6}$ Institute of Space and Astronautical Science (ISAS), Japan Aerospace Exploration Agency (JAXA) Kanagawa 252-5210, Japan\\ 
$^{7}$ Astrophysics Science Division, X-ray Astrophysics Laboratory, NASA Goddard Space Flight Center, Greenbelt, MD 20771, USA\\ 
$^{8}$ Department of Physics and Astronomy, University of Alabama in Huntsville, Huntsville, AL 35899, USA\\
$^{9}$ Harvard-Smithsonian Center for Astrophysics, 60 Garden Street, Cambridge, MA, 02138, USA
}

\date{Accepted XXX. Received YYY; in original form ZZZ} 
\pubyear{2018} 
\begin{document}
 \label{firstpage}
\pagerange{\pageref{firstpage}--\pageref{lastpage}}
\maketitle 
\begin{abstract}
There are few direct measurements of ICM velocity structure, despite its importance for understanding clusters. We present a detailed analysis of the velocity structure of the Centaurus cluster using {\it XMM-Newton} observations. Using a new EPIC-pn energy scale calibration, which uses the Cu K$\alpha$ instrumental line as reference, we are able to obtain velocity measurements with uncertainties down to $\Delta v \sim 79$ km/s. We create 2D spectral maps for the velocity, metallicity, temperature, density, entropy and pressure with an spatial resolution of 0.25$'$. We have found that the velocity structure of the ICM is similar to the velocity structure of the main galaxies while the cold fronts are likely moving in a plane perpendicular to our line of sight with low velocity.  Finally, we have found a contribution from the kinetic component of $<25\%$ to the total energetic budget for radius $>30$ kpc.
\end{abstract} 


\section{Introduction}\label{sec_in} 
  
Theoretical simulations predict that the intracluster medium (ICM) should contain turbulent, or random motions, and bulk flows caused by the merger of other clusters and subcomponents \citep{lau09,vaz11,sch17,haj18,lim18,vaz21}.  In addition there can be relative bulk motions of $\sim  500$ km/s due to sloshing of the ICM in the potential well, generated by merging substructures \citep{asc06,ich19,vaz18,zuh18}. The action of the relativistic jets and inflation of bubbles by the central AGN also likely generates motions of a few hundred km/s \citep{bru05,hei10,ran15,yan16,bam19}. This is also important for several other reasons. Turbulent motions affects calculations of hydrostatic equilibrium and cluster mass estimates given that they provide additional pressure support, particularly at large radii \citep[e.g. ][]{lau09}. Measuring the velocities would help to constrain AGN feedback models because the distribution of energy within the bulk of the cluster depends on the balance between turbulence and shocks or sound waves \citep[see][for a review]{fab12b}. Velocity is an excellent probe of the microphysics of the ICM, such as viscosity, which acts to smooth velocity structure. Simulations predict that there is a close connection between the velocity power spectra and the overall ICM physical state \citep[which is identified by its temperature, density, pressure and entropy, see e.g.][]{gas14}. Motions will also cause transport of metals within the ICM, due to uplift and sloshing of metals by AGN \citep[e.g.][]{sim08,wer10}. In addition, measuring velocities should directly measure the sloshing of gas in cold fronts, which can remain for several Gyr \citep{roe12,roed13,wal18}. 

Despite its importance, the velocity structure of the ICM remains poorly observationally constrained. {\it Suzaku} observations were used to obtain velocities in several systems by measuring the Fe-K line. \citet{tam11,tam14} placed upper limits on relative velocities of $300$ km/s over scales of $400$ kpc in Perseus. \citep{ota16} examined several clusters with {\it Suzaku}, although systematic errors from the {\it Suzaku} calibration were likely around $300$ km/s and its PSF was large. Other direct methods of measuring velocities include using {\it XMM-Newton} RGS spectra to measure line broadening and resonant scattering, showing low turbulence motion with velocities between $100-300$ km/s \citep{san13,pin15,ogo17}, but limited to the cluster core. Finally, indirect measurements of velocity structure in clusters include looking at the power spectrum of density fluctuations and linking this via simulations to the velocity spectrum \citep[e.g. ][]{zhu14} or examining the magnitude of thermodynamic perturbations \citep{hof16}. However, these methods are model-dependent.

Random and bulk motions in the ICM were directly measured by the microcalorimeter Soft X-ray Spectrometer (SXS) X-ray detector on board of the {\it Hitomi} observatory using the Fe-K emission lines. In the core of the Perseus cluster it measured a gradient of $150$ km/s bulk flow across $60$ kpc of the cluster core and a line-of-sight velocity dispersion of $164 \pm 10$ km/s between radii of $30-60$ kpc \citep{hit16}. They indicate that the measured level of turbulence may be sufficient to offset radiative cooling if driven on scales comparable with the size of the largest bubbles in the field (about 20–30 kpc). On the other hand, the low level of turbulence in the system indicates that it might not be fast enough to replenish heating in the short time necessary to balance cooling, although shocks or sound waves could plausibly do this \citep{fab17}. Gravity waves, which could replenish turbulence, do not propagate efficiently radially \citep{fab17}. Simulations analyzed by \citet{bam18} indicated that in the presence of large-scale magnetic fields the resulting production of turbulence is inefficient, even when able to preserve AGN-drive bubbles. In their analysis of {\it XMM-Newton} observations of the cool core clusters RXCJ1504.1-0248 and Abell 1664 done by \citet{liu21}, they found that the turbulent energy density is less than 9\% and 27\%, respectively, thus insufficient for AGN heating to propagate the cool core via turbulence. Unfortunately, due to the loss of {\it Hitomi} we will not be able to make further measurements in other clusters or in different regions of Perseus. The next planned observatories with instruments capable of measuring velocities are likely {\sc XRISM} \citep{xri20}, to be launched in 2023 and {\sc Athena} \citep{bar18} in 2030.

  \begin{figure} 
\centering  
\includegraphics[width=0.45\textwidth]{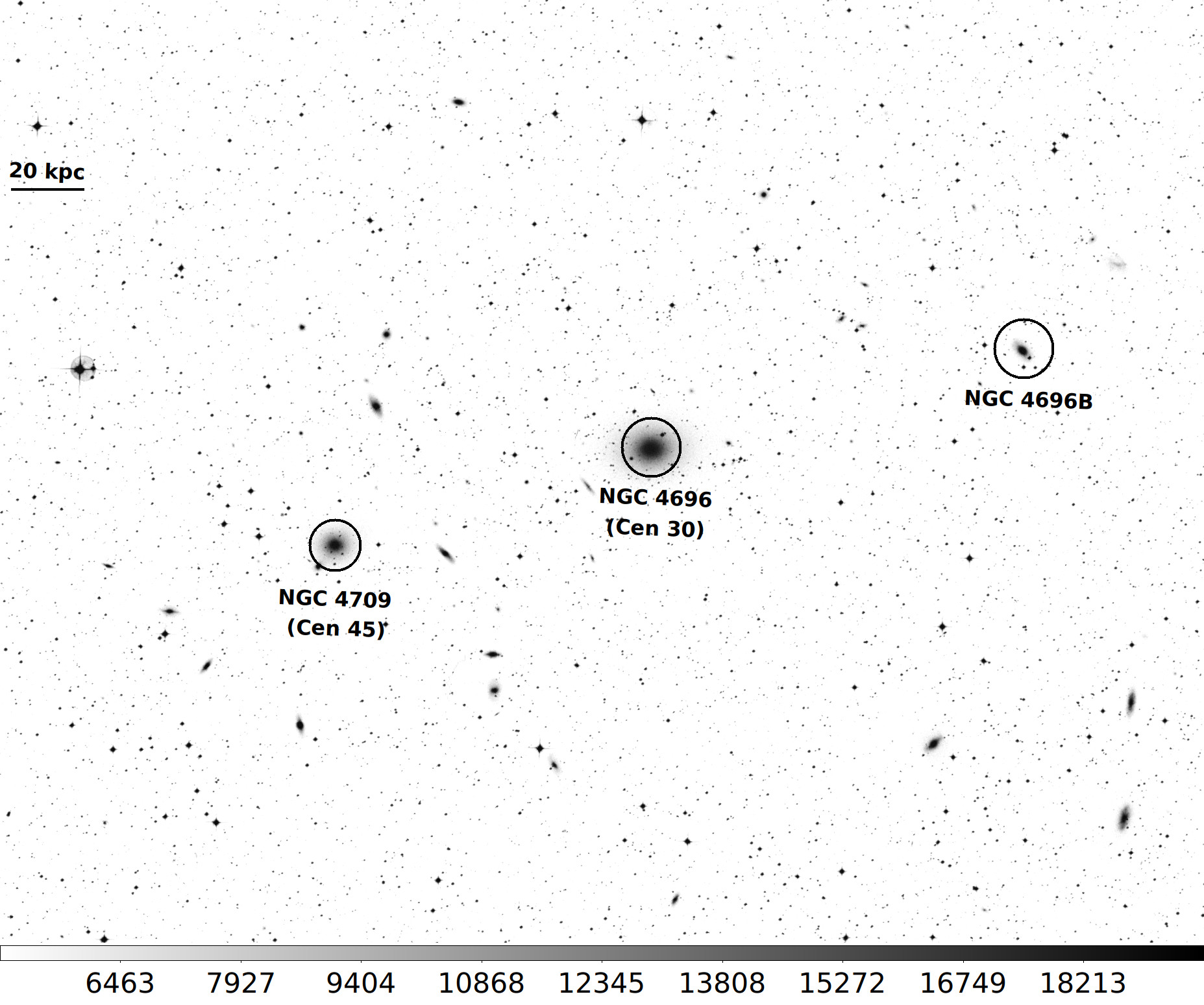}  
\caption{Digitized Sky Survey (DSS) image of the Cen 45 (NGC 4709) and Cen 30 (NGC 4696) merging systems. Cen~30 is the brightest cluster galaxy (BCG) of the Centaurus cluster. Units are arbitrarily scaled}\label{fig_optical_image} 
\end{figure}

\citet{san20} present a novel technique to improve velocity measurements calibrating the absolute energy scale of the detector to better than $150$ km/s at Fe-K by using background X-ray lines seen in the spectra of the {\it XMM-Newton} EPIC-pn detector. Using this technique, \citep{san20} mapped the bulk velocity distribution of the ICM over a large fraction of the central region of two nearby clusters of galaxies, Coma and Perseus. For the Coma cluster, they found that the velocity of the gas is close to the optical velocities of the two central galaxies, NGC 4874 and NGC 4889, respectively. In the case of the  Perseus cluster, they detect evidence for sloshing associated with a cold front. Following the same methodology, \citet{gat21} analyzed the velocity structure of the Virgo cluster. They identified a complex high velocity structure, most likely indicating the presence of an outflow near the AGN. Moreover, they found that the hot gas located within the radio flows move in opposite directions (i.e. blueshifted, redshifted) possibly driven by the radio jets or bubbles and with velocities of $\sim 331$ km/s $\sim -258$ km/s, with respect to M87.

The Centaurus cluster is an excellent target to apply such technique due to it being one of the X-ray brightest in the sky, its proximity allowing measurements on small scales \citep[$z=0.0104$,][]{luc86a}, high metallicity giving a strong Fe line and being a system with a merger and AGN feedback. Using optical observations \citet{luc86a} identified two subgroups in the Centaurus cluster, the main one centered on the galaxy NGC~4696 (refereed as Cen~30) and a second one centered on the galaxy NGC~4709 (refereed as Cen~45) 5 arcmin to the east with a line-of-sight (see Figure~\ref{fig_optical_image}). The heliocentric velocities of the main galaxies are NGC~4696: $2958\pm 15$ km/s \citep{dev91}, NGC~4709: $4678\pm 4$ \citep{smi00} and NGC~4696B: $3111\pm 7$ \citep{smi00}. \citet{wal13a} found evidence for a shock-heated region between the two systems, which could be explained by a simple shock-heating model. Despite this, \citet{ota16} found that the velocity of Cen~45 was consistent with the main cluster velocity, with a velocity difference of $<750$ km/s (90\% limit). However, {\it Suzaku} had a large PSF and the spatial regions examined were large. The cluster contains three or more cold fronts, likely caused by the ICM sloshing in the potential well. These are seen as discontinuities in surface brightness, temperature and metallicity \citep{san16}. Centaurus has probably the clearest cold front morphology seen in any nearby galaxy cluster. There is a possible large-scale Kelvin-Helmholtz instability in the core \citep{wal17}. Moreover, by analyzing {\it Chandra} observations, \citet{san16} have shown that the central AGN appears to have been repeatedly active over long timescales with periods of 10s of Myr. This is seen in edge-filtered Chandra X-ray images which show multiple cavity-like structures and possible sound waves. There is also extended low frequency radio emission associated with some of these cavities. Finally, in the core of the cluster there are also two weak shocks, surrounding the nucleus and inner cavities. Given all these highly energetic processes, we expect to find high velocities and, perhaps, velocity structure.

We present an analysis of the ICM velocity structure in the Centaurus galaxy cluster using {\it XMM-Newton} observations. The outline of this paper is as follows. We describe the data reduction process in Section~\ref{sec_dat}. The fitting procedure is shown in Section \ref{sec_fits} while a discussion of the results is included in Section~\ref{sec_dis}. Finally, Section~\ref{sec_con} presents the conclusions and summary. Throughout this paper we assume assumed a distance of Centaurus of \citep[$z=0.0104$,][]{luc86a} and a concordance $\Lambda$CDM cosmology with $\Omega_m = 0.3$, $\Omega_\Lambda = 0.7$, and $H_{0} = 70 \textrm{ km s}^{-1}\ \textrm{Mpc}^{-1} $.

\begin{table} 
\footnotesize
\caption{\label{tab_obsids}{\it XMM-Newton} observations of the Centaurus cluster.}
\centering 
\begin{tabular}{cccccccc}   
\hline
ObsID & RA & DEC & Date & Exposure \\
 &&&Start-time& (ks)\\
\hline
 0046340101 &192.20& 	-41.31&	2002-01-03 &	47.7\\
 0504360101 &192.20& 	-41.31&	2006-07-24 &	124.3\\
 0823580101 &192.51& 	-41.38&	2007-07-27 &	43.1\\
 0823580301 &191.75& 	-41.34&	2007-12-27 &	34.9\\
 0823580501 &192.44& 	-41.29&	2018-07-03 &	11.4\\
 0823580701 &192.26& 	-41.23&	2018-07-13 &	113.9\\
 0406200101 &192.08& 	-41.16&	2018-07-20 &	117.0\\
 0504360201 &192.13& 	-41.42&	2018-07-26 &	116.9\\
 0823580201 &191.96& 	-41.34&	2018-07-30 &	116.9\\
 0823580401 &192.31& 	-41.48&	2018-08-09 &	114.9\\
 0823580601 &192.47& 	-41.27&	2019-01-22 &	140.8\\
 0823580801 &192.26& 	-41.45&	2019-06-30 &	137.9\\
 \\
 \hline
\end{tabular}
\end{table}
 
   \begin{figure*}
        \centering
        \begin{subfigure}{0.50\textwidth}
            \includegraphics[scale=0.185]{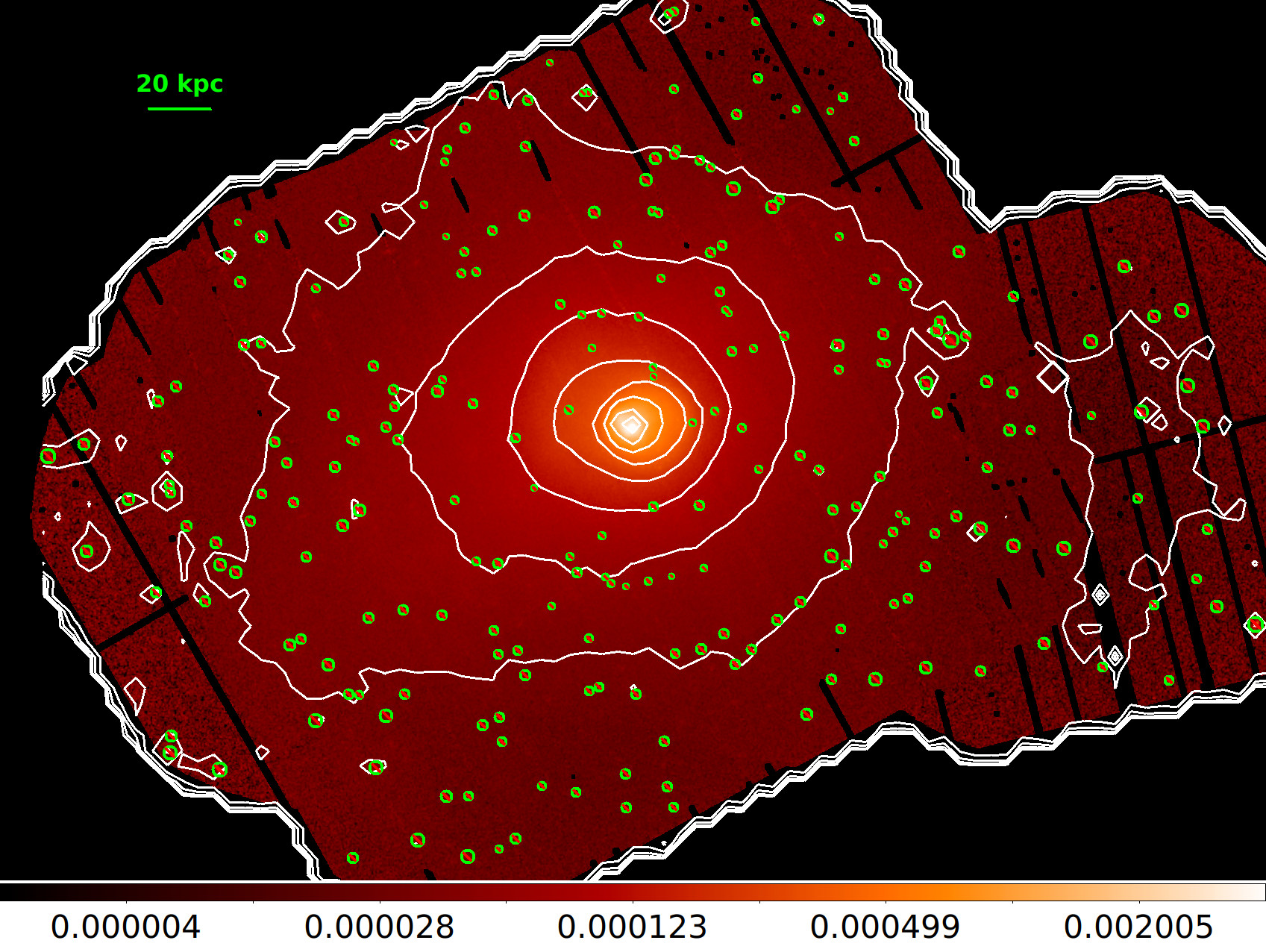}
        \end{subfigure}
          \centering
        \hspace{2.3cm}          
        \begin{subfigure}{0.32\textwidth}
        	\vspace{-0.2cm}
           \includegraphics[scale=0.09]{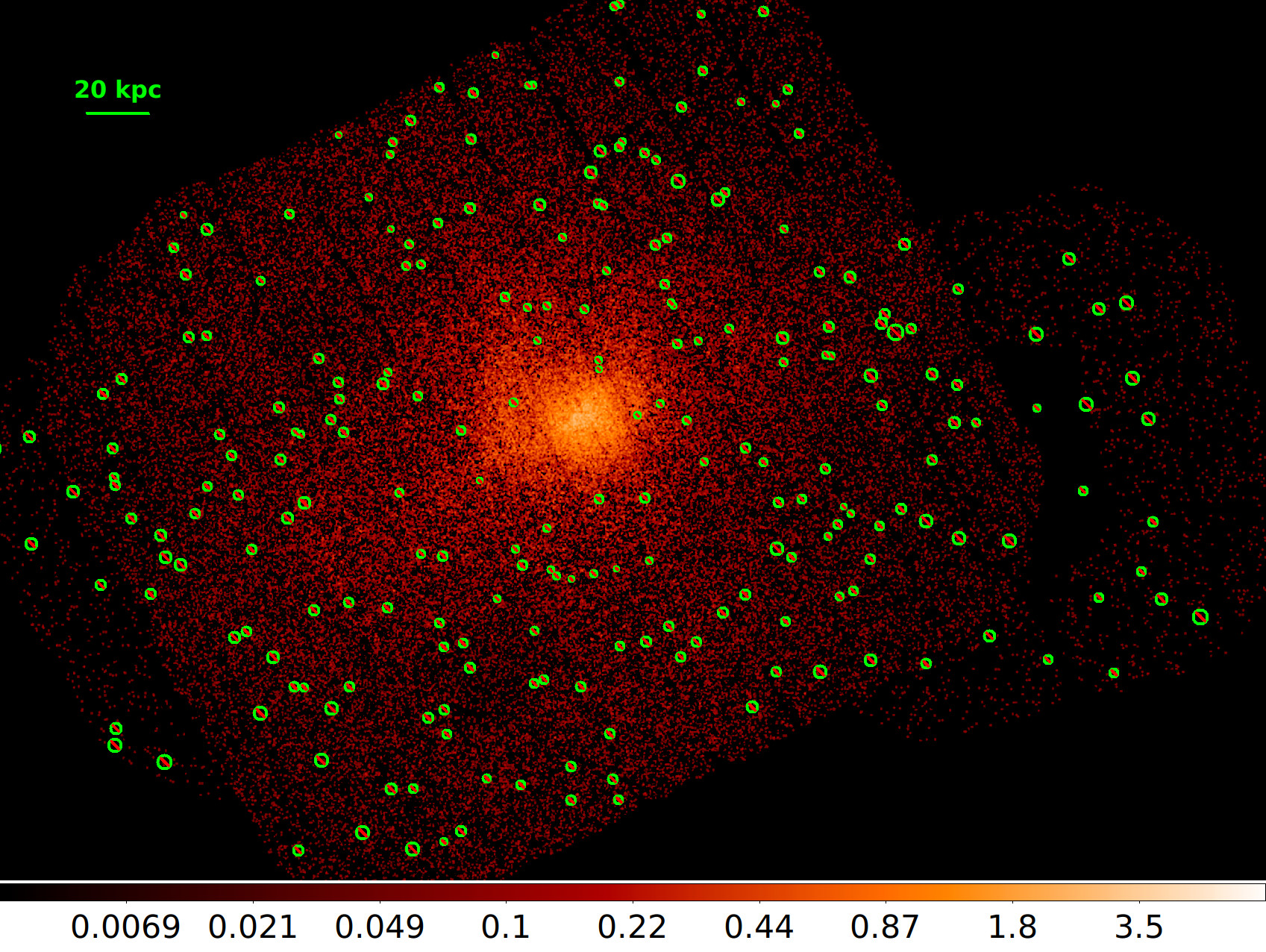}\\
           \includegraphics[scale=0.09]{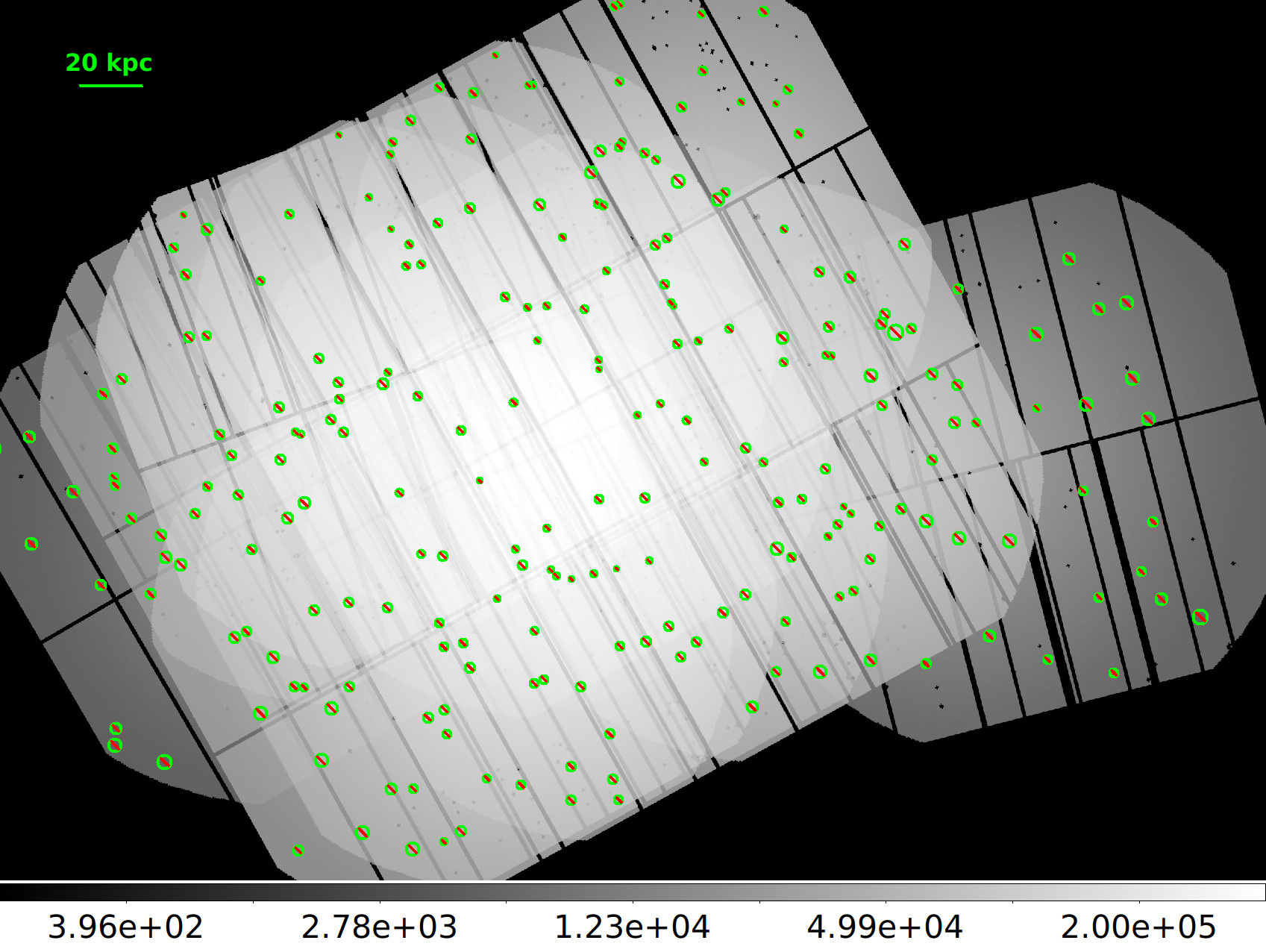}
        \end{subfigure}
        \caption{\emph{Left panel:} X-ray surface brightness exposure corrected of the Centaurus cluster in the 0.5 to 9.25 keV energy range. Green circles indicate the point-like sources which were excluded in the analysis. The contours of the X-ray image are shown in white. \emph{Top right panel:} Fe-K count map, showing the number of counts in each 1.59 arcsec pixel in the Fe-K complex. \emph{Bottom right panel:} total exposure time (s) in the 4.0-9.25 keV energy range. } \label{fig_xray_maps} 
    \end{figure*}

     \begin{figure} 
\centering 
\includegraphics[width=0.485\textwidth]{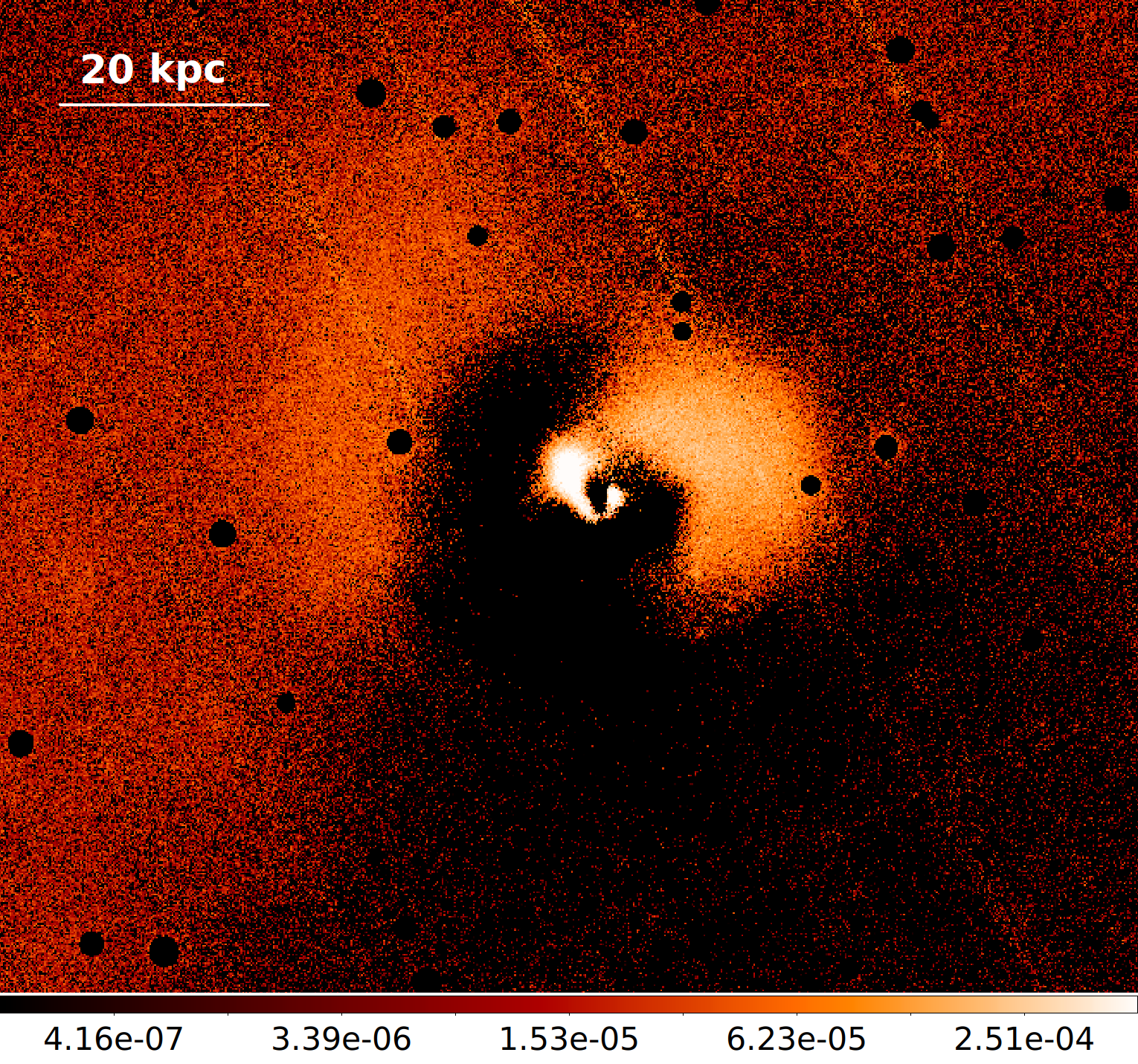}
\caption{Subtracted fractional difference from the average at each radius for the 0.5-7 keV X-ray surface brightness. } \label{fig_xra_sb} 
\end{figure}

     \begin{figure} 
\centering 
\includegraphics[width=0.485\textwidth]{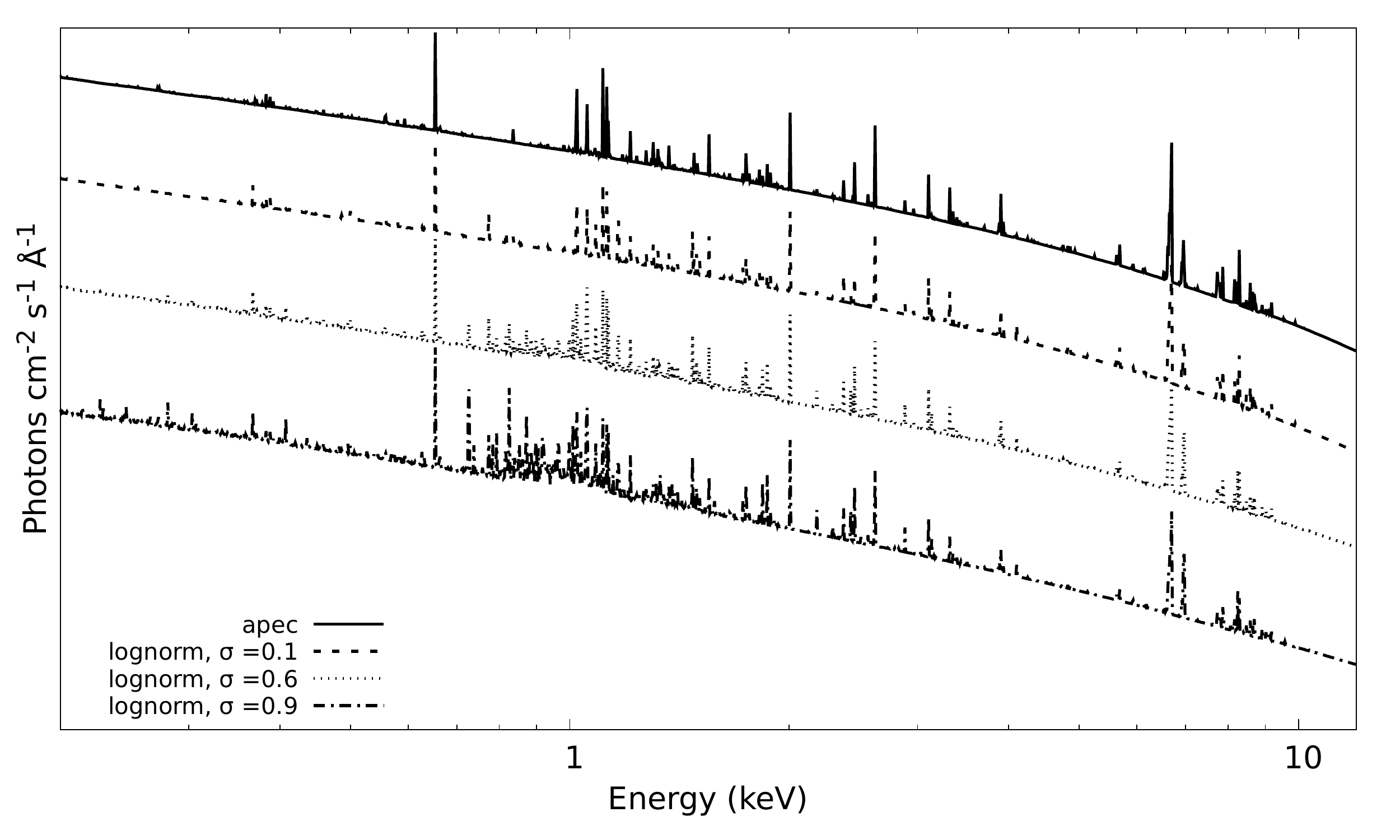} 
\caption{A comparison  between the {\it apec} and {\it lognorm} models. For both models the temperature is fixed to $\log{T}=4$ keV and metallicity to $Z=1$ while for {\it lognorm} the $\sigma$ parameter varies. Normalizations have been chosed arbitrarily for illustrative purposes.} \label{fig_lognorm_model} 
\end{figure}
    
\section{Data reduction}\label{sec_dat}
Information on the Centaurus cluster observations analyzed in this paper is provided in Table~\ref{tab_obsids}, including IDs, coordinates, dates and clean exposure times. We reduced the {\it XMM-Newton} European Photon Imaging Camera \citep[EPIC,][]{str01} spectra with the Science Analysis System (SAS\footnote{\url{https://www.cosmos.esa.int/web/xmm-newton/sas}}, version 18.0.0). Each observation was processed with the {\tt epchain} SAS tool, including only single-pixel events (PATTERN==0) and filtering the data with FLAG==0 to avoid regions close to CCD edges or bad pixels. Bad time intervals were filtered from flares applying a 1.0 cts/s rate threshold. 

In order to obtain velocity measurements down to 150 km/s at Fe-K, we have applied the technique described in \citet{san20} to calibrate the absolute energy scale of the EPIC-pn detector using the background X-ray lines identified in the spectra of the detector. This calibration method include corrections (1) to the average gain of the detector during the observation, (2) to the spatial gain variation across the detector with time and (3) to the energy scale as function of detector position and time. In particular, we used the calibration files computed by \citet{gat21}. However, we do not expect changes with new caldb as observations are older than 2020.

Figure~\ref{fig_xray_maps} shows the X-ray image obtained in the 0.5 to 9.25~keV energy band (left panel). We have identified point sources using the SAS task {\tt edetect\_chain}, with a likelihood parameter {\tt det\_ml} $> 10$. In the following analysis we have excluded the point sources. Figure~\ref{fig_xray_maps} top right panel shows the number of counts in each 1.59 arcsec pixel in the Fe-K complex (6.50 to 6.90 keV, in restframe), after subtracting neighbouring scaled continuum images (6.06 to 6.43 and 6.94 to 7.16 keV, in restframe). A gaussian smoothing of $\sigma = 4$ pixels was applied for illustrative purposes. The image shows that a large number of counts in the multiple directions around the cluster center, due to the spatial offset in the observations. Figure~\ref{fig_xray_maps} bottom right panel shows the total exposure time in the 4.0-9.25 keV energy range. Figure~\ref{fig_xra_sb} shows the fractional difference in 0.5 to 9.25 keV surface brightness from the average at each radius. The central region of the cluster exhibits a complex structure, no clear center and a very faint point source at the location of the AGN \citep{san16}, thus preventing us from perform a velocity analysis using RGS spectra.

\begin{figure*} 
\centering 
\includegraphics[width=0.485\textwidth]{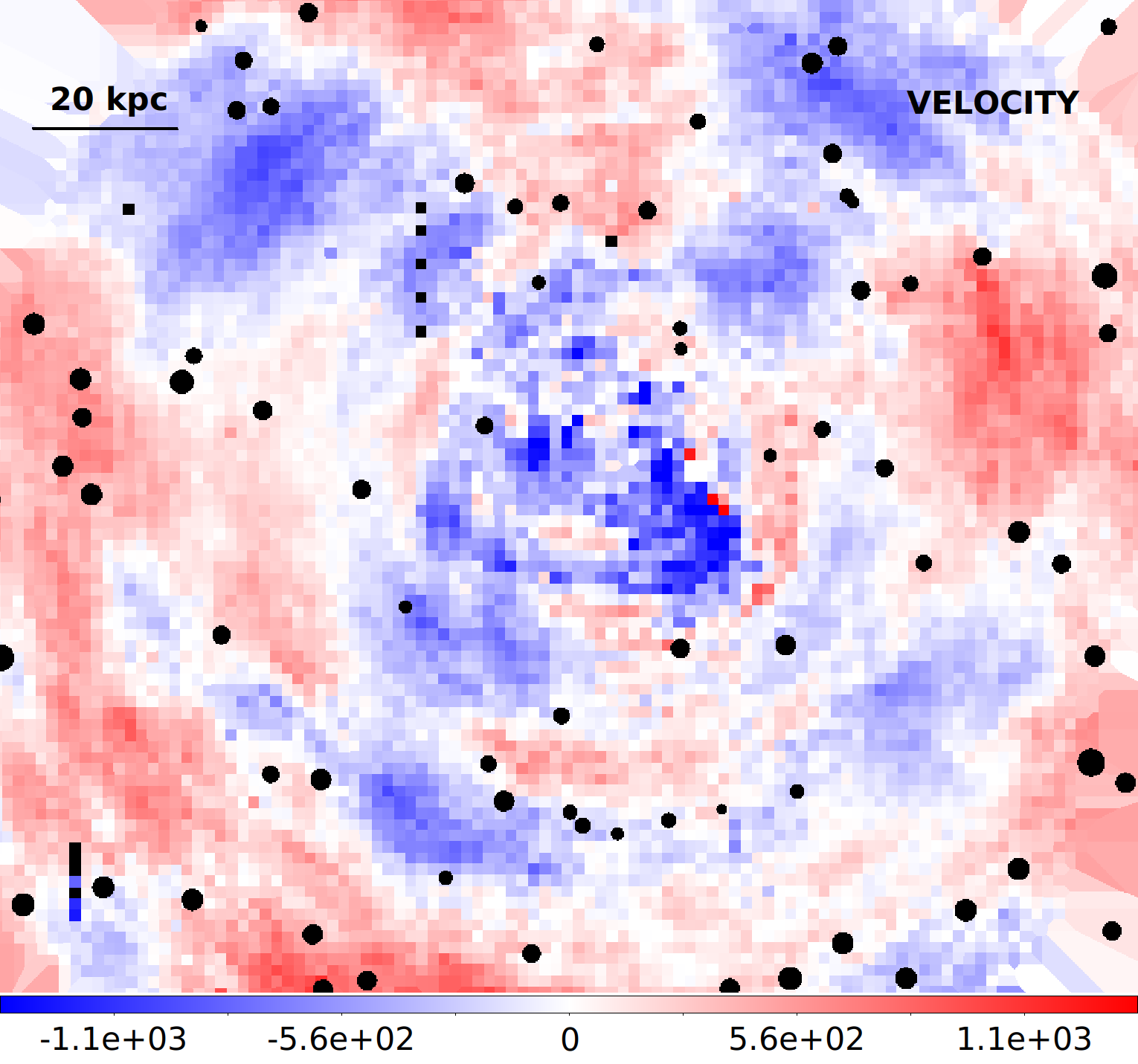} 
\includegraphics[width=0.485\textwidth]{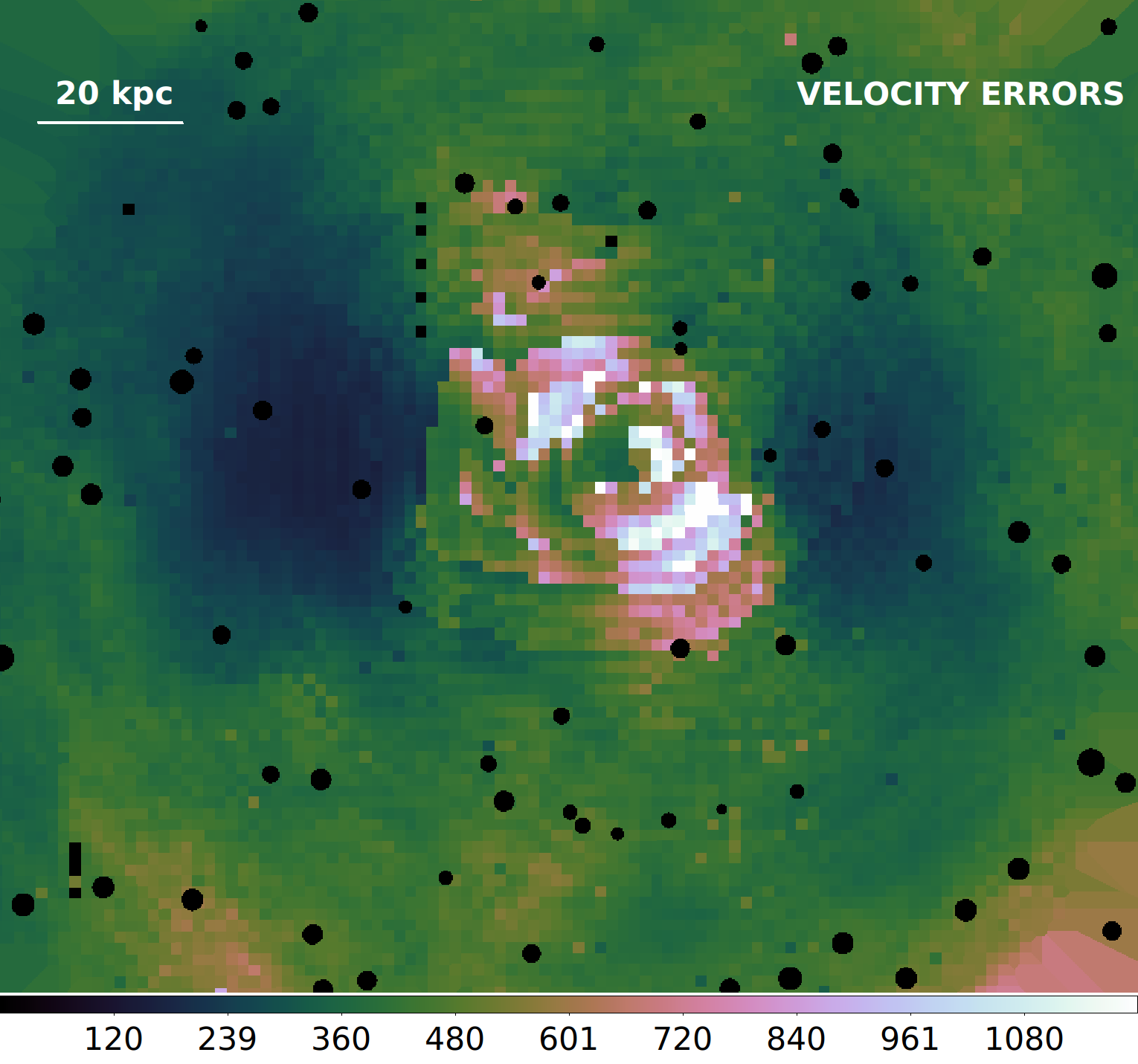} 
\caption{\emph{Left panel:} velocity map (km/s) relative to the Centaurus cluster ($z=0.0104$). Black circles correspond to point sources which were excluded from the  analysis. Maps were created by moving 2:1 elliptical regions (i.e. rotated to lie tangentially to the nucleus) containing $\sim$750 counts in the Fe-K region.  \emph{Right panel:}  1$\sigma$ statistical uncertainty on the velocity map (km/s). } \label{fig_velocity_ellipses1} 
\end{figure*}

 \begin{figure} 
\centering 
\includegraphics[width=0.46\textwidth]{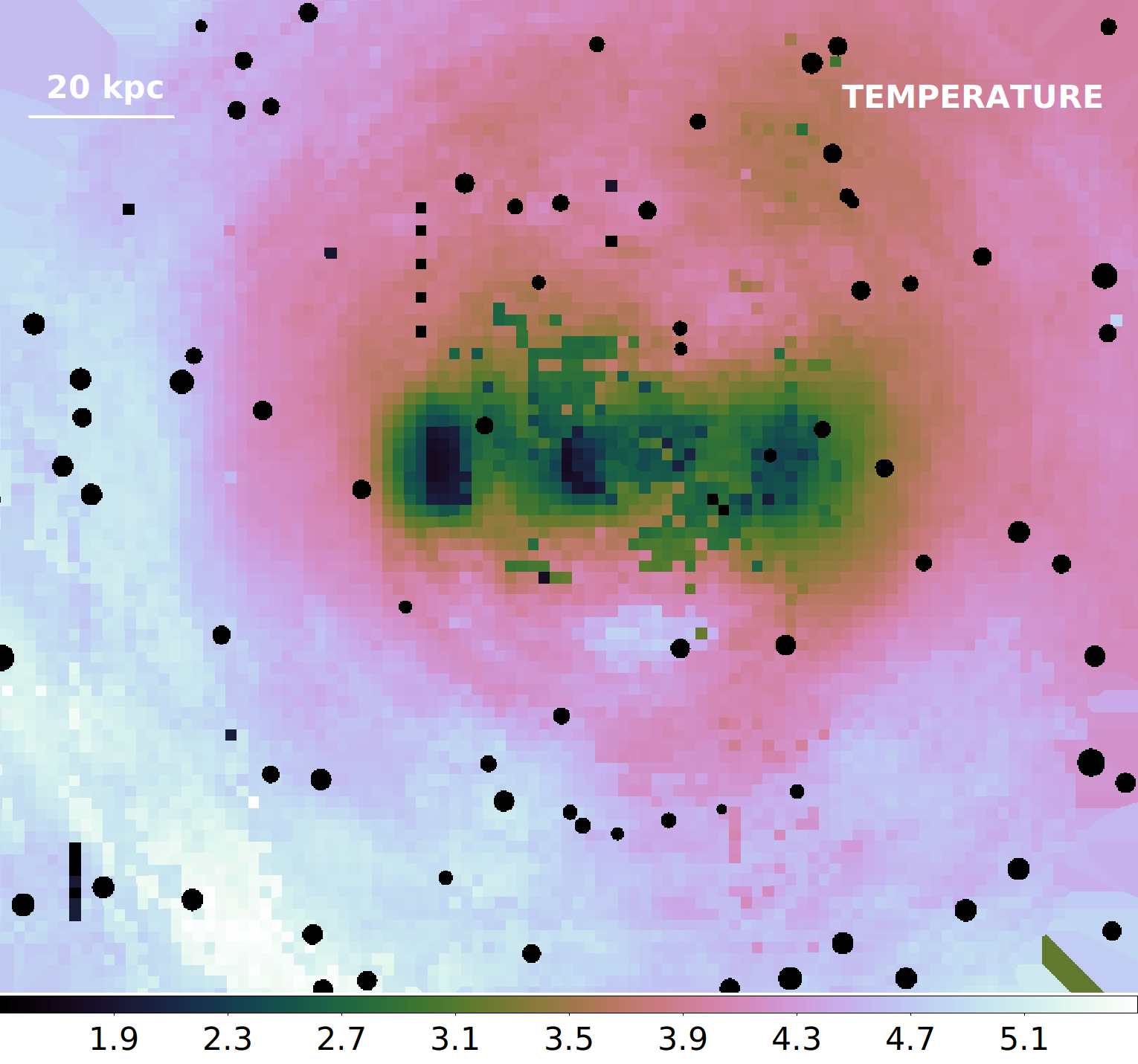}\\
\includegraphics[width=0.46\textwidth]{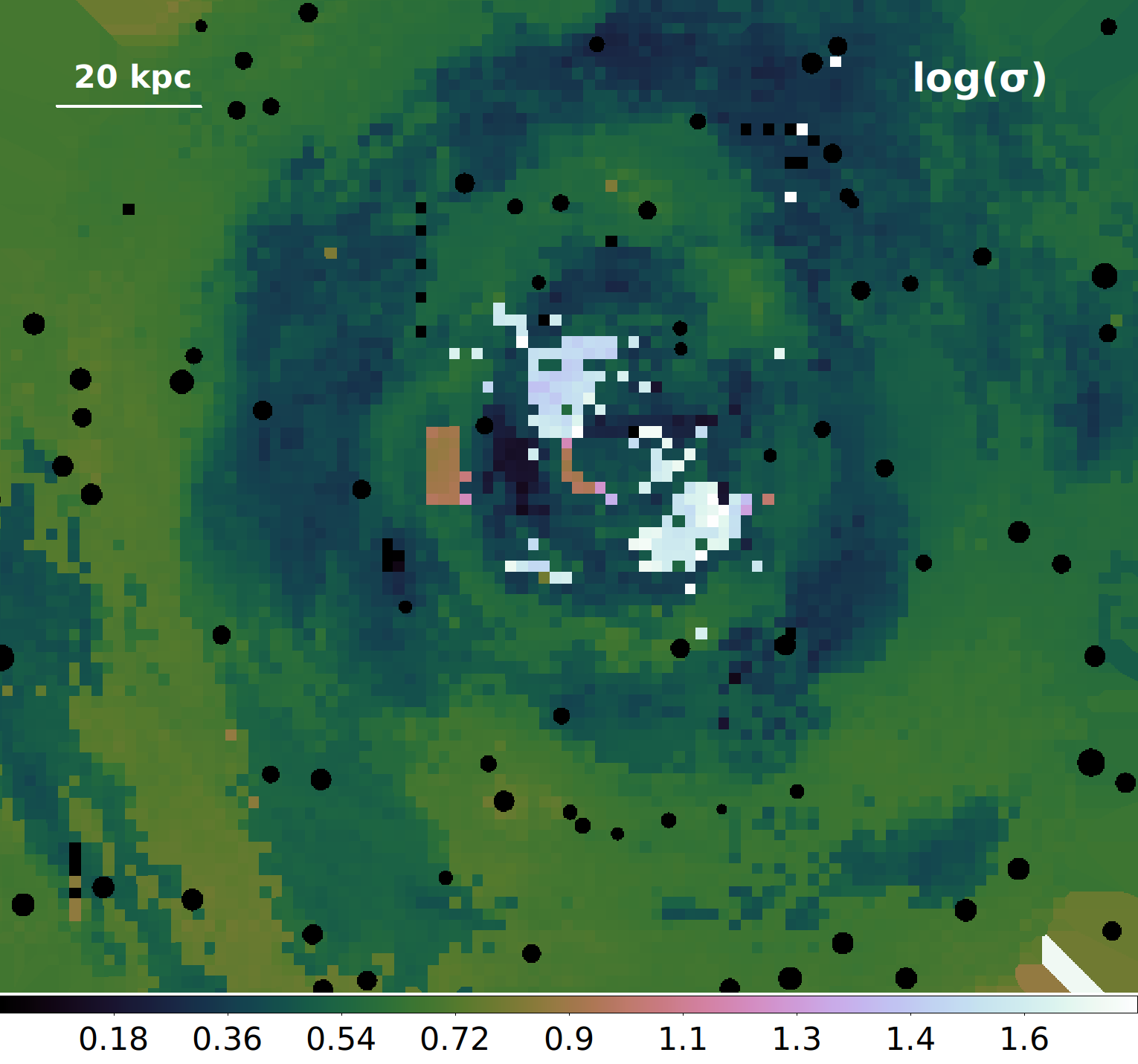}\\
\includegraphics[width=0.46\textwidth]{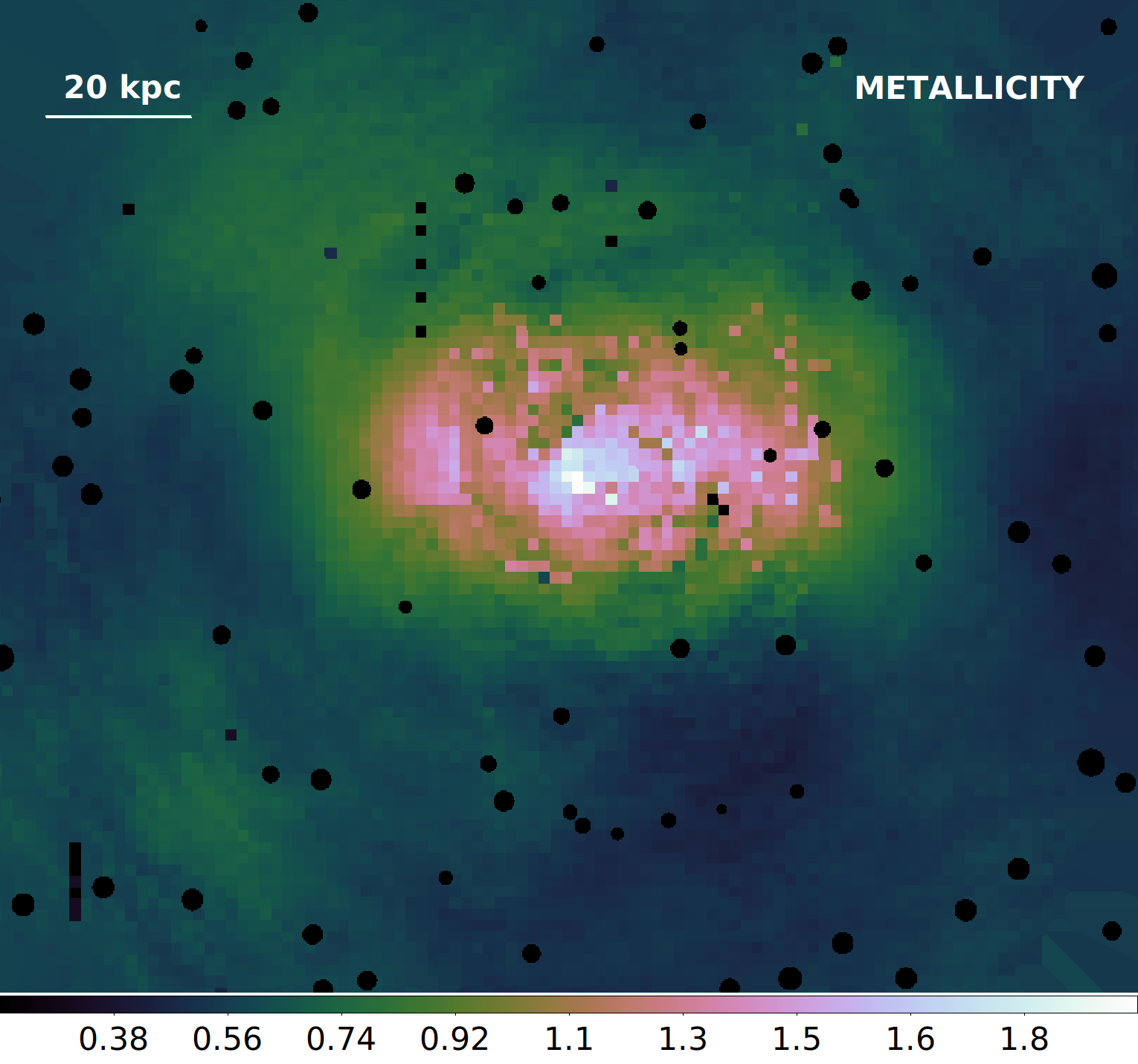}
\caption{\emph{Top panel:} temperature map in units of KeV. \emph{Middle panel:} $\log(\sigma)$ map. \emph{Bottom panel:} abundance map relative to solar abundances from \citet{lod09}. } \label{fig_velocity_ellipses2} 
\end{figure}

 \begin{figure} 
\centering  
\includegraphics[width=0.43\textwidth]{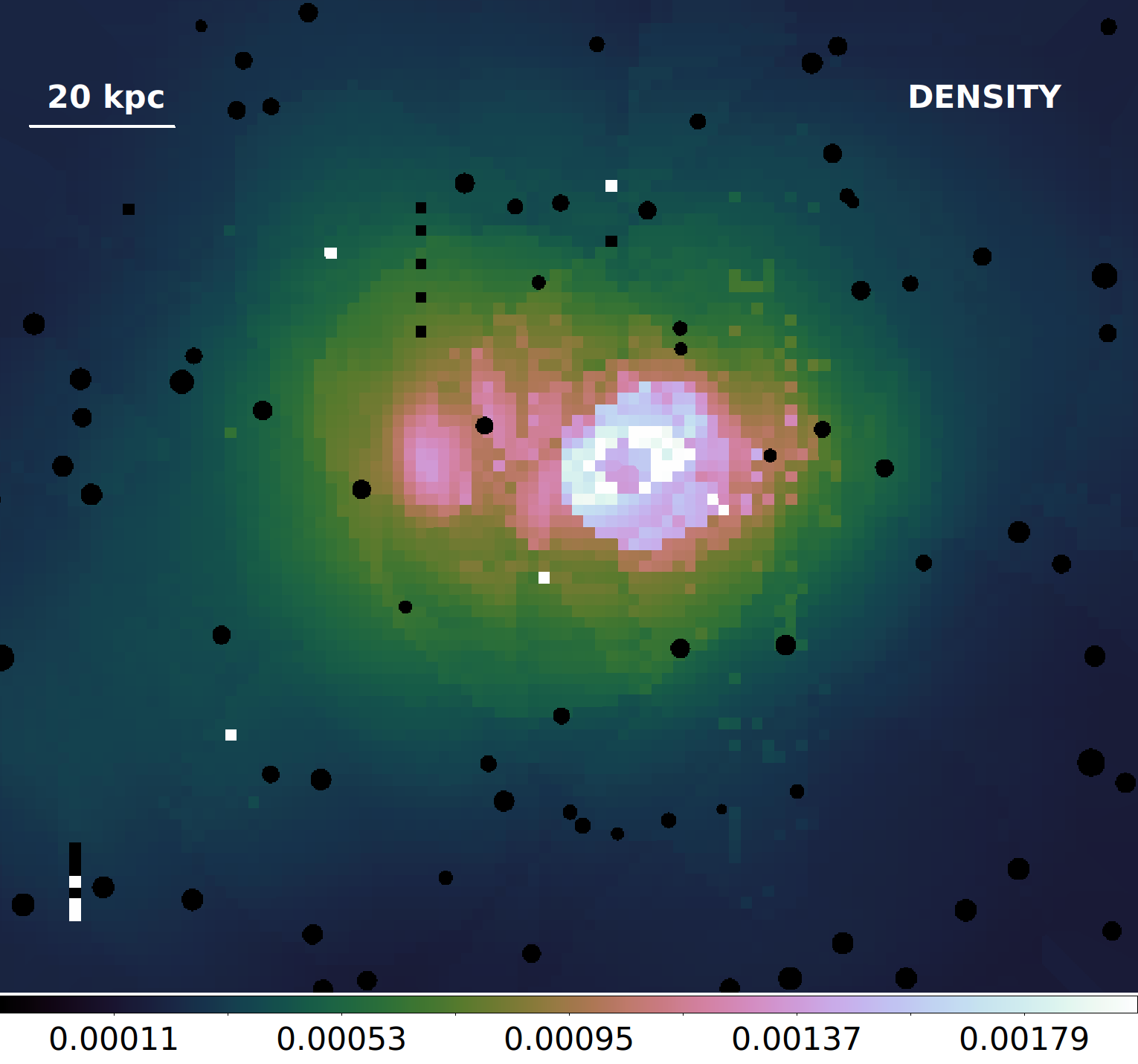} \\
\includegraphics[width=0.43\textwidth]{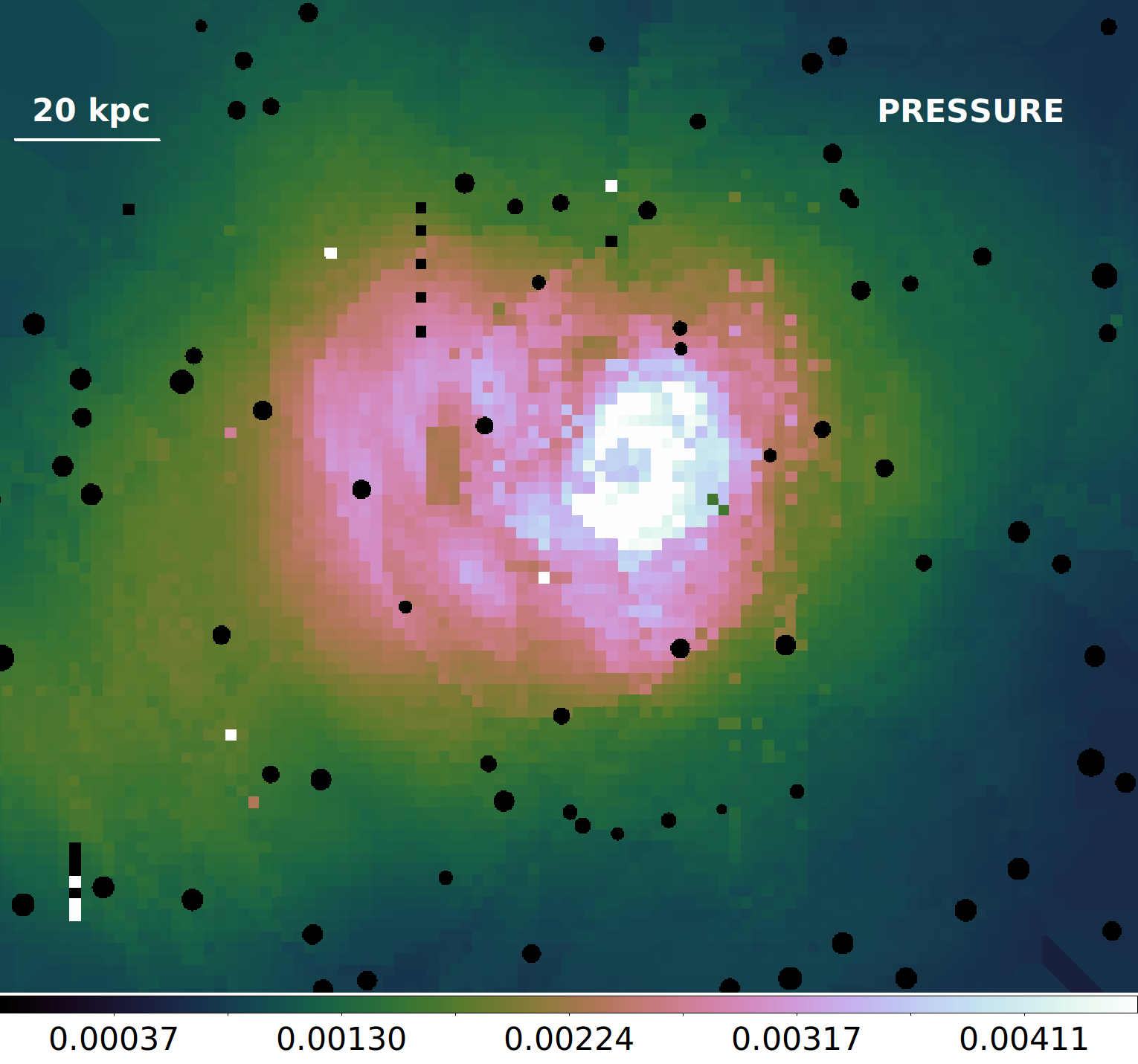} \\
\includegraphics[width=0.43\textwidth]{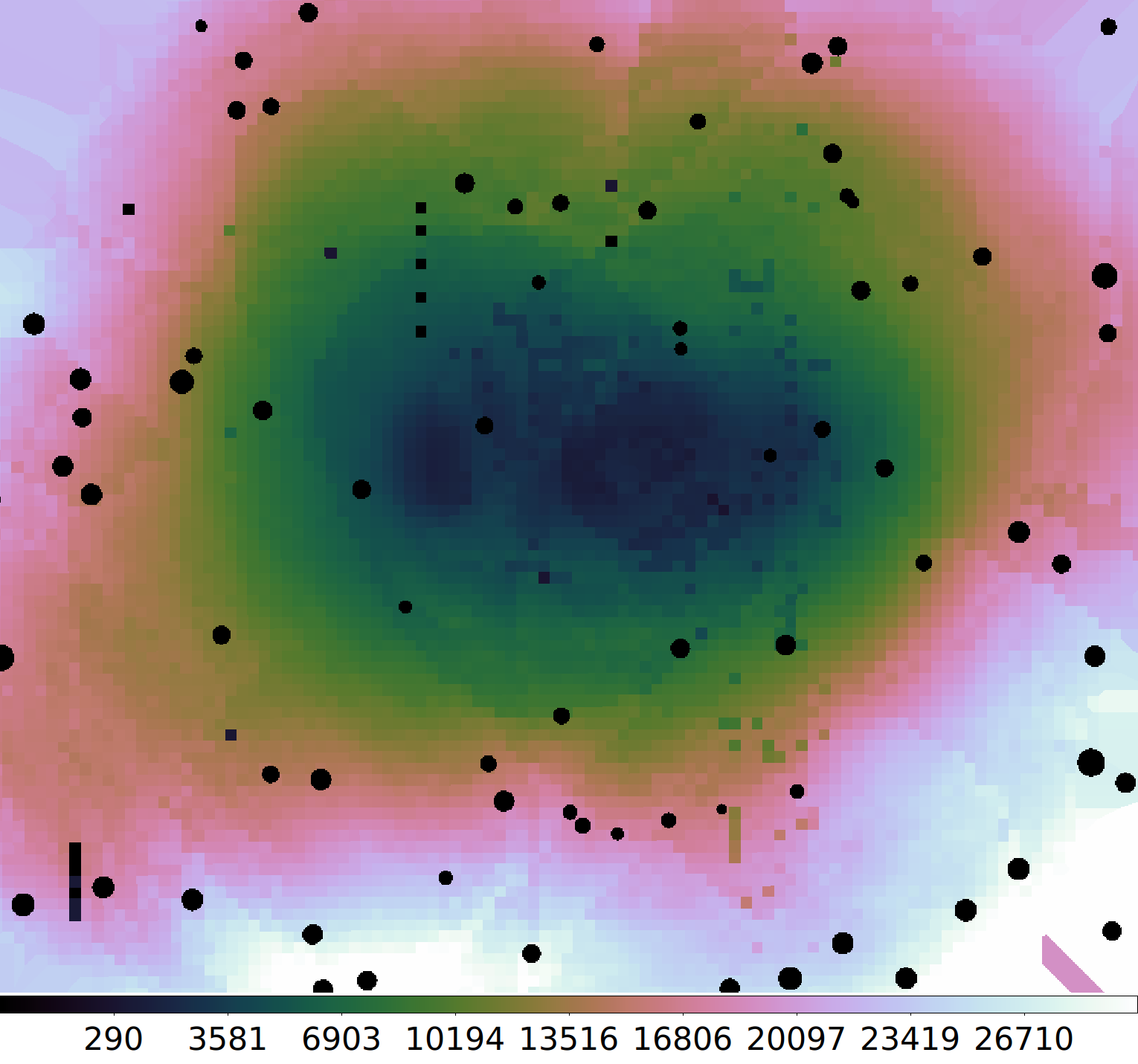}
\caption{\emph{Top panel:} density map (cm$^{-3}$). Black circles correspond to point sources which were excluded from the  analysis. \emph{Middle panel:} pressure map (keV cm$^{-3}$). \emph{Bottom panel:} entropy map (keV cm$^{2}$). } \label{fig_entropy_ellipses} 
\end{figure}

\section{Data analysis}\label{sec_fits} 
In order to account for the possible presence of a multi-temperature plasma, we constructed a model to describe a log-normal temperature distribution (namely {\tt lognorm}). Such a log-normal temperature distribution associated with galaxy clusters was identified by \citet{fra13} in their analysis of 62 galaxy clusters using {\it XMM-Newton} spectra. The model takes as input a central temperature ($T$), width of the temperature distribution in log space ($\sigma$), metallicities, redshift and a normalization.  The output spectrum is constructed by summing a number of {\it apec} components with log-normal relative normalisations assuming they have the same redshift and metallicities. The relative normalisation of each component is given by 

\begin{equation}
exp( -((\log T_{i}- \log T)/\sigma)^2/2 )
\end{equation}

where $\log{T_{i}}$ varies between $\log{T} - N\sigma$ to $\log{T} + N\sigma$ in $n$ steps (where $N=3$ and $n=21$ by default). The relative normalizations are scaled so that the total is the overall normalization. Figure~\ref{fig_lognorm_model} shows a comparison between the {\it apec} and {\it lognorm} models for different $\sigma$ parameter values.

We included a {\tt tbabs} component \citep{wil00} to account for the Galactic X-ray absorption. For each spatial region analyzed, we combine the spectra from different observations together and rebinned the spectra to have at least 1 count per channel. Then, we load the data twice in order to fit separately, but simultaneously, the 1.5-4.0 keV and hard 4.0-10 keV energy bands. The free parameters in the model are the metallicity, log($\sigma$) of the temperature distribution, temperature and normalization. The redshift is a free parameter only for the 4.0-10 keV energy band given that for lower energies the new EPIC-pn energy calibration scale cannot be applied. We note that the inclusion of the lower energy band data, for which the redshift parameter is fixed to the cluster value, leads to a better constrain for the temperatures and metallicites. We fixed the column density to $8.10\times 10^{20}$ cm$^{-2}$, although it is important to note that in the energy range analyzed the absorbing component has a weak effect in the modeling. For the data analysis we used the {\it xspec} spectral fitting package (version 12.10.1\footnote{\url{https://heasarc.gsfc.nasa.gov/xanadu/xspec/}}). We assumed {\tt cash} statistics \citep{cas79}. Errors are quoted at 1$\sigma$ confidence level unless otherwise stated. Finally, abundances are given relative to \citet{lod09}. As background components we have included Cu-$K\alpha$, Cu-$K\beta$, Ni-$K\alpha$ and Zn-$K\alpha$ instrumental emission lines, and a powerlaw component with its photon index fixed at 0.136 \citep[the average value obtained from the archival observations analyzed in ][]{san20}.

\subsection{Spectral maps}\label{spec_maps} 
We created a velocity map of the cluster following the method shown in \citet{san20} and \citet{gat21}. We created elliptical regions with a 2:1 axis ratio, rotating them such that the longest axis lay tangentially to a vector pointing to the central core. In order to reduce uncertainties on the velocities, the radii of the ellipses changed adaptively to have $\sim$ 500 counts in the Fe-K complex after continuum substraction.  We moved in grid with a spacing of 0.25 arcmin. For each region we extracted the spectra for all observations and performed a combined fit. Then, we created weighted-average ancillary responses files (arf) for each ellipse from the individual ones for each observation but used the same response matrix for all the pixels. 

Figure~\ref{fig_velocity_ellipses1} shows the velocity maps obtained. In order to improve the clarity we use a red-blue color scale to indicate whether the gas is moving away from the observer (red) or towards (blue). The map shows the complexity of the velocity structure in the system. Interestingly, there is no clear spiral pattern associated with gas sloshing \citep[see for example Figure~18 in][]{san20} nor an opposite blueshifted-redshifted structure around the cluster center \citep[see for example Figure~11 in][]{gat21}. Such features could point out that our line-of-sight is perpendicular to the sloshing plane of the system \citep[see for example ][]{zuh16}. There is a clear blueshifted structure along the south-west direction from the cluster center with velocities larger than $1000$ km/s, which could be due to the impact of AGN outflows from the NGC~4696 system. This could be a hint of azimuthal structure in the velocity. However, such structure corresponds to a region with large velocity uncertainties, which could also be a feature of gas sloshing \citep[see for example Figure 19 in][]{gat21}. The velocity spectral map shows a large blueshifted region southward from the Galactic cluster core. Such velocity increasing could be associated to gas sloshing signatures \citep[see Figures 2 and 3 in][]{zuh18}. The blueshifted gas displays an X-shape, which we have not identified in magnetohydrodynamic simulations. It is important to note that the resolution obtained in these spectral maps ($\sim$10 kpc for the smallest ellipse) is larger than the radio observations \citep[see for example ][]{tay06}, therefore preventing us from performing a direct analysis of the velocity structure within the extended component of the central radio source. Moreover, \citet{tay06} shown that the radio emission seems initially directed north-south whit hints of interaction with the thermal gas which leads to a east-west collimation. However, the main velocity field along the structure of the radio source is fairly close to the plane of the sky. Figure~\ref{fig_velocity_ellipses2} shows the temperature, $\log(\sigma)$ and metallicity spectral maps. Near the Centaurus cluster core there is a  high-metallicity region with low-temperature. A similar structure was identified, with higher resolution due to the {\it Chandra} instruments, by \citet[][see Figure~3]{san16}. Outside the central region the temperature and metallicities changes smoothly, with the temperature increasing and the metallicity decreasing as we move to the outskirts of the cluster. We note that the two opposite regions with large velocity uncertainties located NE and SW from the cluster core, correspond to regions with large $\log(\sigma)$ and large temperature uncertainties where the Fe-K line is weak and broad.

\begin{table*}
\scriptsize 
\caption{\label{tab_circular_fits}Centaurus cluster best-fit parameters for Case 1 extracted regions. }
\centering
\begin{tabular}{ccccccc}
\\
Region &\multicolumn{6}{c}{{\tt lognorm} model}  \\
\hline
 &$kT$&$\sigma$& Z& $z$   & $norm$   & cstat/dof\\ 
  & & & &  ($\times 10^{-3}$) &   ($\times 10^{-3}$)  \\ 
1&$2.40\pm 0.03$&$0.42\pm 0.02$&$1.76\pm 0.07$&$9.59\pm 0.55$&$15.09\pm 0.30$&$1953/1687$\\
2&$2.32\pm 0.02$&$0.41\pm 0.02$&$1.70\pm 0.05$&$9.55\pm 0.38$&$27.55\pm 0.36$&$2035/1688$\\
3&$2.63\pm 0.02$&$0.42\pm 0.02$&$1.52\pm 0.04$&$10.17\pm 0.34$&$30.54\pm 0.31$&$2024/1688$\\
4&$2.89\pm 0.02$&$0.39\pm 0.02$&$1.34\pm 0.03$&$9.97\pm 0.30$&$31.10\pm 0.25$&$2250/1688$\\
5&$3.03\pm 0.03$&$0.38\pm 0.03$&$1.17\pm 0.02$&$10.64\pm 0.30$&$19.83\pm 0.14$&$2046/1688$\\
6&$2.94\pm 0.02$&$0.49\pm 0.02$&$1.06\pm 0.02$&$9.98\pm 0.30$&$19.82\pm 0.13$&$2376/1688$\\
7&$3.22\pm 0.02$&$0.47\pm 0.02$&$0.87\pm 0.02$&$10.19\pm 0.32$&$16.85\pm 0.10$&$2114/1688$\\
8&$3.57\pm 0.04$&$0.34\pm 0.04$&$0.75\pm 0.02$&$10.02\pm 0.36$&$12.84\pm 0.07$&$1891/1688$\\
9&$3.79\pm 0.05$&$0.43\pm 0.03$&$0.70\pm 0.02$&$10.12\pm 0.40$&$12.67\pm 0.07$&$1912/1688$\\
10&$4.01\pm 0.05$&$0.43\pm 0.04$&$0.63\pm 0.02$&$10.80\pm 0.43$&$12.85\pm 0.07$&$1900/1688$\\
11&$4.07\pm 0.05$&$0.45\pm 0.04$&$0.60\pm 0.02$&$10.93\pm 0.46$&$12.08\pm 0.06$&$1968/1688$\\
12&$4.15\pm 0.05$&$0.45\pm 0.05$&$0.55\pm 0.02$&$10.16\pm 0.52$&$11.52\pm 0.07$&$1973/1688$\\
13&$4.38\pm 0.08$&$0.51\pm 0.06$&$0.55\pm 0.02$&$10.05\pm 0.58$&$11.49\pm 0.08$&$2180/1688$\\
14&$4.40\pm 0.10$&$0.60\pm 0.05$&$0.53\pm 0.02$&$11.39\pm 0.66$&$13.80\pm 0.10$&$2161/1688$\\
15&$4.25\pm 0.05$&$0.62\pm 0.03$&$0.48\pm 0.02$&$11.20\pm 0.59$&$21.91\pm 0.14$&$2742/1688$\\
\\ 
 \hline
\end{tabular}
\end{table*}

   \begin{figure}
   \centering
\includegraphics[width=0.46\textwidth]{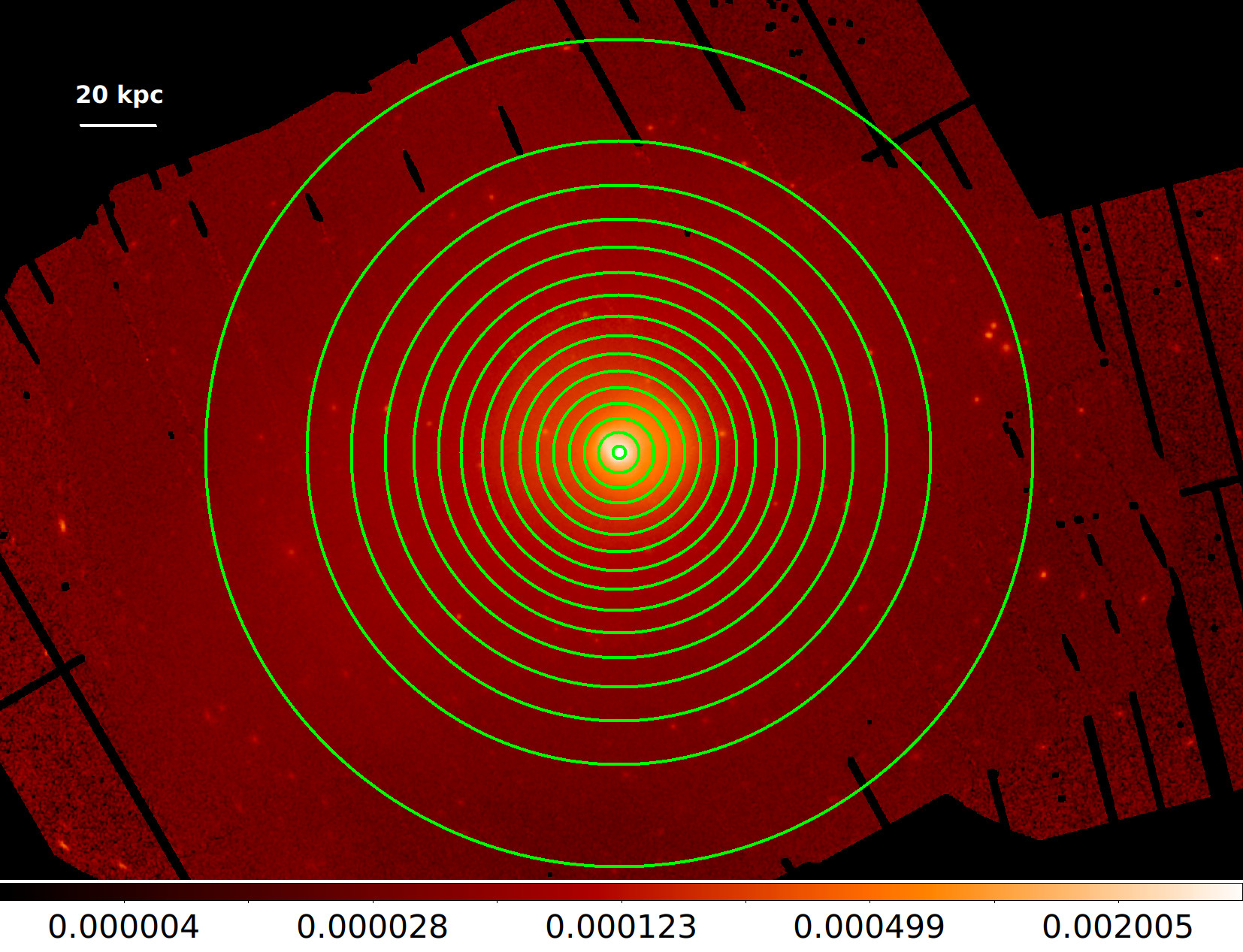}\\
\includegraphics[width=0.46\textwidth]{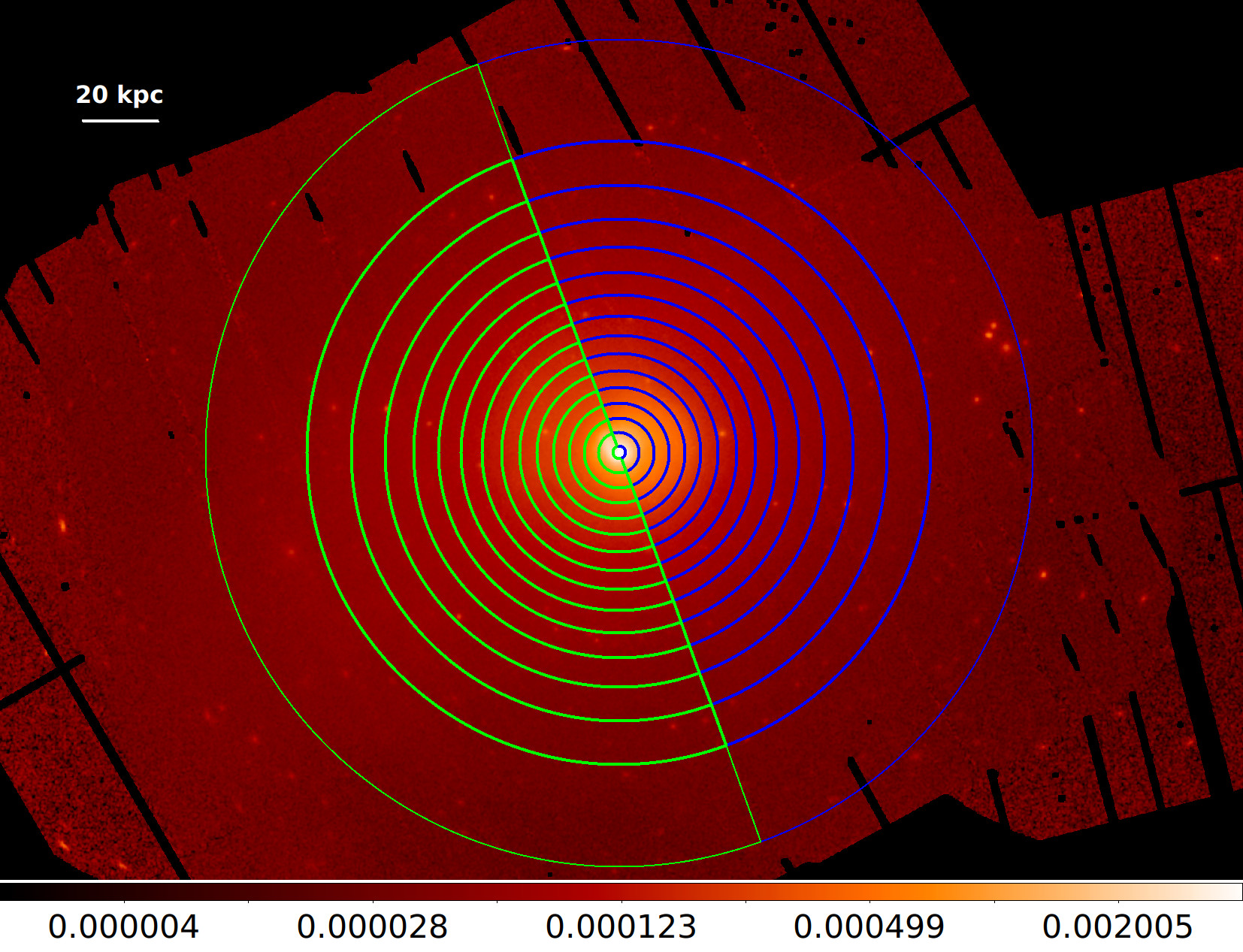}
\caption{Centaurus cluster extracted regions for Case 1 (top panel) and Case 2 (bottom panel). } \label{fig_cas1_region} 
    \end{figure}

\begin{figure} 
\centering 
\includegraphics[width=0.49\textwidth]{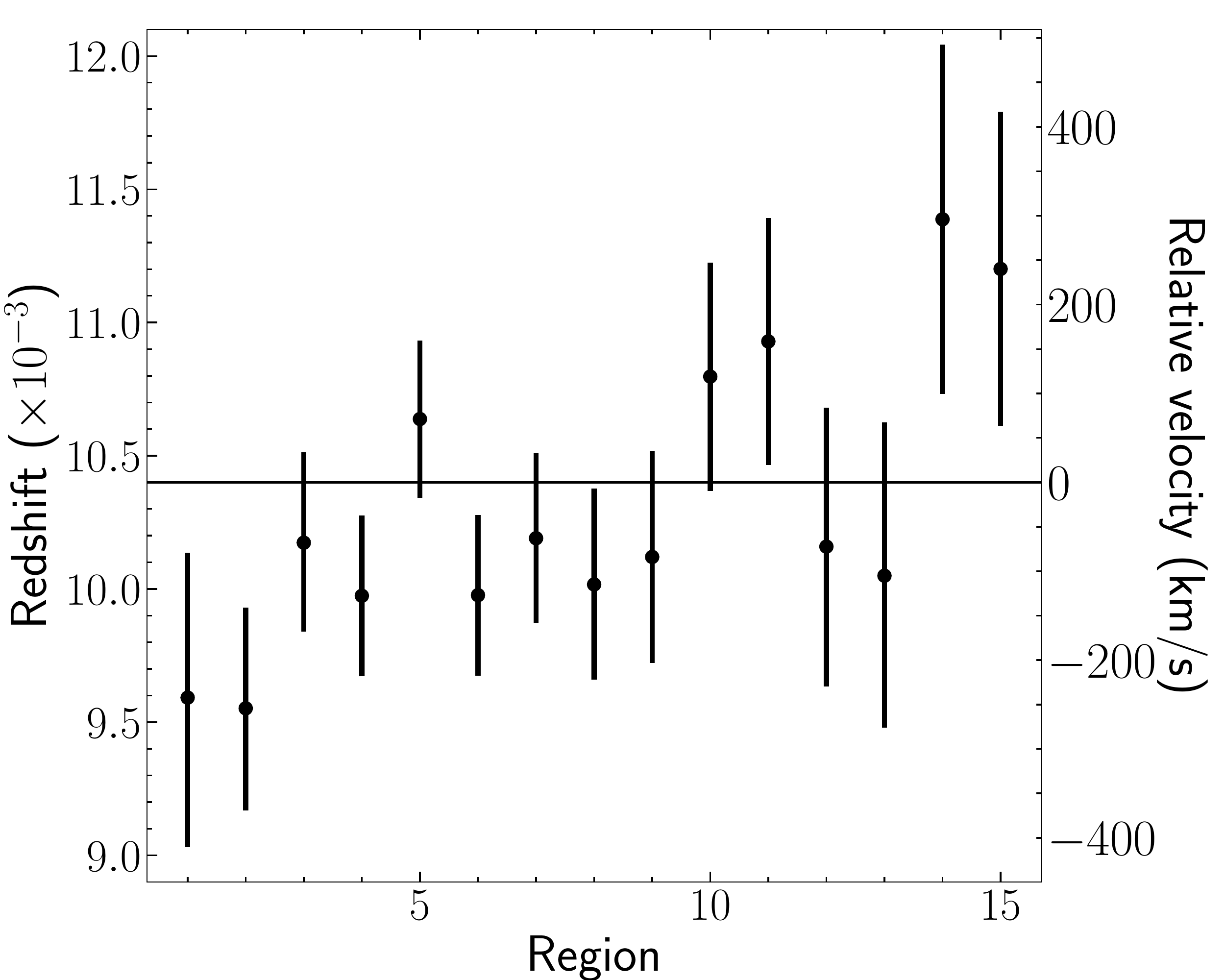} 
        \caption{Velocities obtained for each region for Case 1 (numbered from the center to the outside). The Centaurus redshift is indicated with a horizontal line.} \label{fig_cas1_resultsa} 
\end{figure}

\begin{figure} 
\centering  
\includegraphics[width=0.49\textwidth]{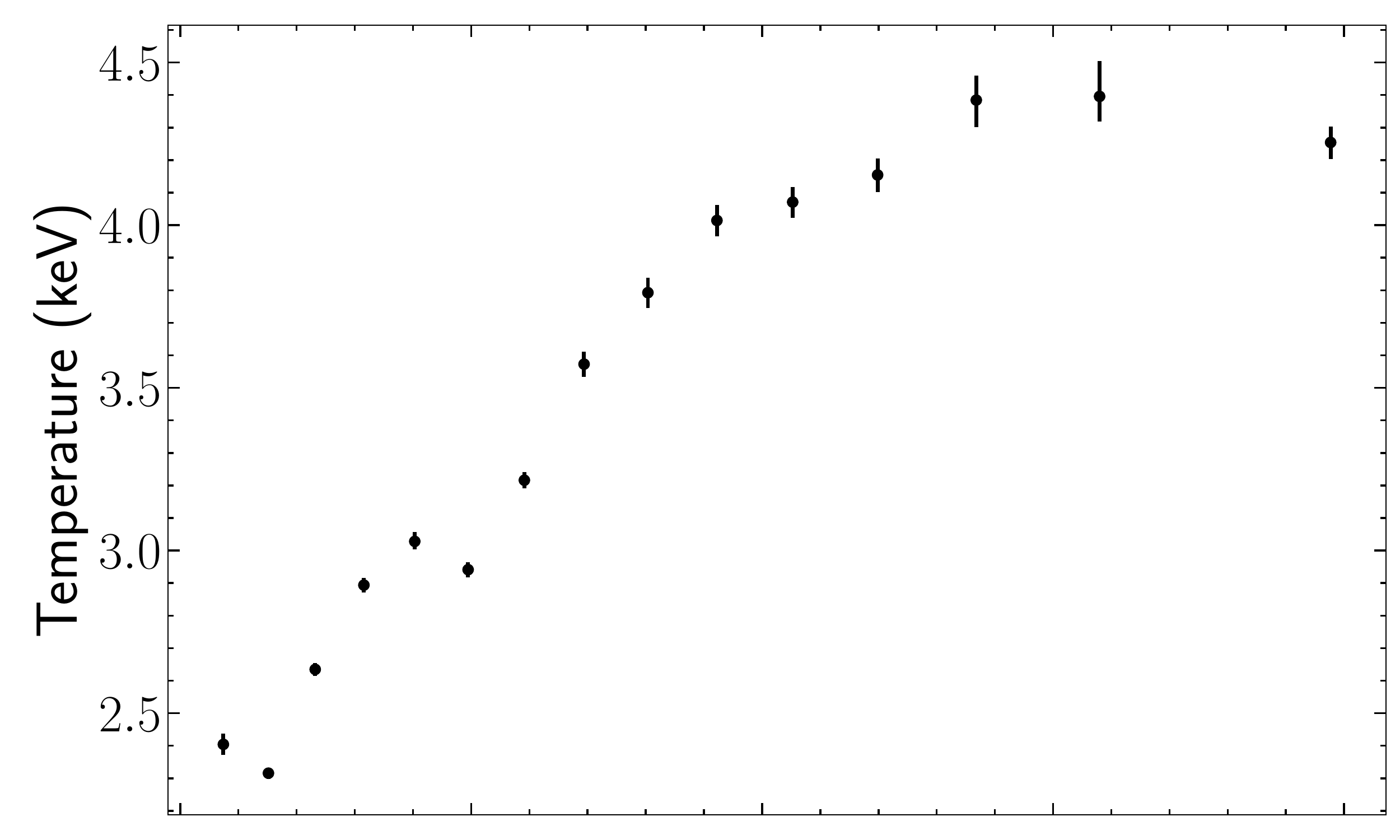}\\
\includegraphics[width=0.49\textwidth]{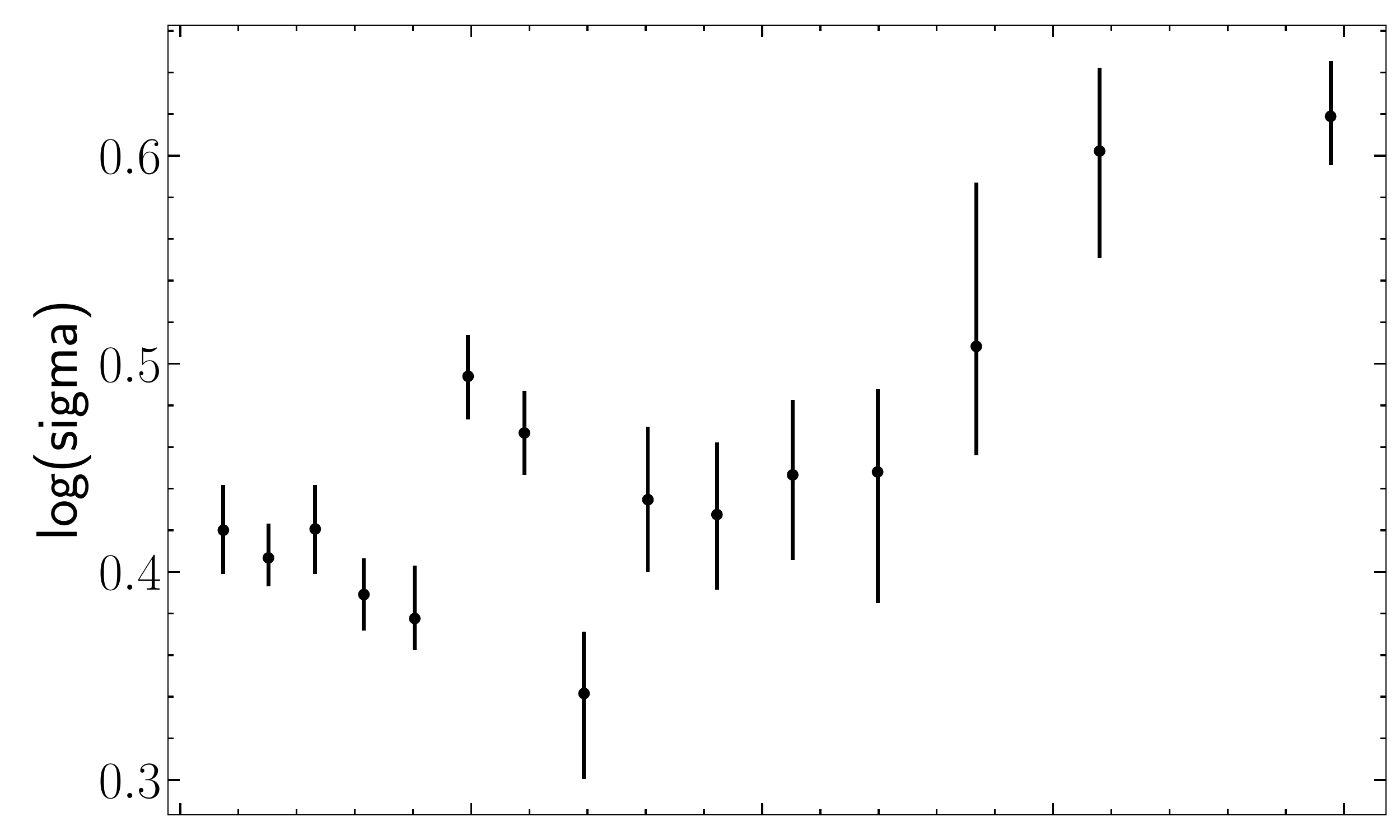}\\
\includegraphics[width=0.49\textwidth]{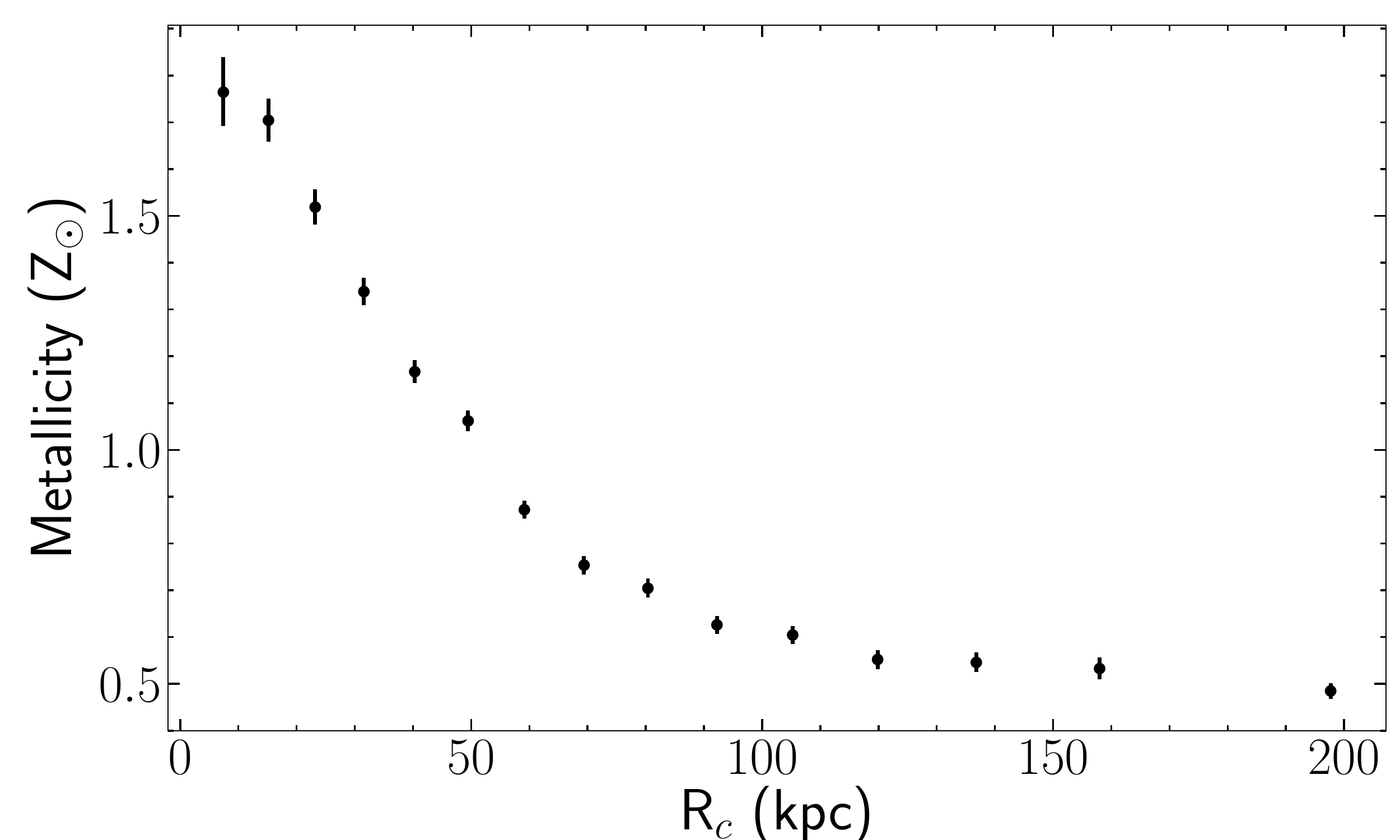} 
        \caption{Temperature (top panel), log(sigma) (middle panel) and metallicity (bottom panel) profiles obtained from the best fit results for Case 1. } \label{fig_cas1_resultsb} 
\end{figure}

\begin{figure} 
\centering 
\includegraphics[width=0.49\textwidth]{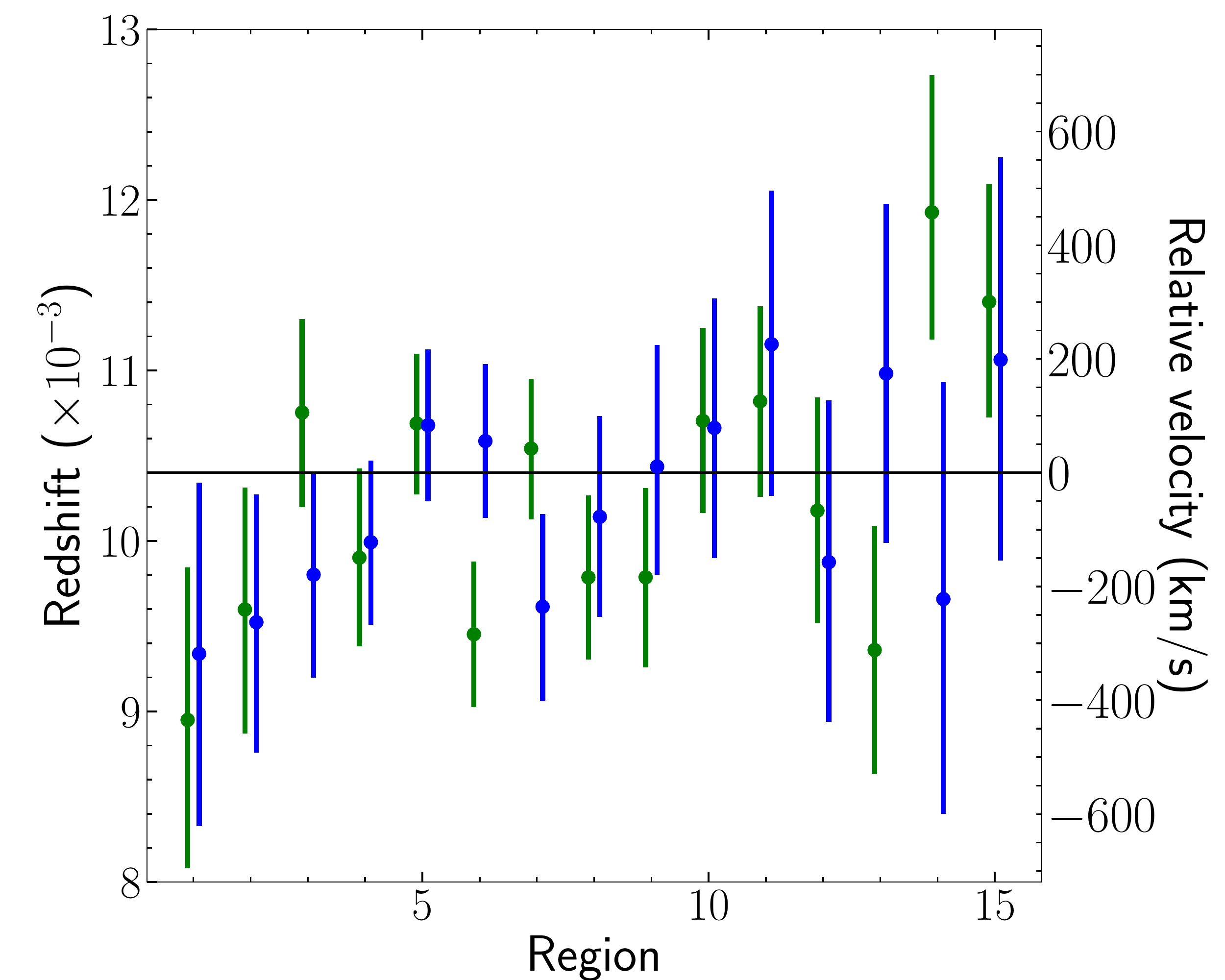} 
        \caption{Velocities obtained for each region for Case 2 (numbered from the center to the outside). The Centaurus redshift is indicated with an horizontal line. Green points correspond to results obtained for regions in the E direction while blue points correspond to regions in the W direction.} \label{fig_cas2_resultsa} 
\end{figure}

\begin{figure} 
\centering  
\includegraphics[width=0.49\textwidth]{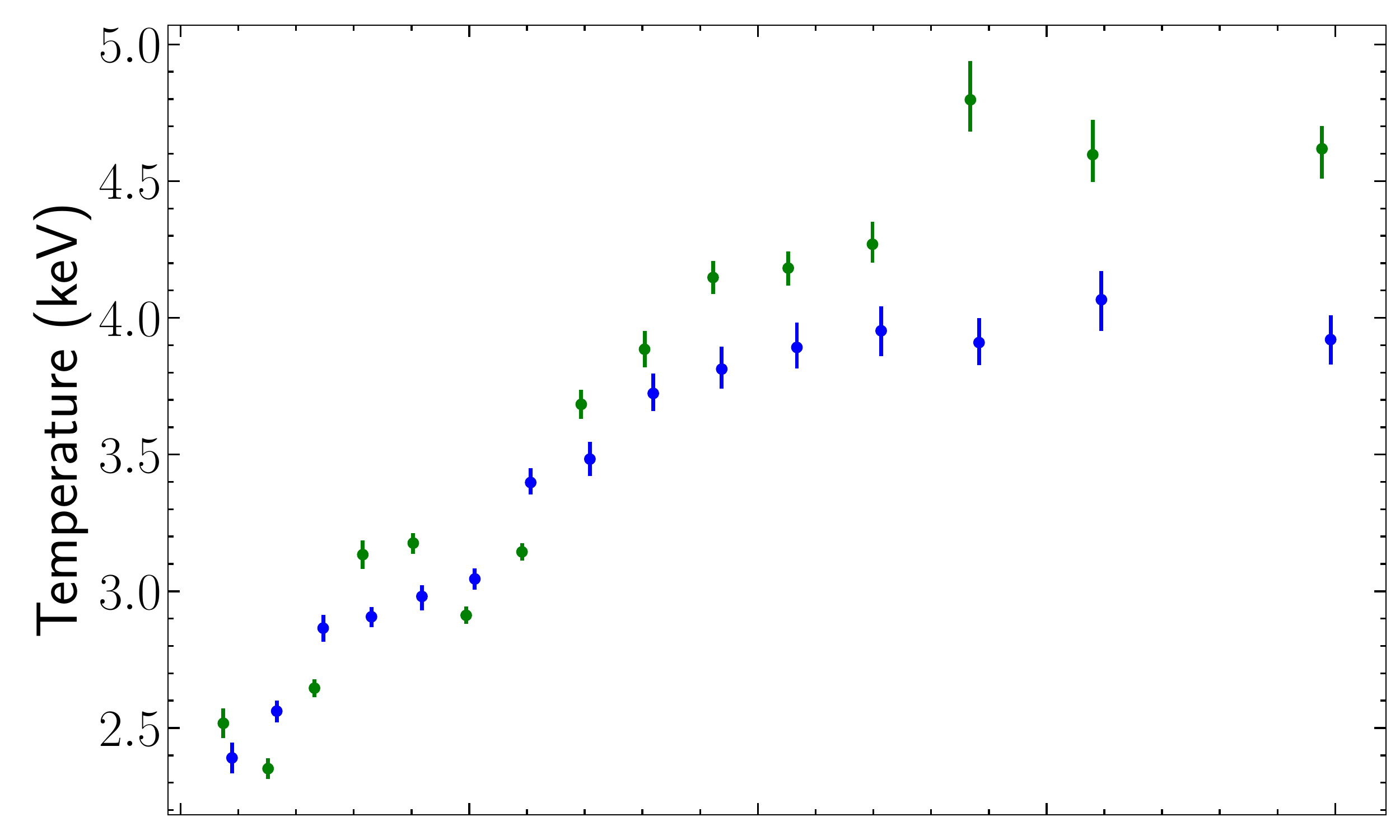}\\
\includegraphics[width=0.49\textwidth]{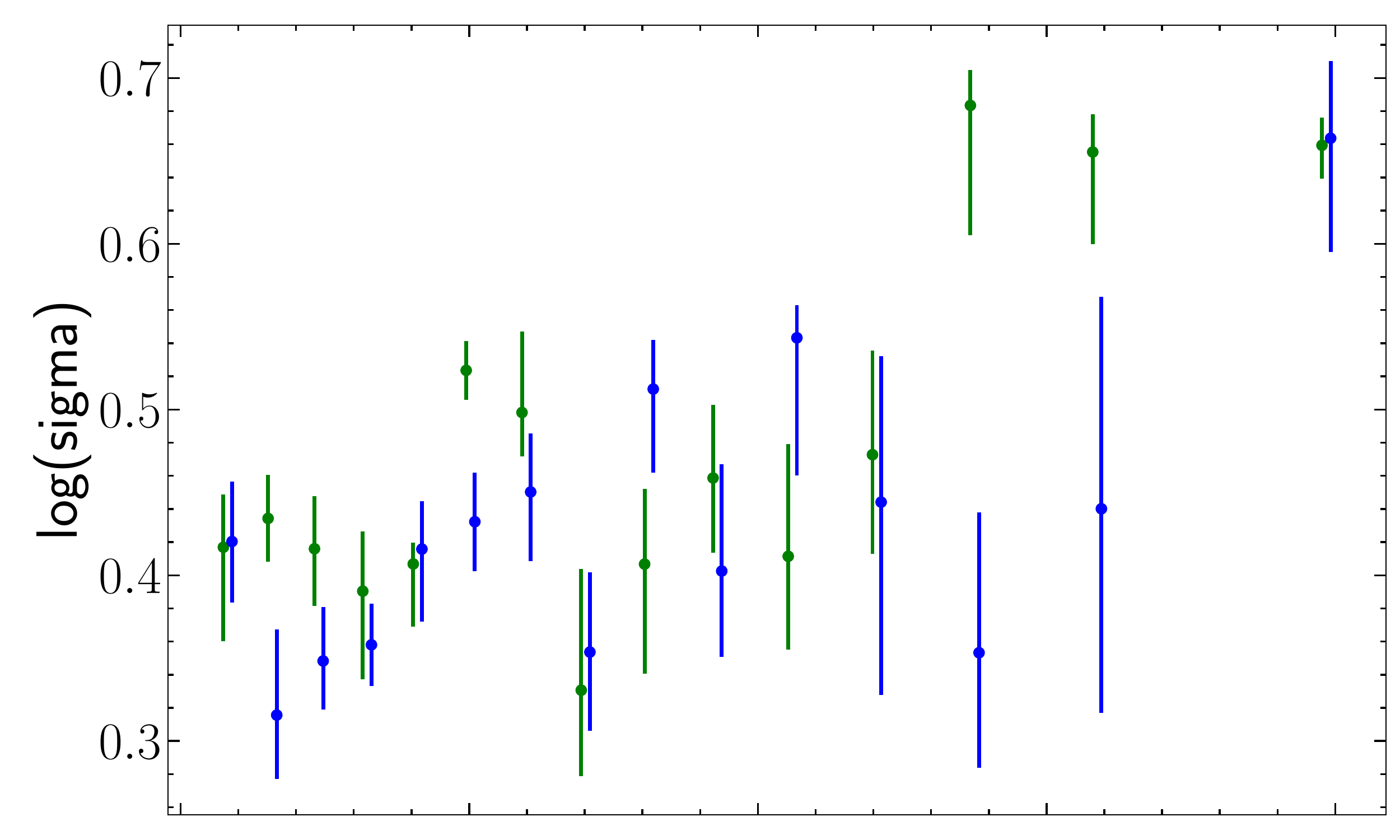}\\
\includegraphics[width=0.49\textwidth]{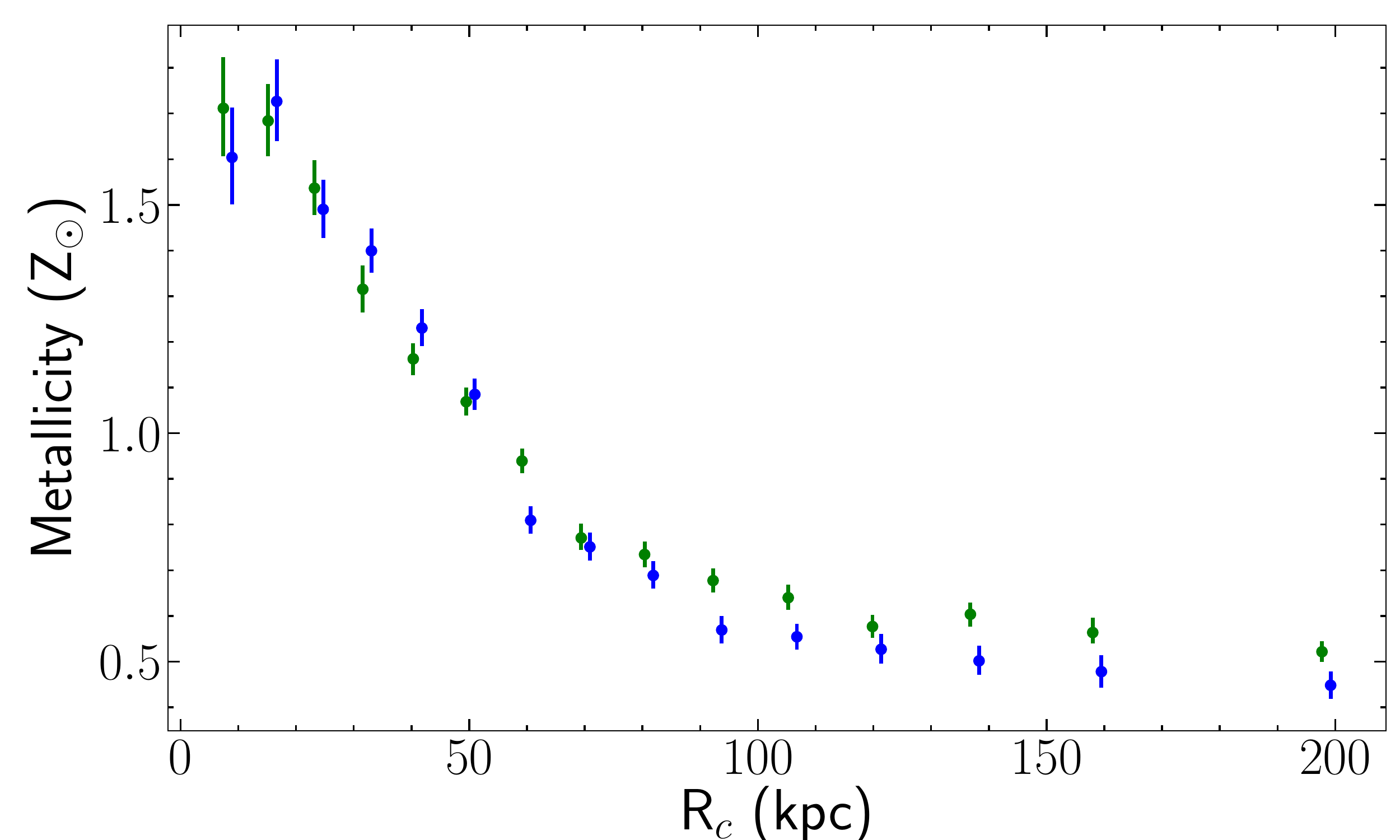} 
        \caption{Temperature (top panel), log(sigma) (middle panel) and metallicity (bottom panel) profiles obtained from the best fit results for Case 2. Green points correspond to results obtained for regions in the E direction while blue points correspond to regions in the W direction.} \label{fig_cas2_resultsb} 
\end{figure}

 \begin{table*}
\scriptsize 
\caption{\label{tab_xmm_proposal_half}Centaurus cluster best-fit parameters for the manually selected substructures (See Section~\ref{regions_xmm_proposal}). }
\centering
\begin{tabular}{ccccccc}
\\
Region &\multicolumn{6}{c}{{\tt lognorm} model}  \\
\hline
 &$kT$&$\sigma$& Z& $z$   & $norm$   & cstat/dof\\ 
  & & & &  ($\times 10^{-3}$) &   ($\times 10^{-3}$)  \\ 
1&$2.48\pm 0.03$&$0.39\pm 0.03$&$1.52\pm 0.07$&$11.17\pm 0.60$&$35.82\pm 0.68$&$1769/1617$\\
2&$2.93\pm 0.08$&$0.50\pm 0.05$&$1.89\pm 0.14$&$9.76\pm 0.88$&$10.49\pm 0.34$&$1417/1504$\\
3&$2.32\pm 0.05$&$0.43\pm 0.03$&$1.76\pm 0.10$&$9.65\pm 0.83$&$9.30\pm 0.25$&$1861/1687$\\
4&$2.97\pm 0.05$&$0.39\pm 0.04$&$1.49\pm 0.07$&$10.50\pm 0.64$&$16.65\pm 0.30$&$1729/1613$\\
5&$2.74\pm 0.04$&$0.38\pm 0.02$&$1.40\pm 0.04$&$10.33\pm 0.41$&$26.59\pm 0.31$&$1888/1677$\\
6&$2.80\pm 0.03$&$0.39\pm 0.03$&$1.31\pm 0.04$&$10.12\pm 0.43$&$61.98\pm 0.68$&$1823/1684$\\
7&$3.32\pm 0.05$&$0.45\pm 0.06$&$1.04\pm 0.05$&$11.27\pm 0.67$&$34.90\pm 0.47$&$1751/1678$\\
8&$3.86\pm 0.08$&$0.37\pm 0.06$&$0.83\pm 0.04$&$10.48\pm 0.67$&$12.92\pm 0.13$&$1689/1684$\\
9&$2.97\pm 0.05$&$0.41\pm 0.03$&$1.40\pm 0.06$&$10.08\pm 0.55$&$14.61\pm 0.22$&$1774/1671$\\
10&$2.98\pm 0.04$&$0.47\pm 0.03$&$1.19\pm 0.04$&$10.69\pm 0.47$&$16.74\pm 0.19$&$2001/1683$\\
11&$4.02\pm 0.11$&$0.56_{-0.09}^{+0.05}$&$0.85\pm 0.05$&$12.90\pm 0.99$&$31.14\pm 0.40$&$1804/1688$\\
12&$3.13\pm 0.03$&$0.44\pm 0.03$&$0.92\pm 0.03$&$10.44\pm 0.40$&$17.69\pm 0.13$&$2002/1687$\\
13&$3.53\pm 0.06$&$0.42\pm 0.06$&$0.67\pm 0.03$&$11.50\pm 0.61$&$13.67\pm 0.11$&$1778/1688$\\
14&$3.98\pm 0.10$&$0.63\pm 0.07$&$0.51\pm 0.03$&$10.5_{-1.24}^{+1.33}$&$12.41\pm 0.16$&$2069/1688$\\
15&$3.77\pm 0.06$&$0.54\pm 0.04$&$0.44\pm 0.02$&$12.54_{-0.83}^{+0.89}$&$13.59\pm 0.12$&$2067/1688$\\
16&$4.20\pm 0.09$&$0.59_{-0.10}^{+0.04}$&$0.53\pm 0.03$&$11.03\pm 0.93$&$9.45\pm 0.08$&$1970/1688$\\
17&$3.83\pm 0.12$&$0.54_{-0.02}^{+0.10}$&$0.45\pm 0.03$&$16.27_{-1.72}^{+1.88}$&$10.05_{-0.12}^{+0.17}$&$2275/1688$\\
18&$4.41\pm 0.15$&$0.43\pm 0.10$&$0.55\pm 0.04$&$9.94\pm 1.14$&$10.78\pm 0.13$&$1809/1687$\\
19&$3.69\pm 0.06$&$0.51\pm 0.04$&$0.93\pm 0.04$&$10.09\pm 0.55$&$22.43\pm 0.21$&$1893/1688$\\
20&$3.16\pm 0.03$&$0.58\pm 0.02$&$1.00\pm 0.03$&$10.83\pm 0.41$&$34.74\pm 0.29$&$1997/1688$\\
21&$4.16\pm 0.08$&$0.59\pm 0.03$&$0.68\pm 0.03$&$10.15_{-0.64}^{+0.70}$&$14.77\pm 0.12$&$2029/1688$\\
22&$4.55\pm 0.11$&$0.44_{-0.05}^{+0.11}$&$0.54\pm 0.03$&$11.12\pm 0.85$&$16.69\pm 0.15$&$1979/1688$\\
23&$4.48_{-0.12}^{+0.18}$&$0.71_{-0.05}^{+0.10}$&$0.48\pm 0.03$&$12.72\pm 1.03$&$21.33_{-0.22}^{+0.34}$&$2213/1688$\\
\\ 
 \hline
\end{tabular}
\end{table*}

\begin{figure} 
\centering 
\includegraphics[width=0.49\textwidth]{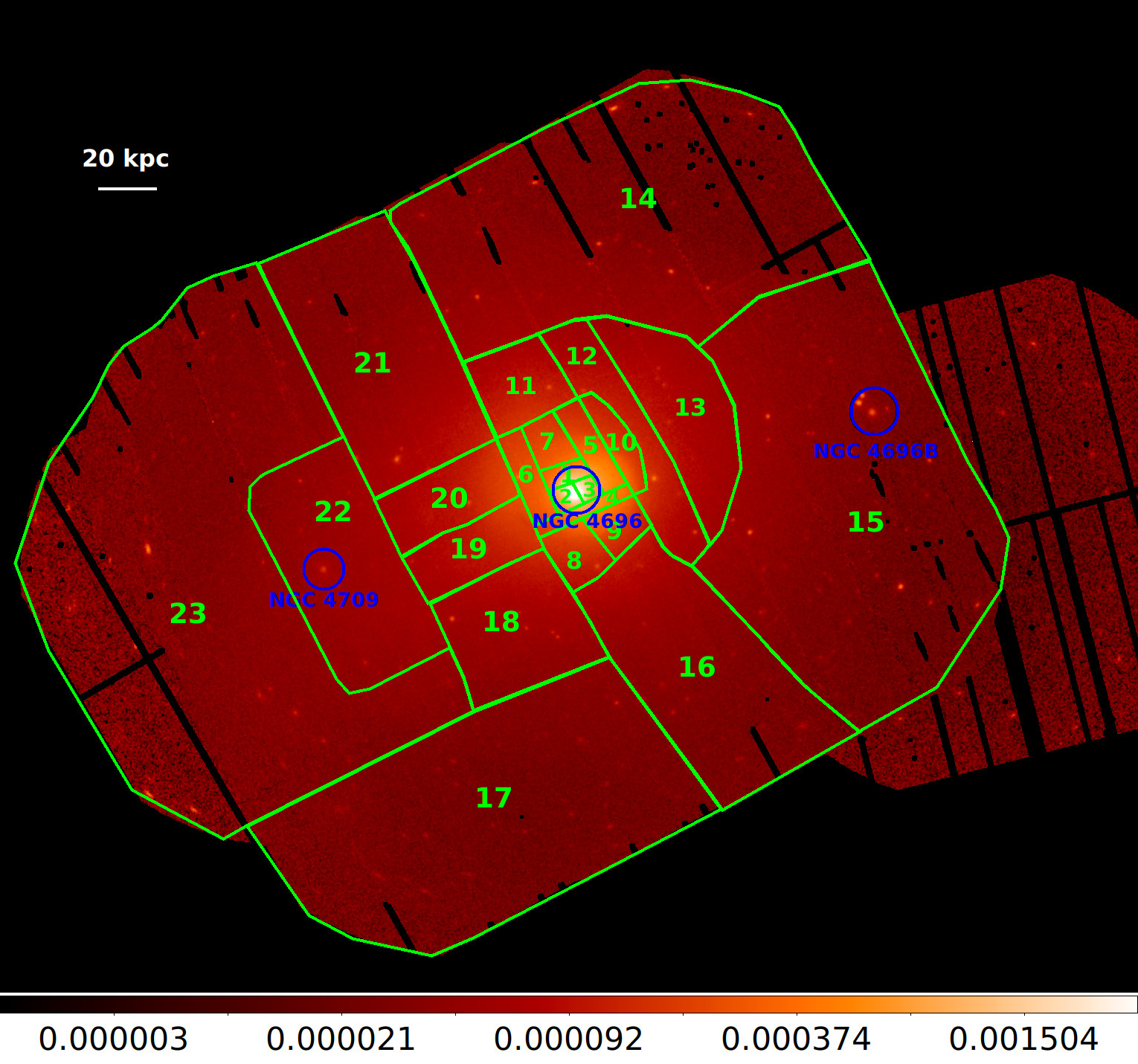}\\ 
\includegraphics[width=0.49\textwidth]{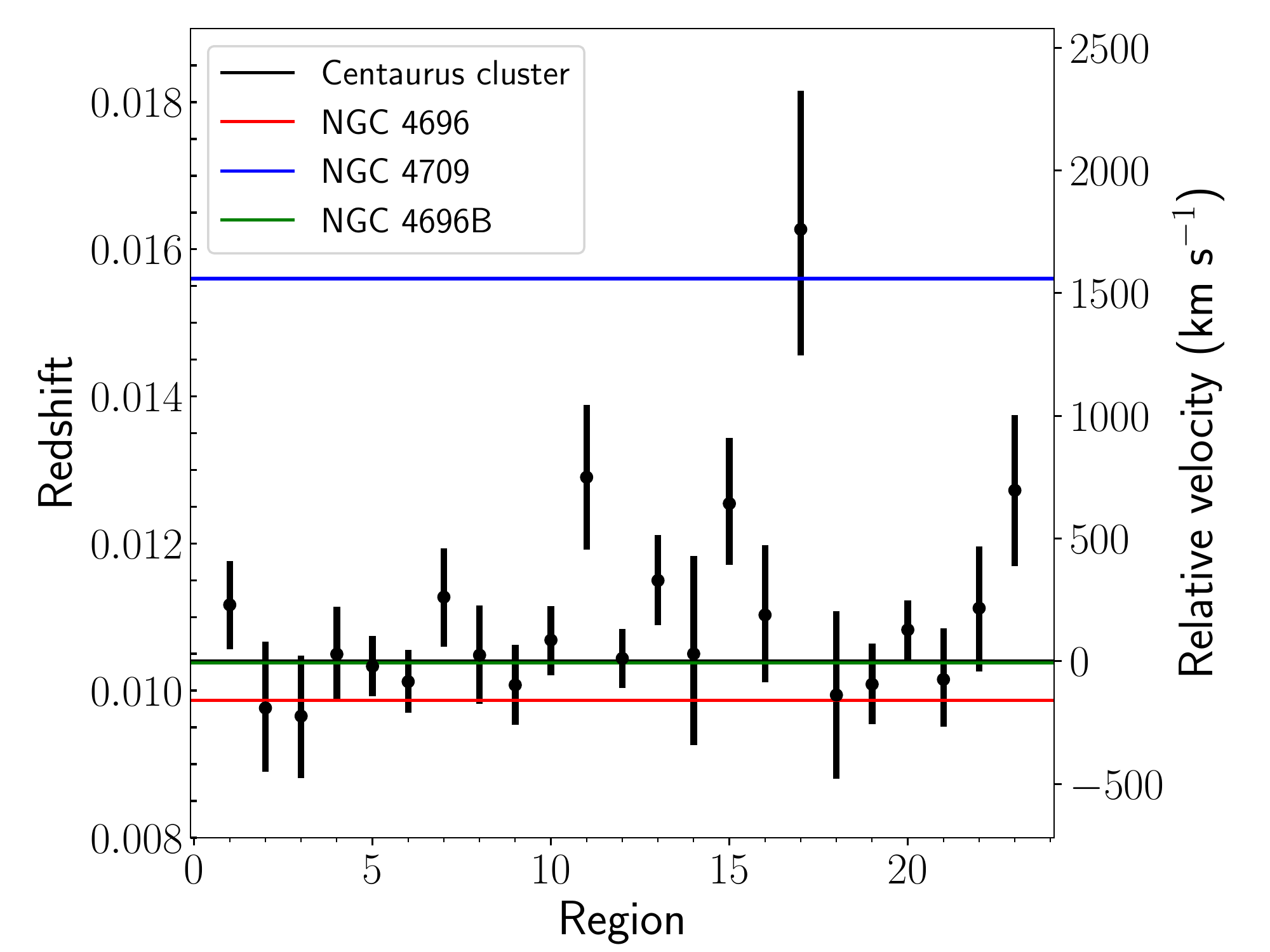}
        \caption{\emph{Top panel:} Centaurus cluster extracted regions for manually selected substructures. The location of the main galaxies, NGC~4696, NGC~4709 and NGC~4696B, are also indicated with blue circles. \emph{Bottom panel:} velocities obtained for each region for Case 2. The Centaurus redshift is indicated with an horizontal line (See Section~\ref{regions_xmm_proposal}). } \label{fig_xmm_proposal_half1} 
\end{figure}

\begin{figure} 
\centering 
\includegraphics[width=0.49\textwidth]{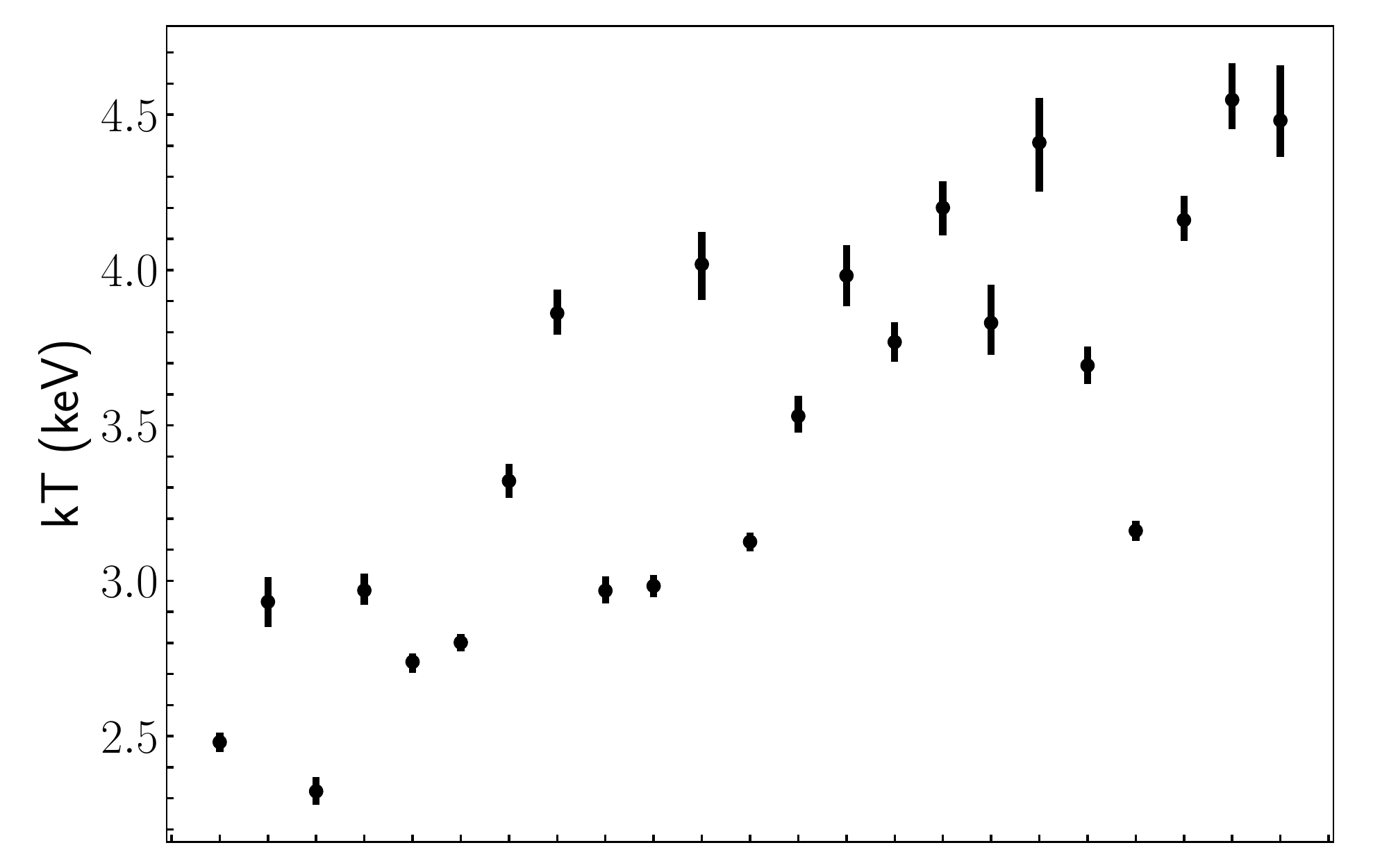}\\
\includegraphics[width=0.49\textwidth]{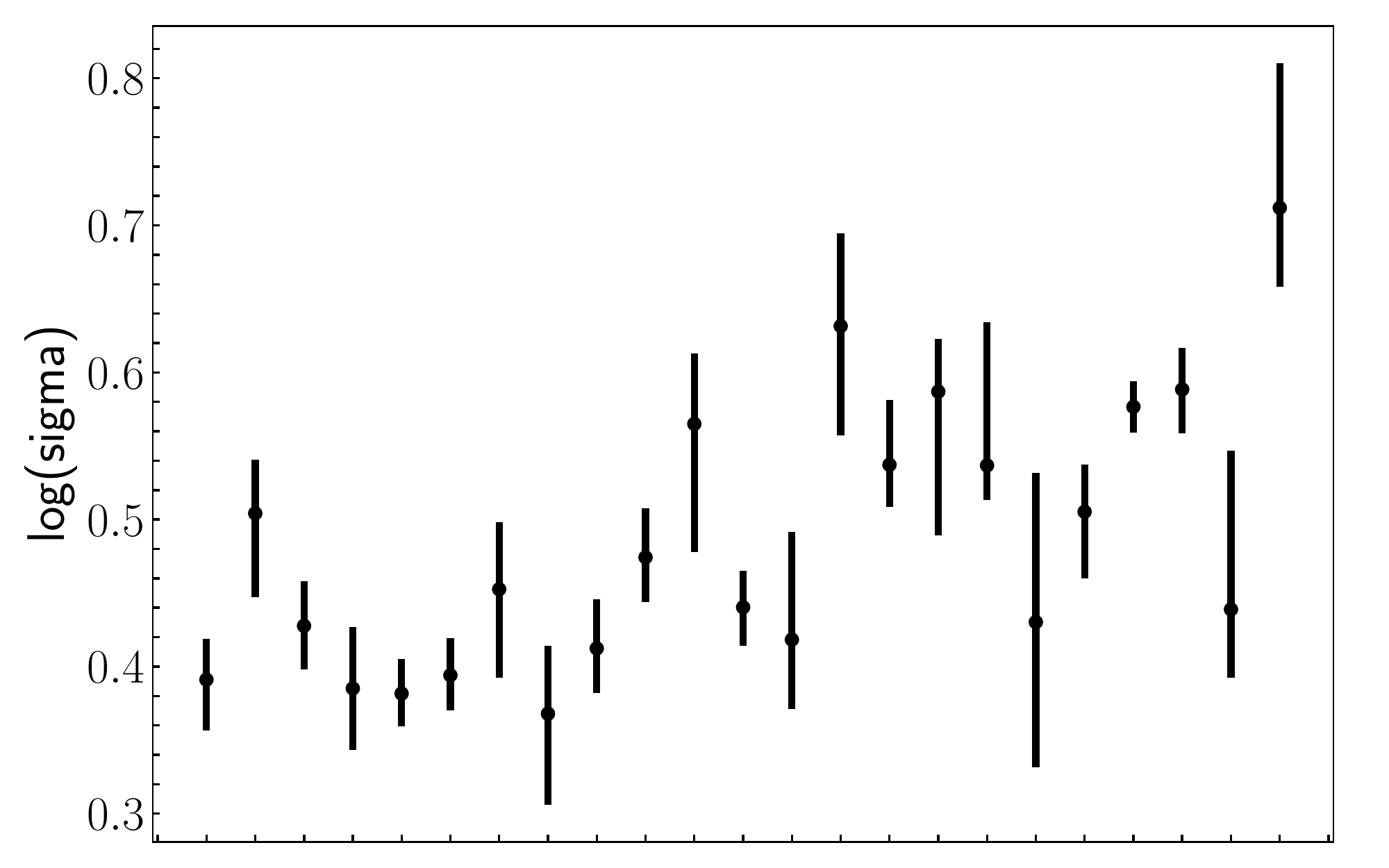}\\
\includegraphics[width=0.49\textwidth]{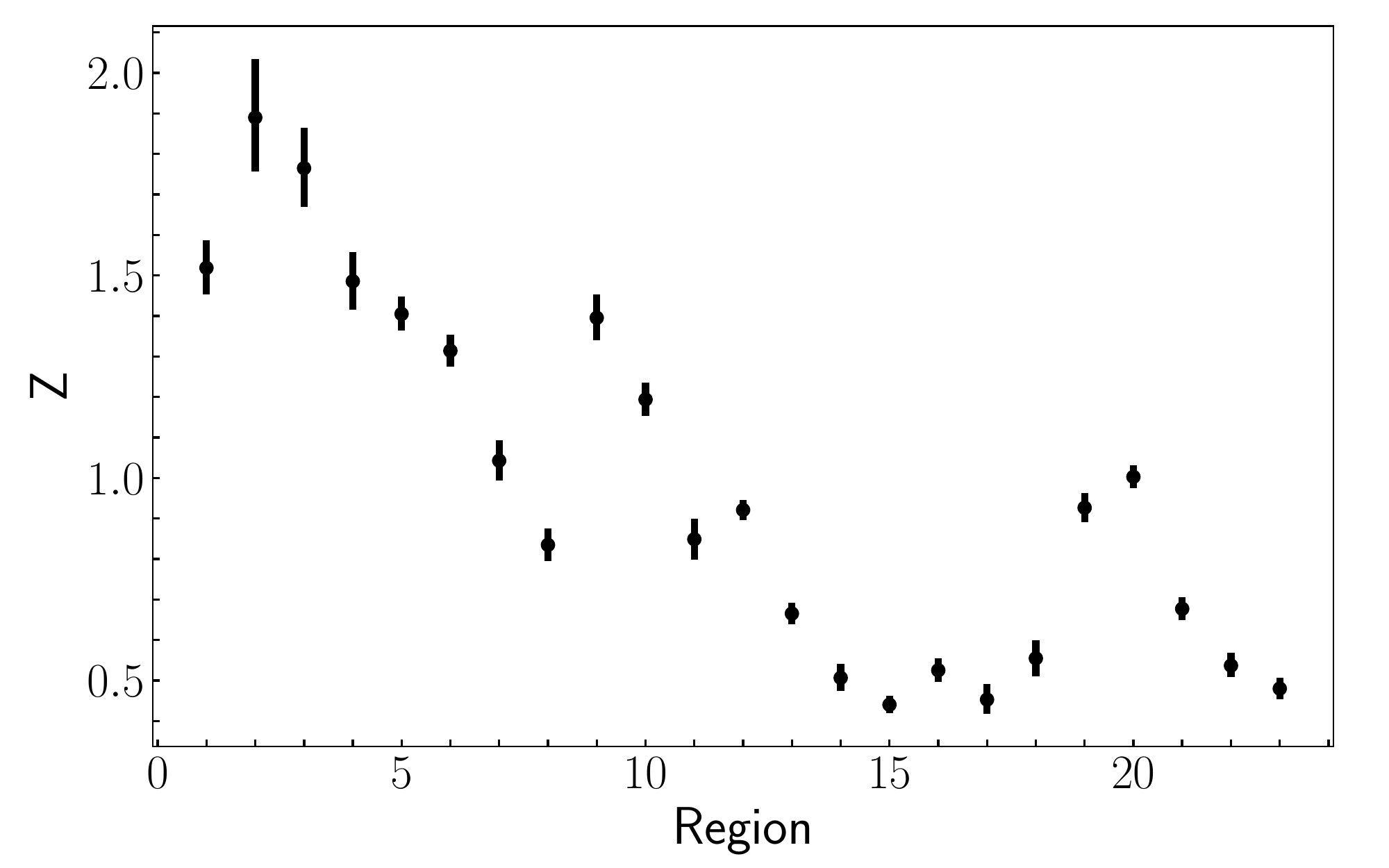}
        \caption{\emph{Top panel:} best-fit temperatures obtained for manually selected substructures. \emph{Middle panel:} best-fit log(sigma) values obtained. \emph{Bottom panel:} best-fit metallicities obtained (See Section~\ref{regions_xmm_proposal}). } \label{fig_xmm_proposal_half2} 
\end{figure}

\begin{table*}
\scriptsize 
\caption{\label{tab_coldfront}Centaurus cluster best-fit parameters for the the cold fronts. }
\centering
\begin{tabular}{ccccccc}
\\
Region &\multicolumn{6}{c}{{\tt lognorm} model}  \\
\hline
 &$kT$&$\sigma$& Z& $z$   & $norm$   & cstat/dof\\ 
  & & & &  ($\times 10^{-3}$) &   ($\times 10^{-3}$)  \\  
1&$3.88\pm 0.05$&$0.34\pm 0.05$&$0.79\pm 0.03$&$9.90\pm 0.46$&$16.44\pm 0.11$&$1870/1688$\\
2&$3.04\pm 0.03$&$0.49\pm 0.02$&$1.03\pm 0.02$&$10.26\pm 0.32$&$24.63\pm 0.17$&$2170/1688$\\
3&$2.49\pm 0.03$&$0.44\pm 0.02$&$1.66\pm 0.06$&$10.07\pm 0.53$&$17.58\pm 0.30$&$1861/1688$\\
4&$2.38\pm 0.05$&$0.45\pm 0.03$&$1.73\pm 0.10$&$9.46\pm 0.88$&$9.05\pm 0.25$&$1843/1688$\\
5&$2.87\pm 0.05$&$0.36\pm 0.04$&$1.21\pm 0.06$&$10.46\pm 0.69$&$8.68\pm 0.14$&$1744/1687$\\
6&$2.83\pm 0.03$&$0.42\pm 0.04$&$1.16\pm 0.03$&$10.54\pm 0.40$&$11.51\pm 0.11$&$2076/1688$\\
\\ 
 \hline
\end{tabular}
\end{table*}

\begin{figure} 
\centering 
\includegraphics[width=0.40\textwidth]{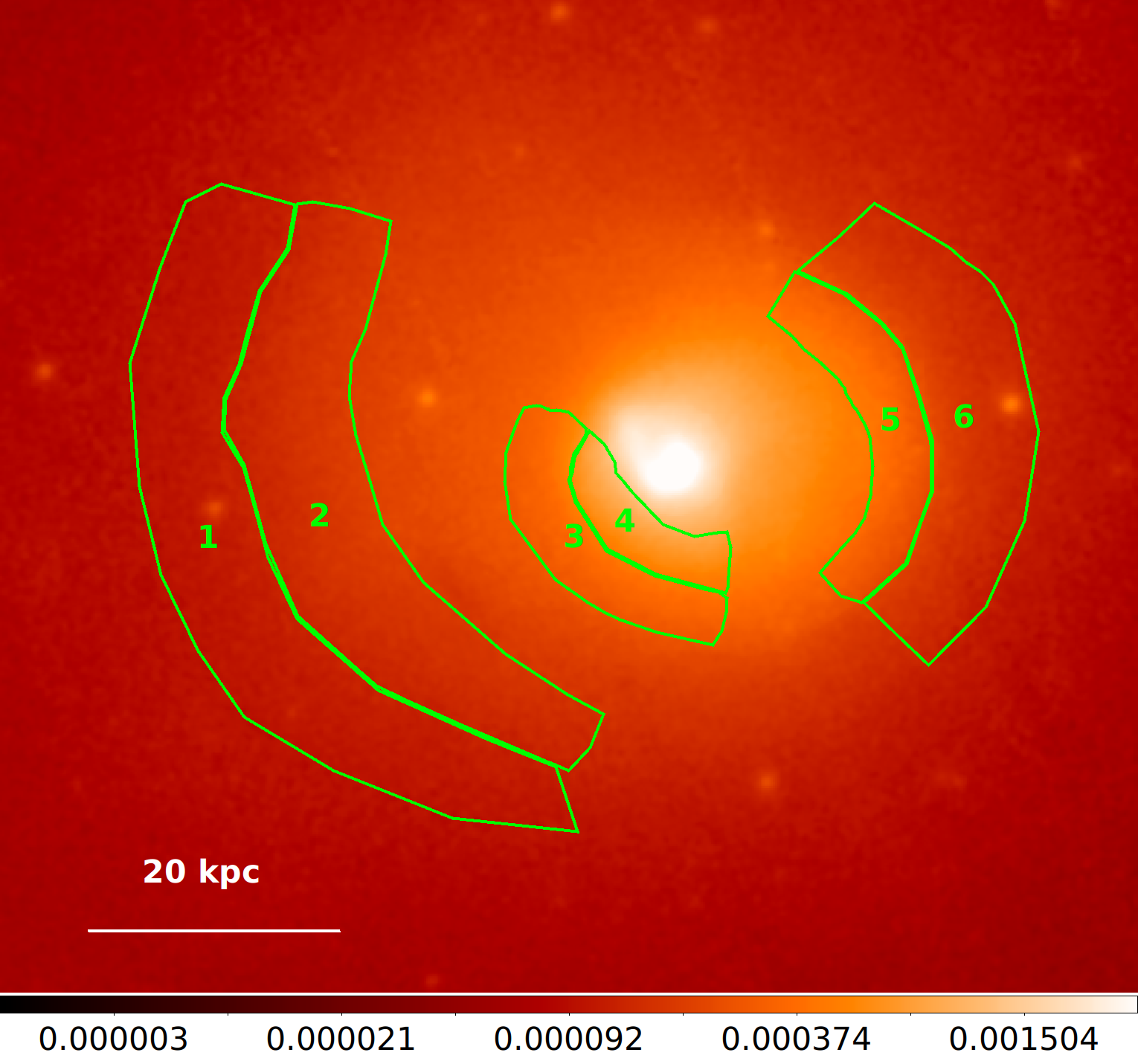}\\
\includegraphics[width=0.45\textwidth]{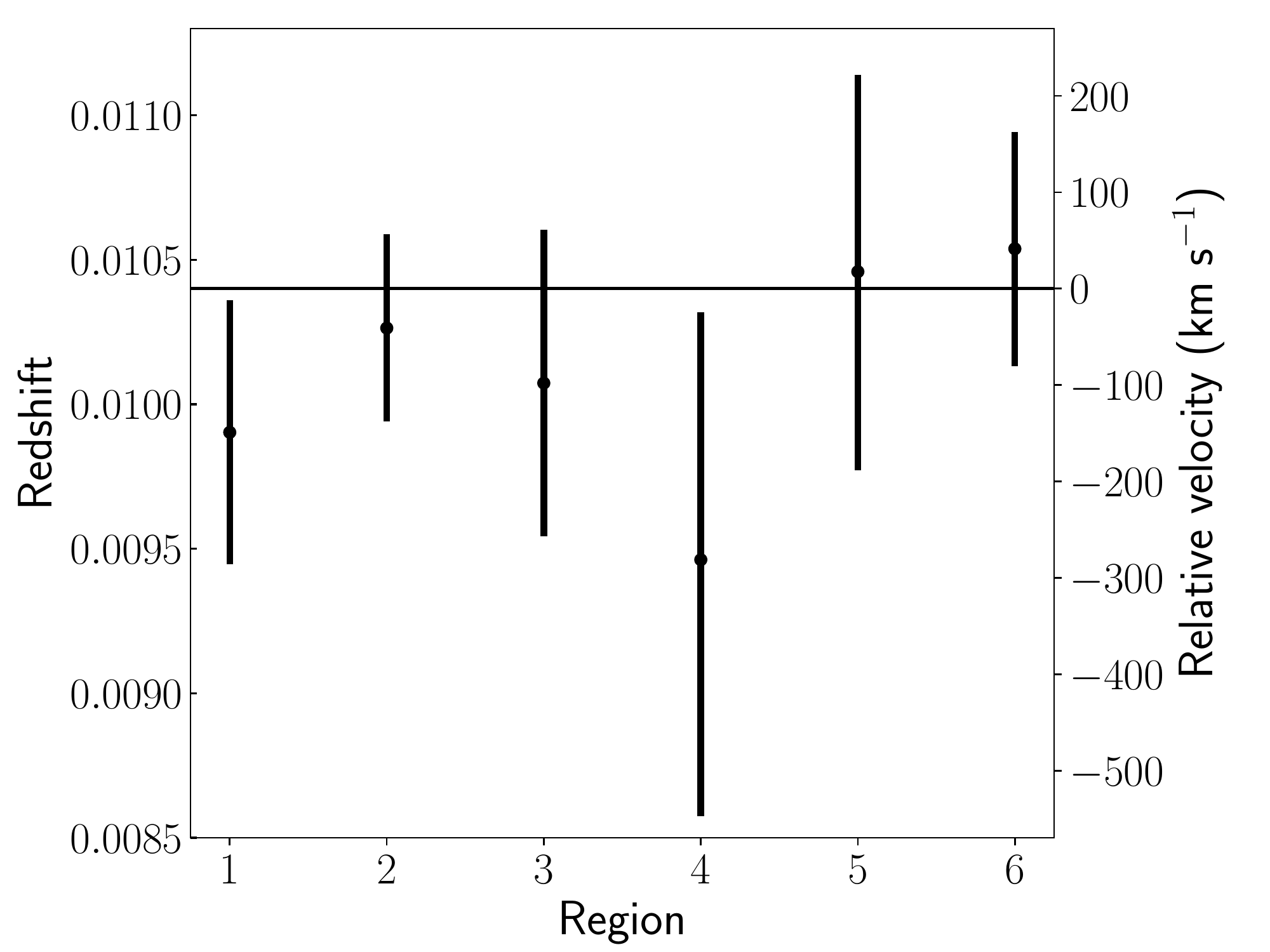}
        \caption{\emph{Top panel:} Centaurus cluster extracted regions for the cold fronts, following the fractional difference in the surface brightness. \emph{Bottom panel:} velocities obtained for each region. The Centaurus redshift is indicated with an horizontal line (See Section~\ref{regions_cold_fronts}).  } \label{fig_cold_frontsa} 
\end{figure}

\begin{figure} 
\centering 
\includegraphics[width=0.49\textwidth]{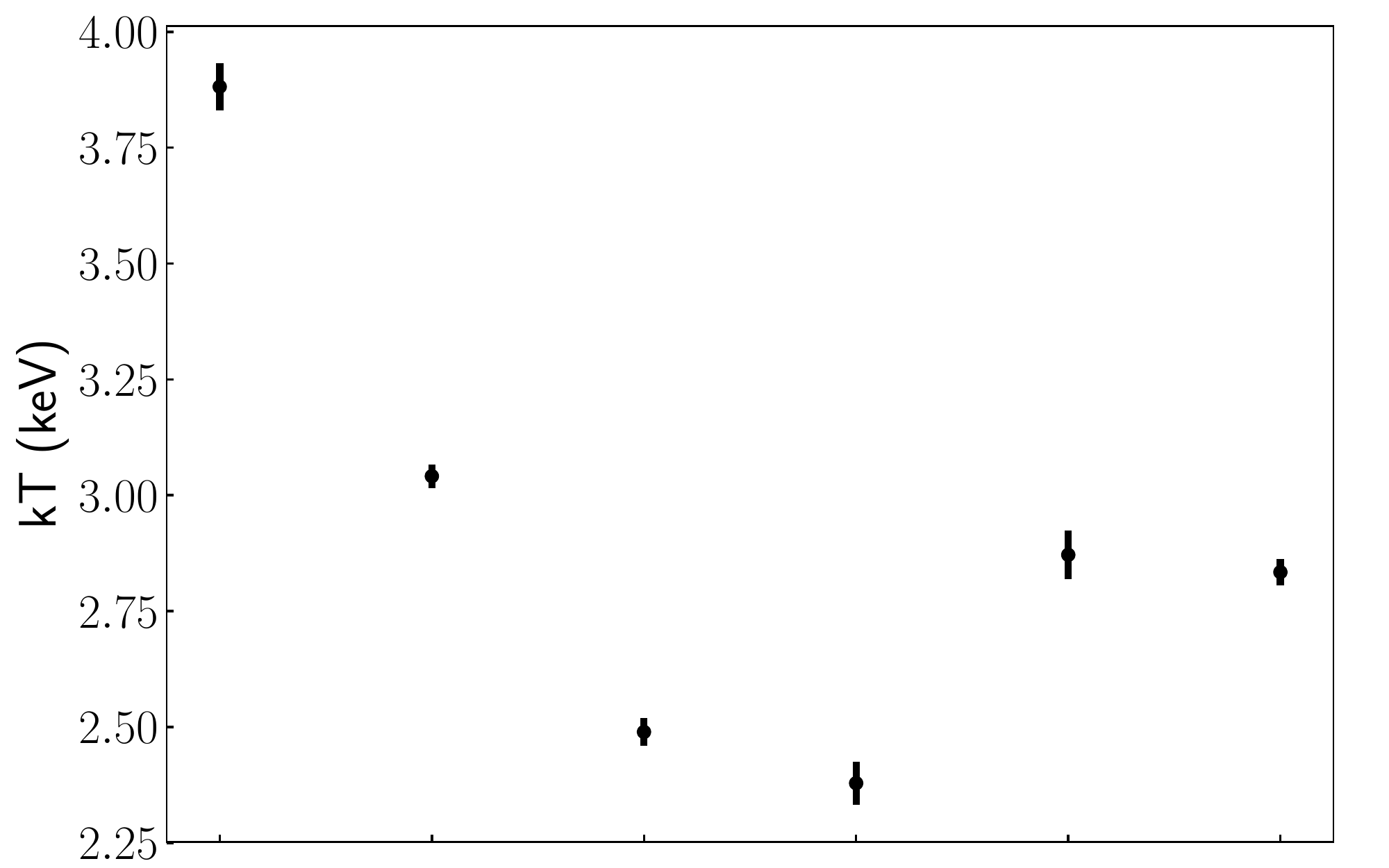}\\
\includegraphics[width=0.49\textwidth]{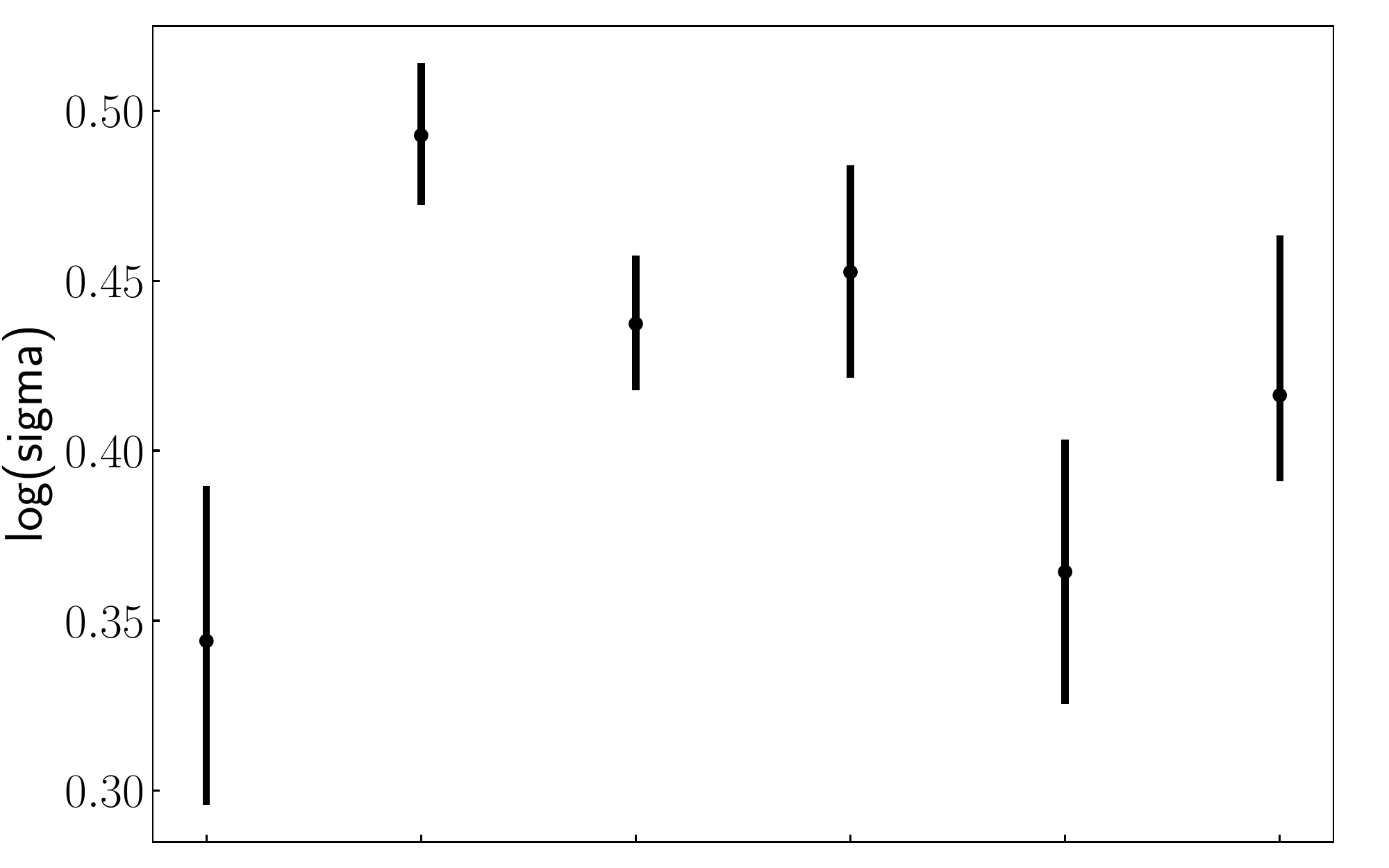}\\
\includegraphics[width=0.49\textwidth]{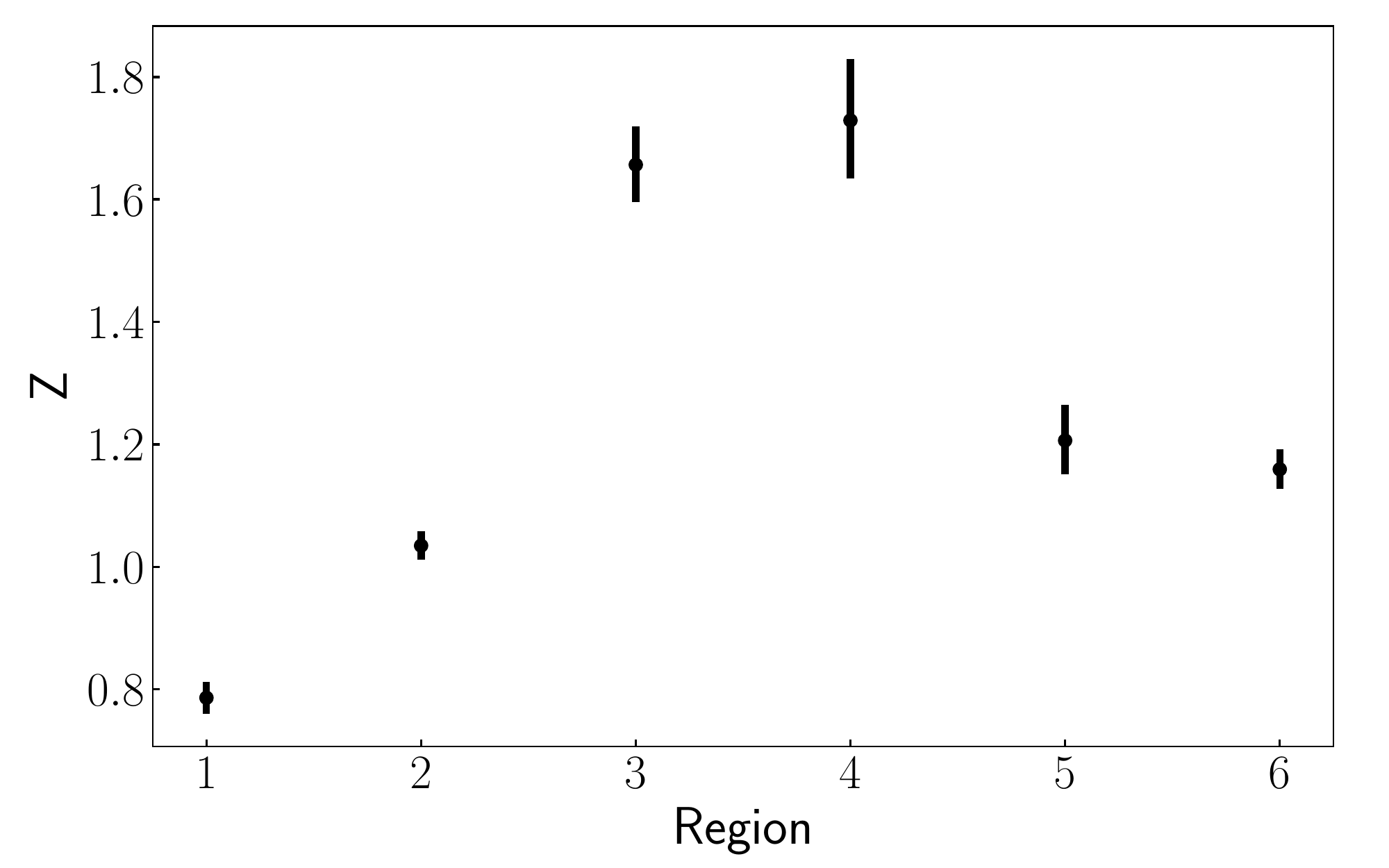}
        \caption{\emph{Top panel:} best-fit temperatures obtained for the cold fronts. \emph{Middle panel:} best-fit log(sigma) values obtained. \emph{Bottom panel:} best-fit metallicities obtained (See Section~\ref{regions_cold_fronts}). } \label{fig_cold_frontsb} 
\end{figure}

Using these spectral maps we calculated projected pseudo-density, pressure and entropy in each spatial bin and assuming a constant line-of-sight depth for all spectral regions. We calculate the pseudo-density as $n\equiv \sqrt{\eta}$, where $\eta$ is the normalization of the {\tt lognorm} model. Then, we estimated pseudo-entropy as $S\equiv n^{-2/3}\times kT$ and pseudo-pressure as $P\equiv n\times kT$ \citep[see][]{hof16}. Figure~\ref{fig_entropy_ellipses} shows the density, pressure and entropy maps created from the best-fit results. We noted that the cluster center displays a high density gas and low entropy, similar to the maps obtained by \citet{san16}. However the region with larger density seems to be displaced in the west direction with respect to the cluster center.

\subsection{Fitting spectra concentric rings}\label{circle_rings}
We studied the velocity structure by analyzing manually extracted regions for two cases:

\begin{itemize}
\item \emph{Case 1:} complete concentric rings, square root spaced and with the center located in the cluster center. 
\item \emph{Case 2:} concentric regions divided in E-W zones (i.e. following the cold fronts location), square root spaced and with the center located in the cluster center.  
\end{itemize}
 
The top panel in Figure~\ref{fig_cas1_region} shows the exact regions analyzed for Case 1. Table~\ref{tab_circular_fits} shows the best-fit results obtained per region. Figures~\ref{fig_cas1_resultsa} and~\ref{fig_cas1_resultsb} show the velocities, temperatures, $\log(\sigma)$ and metallicities obtained from the best-fit per region. We have obtained velocities measurements with uncertainties down to $\Delta v\sim 89$ km/s (for ring 5). The largest blueshift/redshift, with respect to the main cluster redshift, correspond to $-254\pm 114$ km/s and $295\pm 197$ km/s for rings 2 and 14, respectively. Both, the metallicity and temperature profiles, display a discontinuity at $\sim 50$ kpc and at $\sim 100$ kpc. We noted a complex structure in the temperature profile below such discontinuity \citep[similar to results from ][]{wal13a}. Previous analyses of the Centaurus cluster have shown a drop in metallicity near the cluster center within $<10$~kpc \citep[e.g.][]{pan13,lak19}. However our spectra analysis does not include the soft energy band, thus it could be overestimated.

The bottom panel in Figure~\ref{fig_cas1_region} shows the exact regions analyzed for Case 2. Figures~\ref{fig_cas2_resultsa} and~\ref{fig_cas2_resultsb} show the velocities, temperatures, $\log(\sigma)$ and metallicities obtained from the best-fit per region. In the plot green points correspond to results obtained with the E direction while blue points correspond to W direction. We have obtained velocities measurements with uncertainties down to $\Delta v\sim 123$ km/s (for ring 7, east direction). The largest blueshift/redshift, with respect to the main cluster redshift, correspond to $-434\pm 265$ km/s for ring 1 (east direction) and $457\pm 232$ km/s for ring 15 (west direction).  Both, the metallicity and temperature profiles, display a discontinuity at $\sim 30$ kpc. We noted a difference in the metallicity profiles when comparing the outer region (i.e $> 50$ kpc), with larger values in the E direction compared to those rings in the W direction.

\begin{figure} 
\centering 
\includegraphics[width=0.40\textwidth]{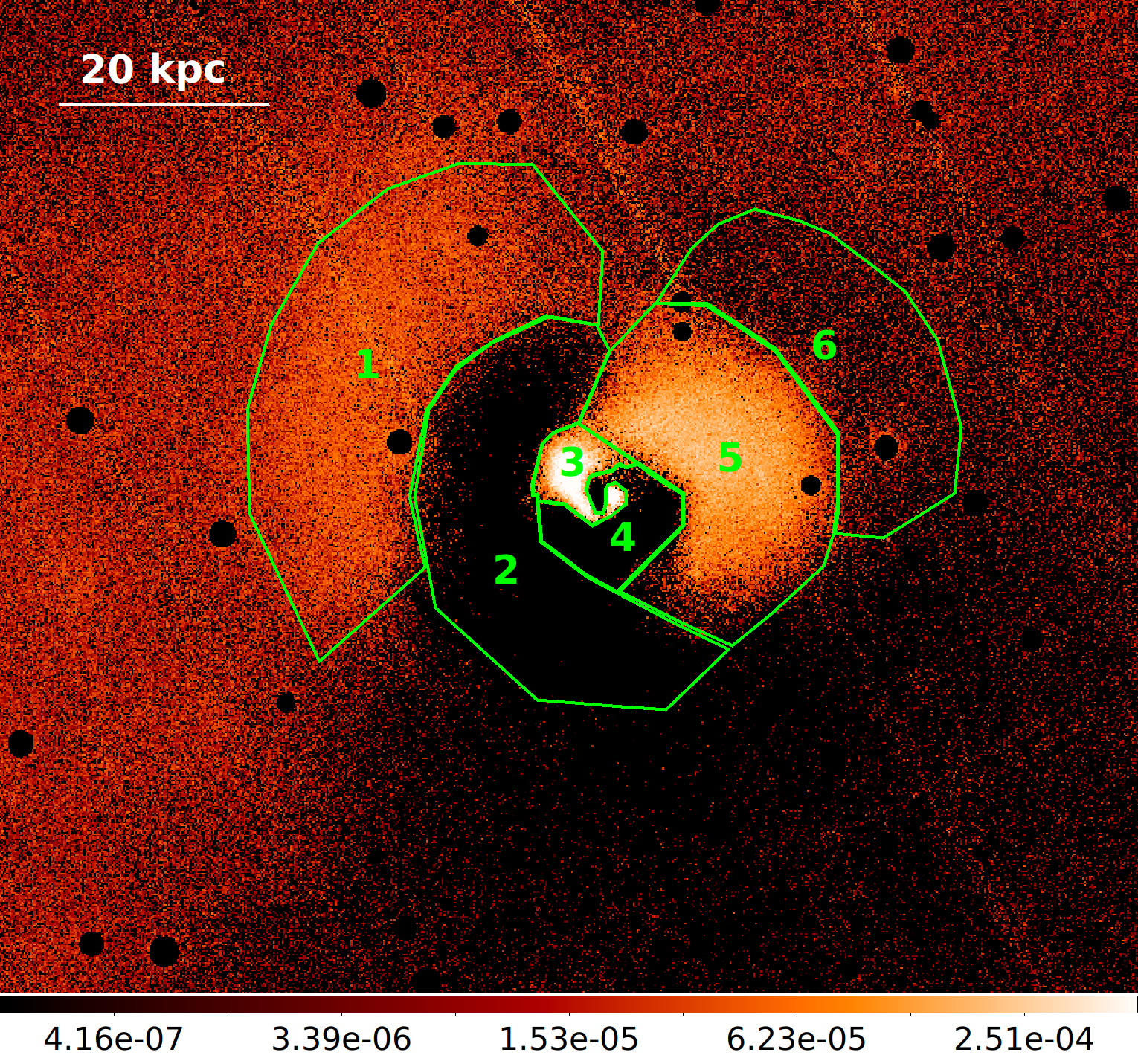}\\
\includegraphics[width=0.45\textwidth]{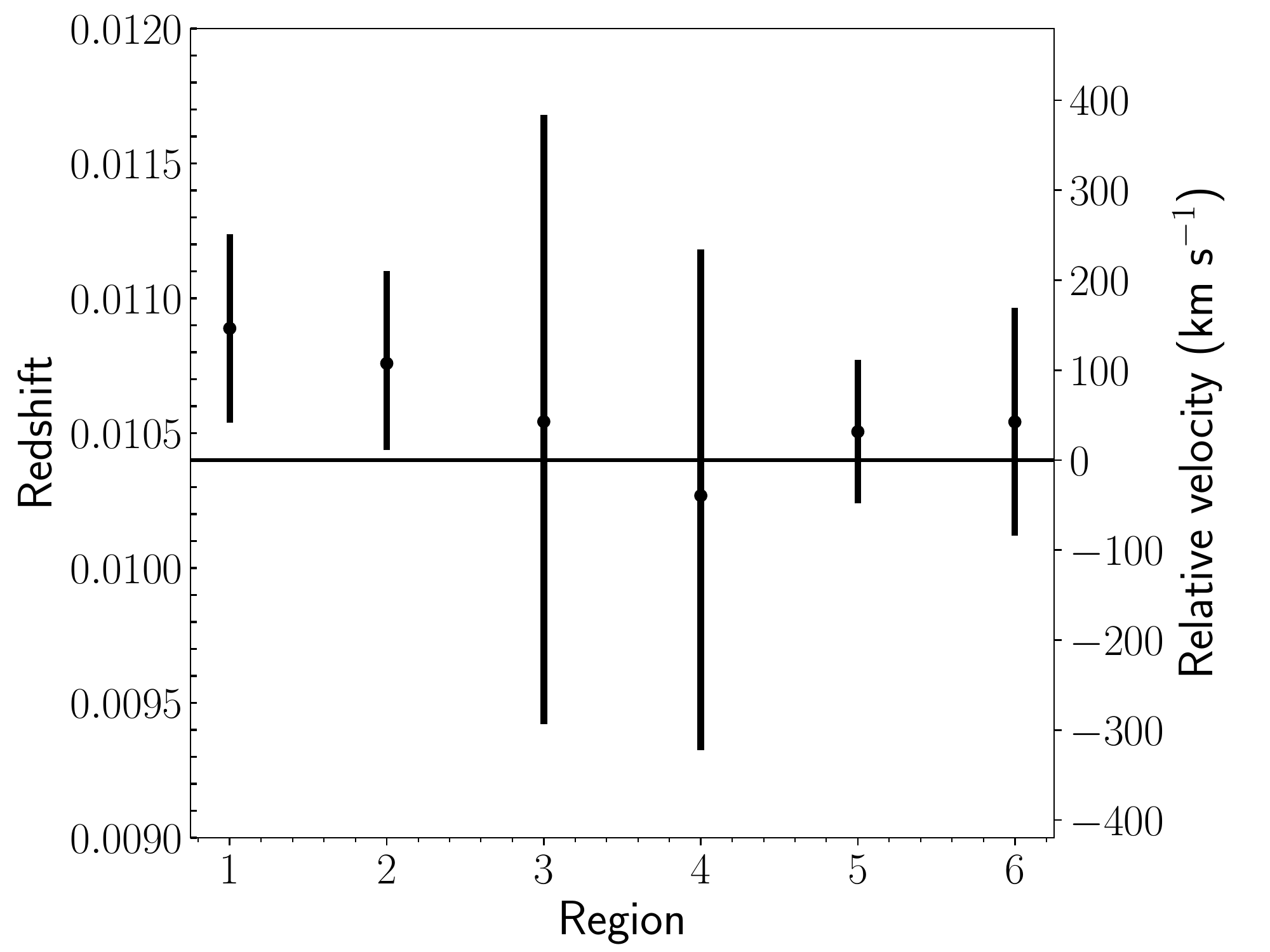}
        \caption{\emph{Left panel:} Centaurus cluster extracted regions for the inner region, following the fractional difference in the surface brightness. \emph{Right panel:} velocities obtained for each region. The Centaurus redshift is indicated with an horizontal line (See Section~\ref{regions_cold_fronts}).  } \label{fig_cold_fronts2a} 
\end{figure}

\begin{figure} 
\centering 
\includegraphics[width=0.49\textwidth]{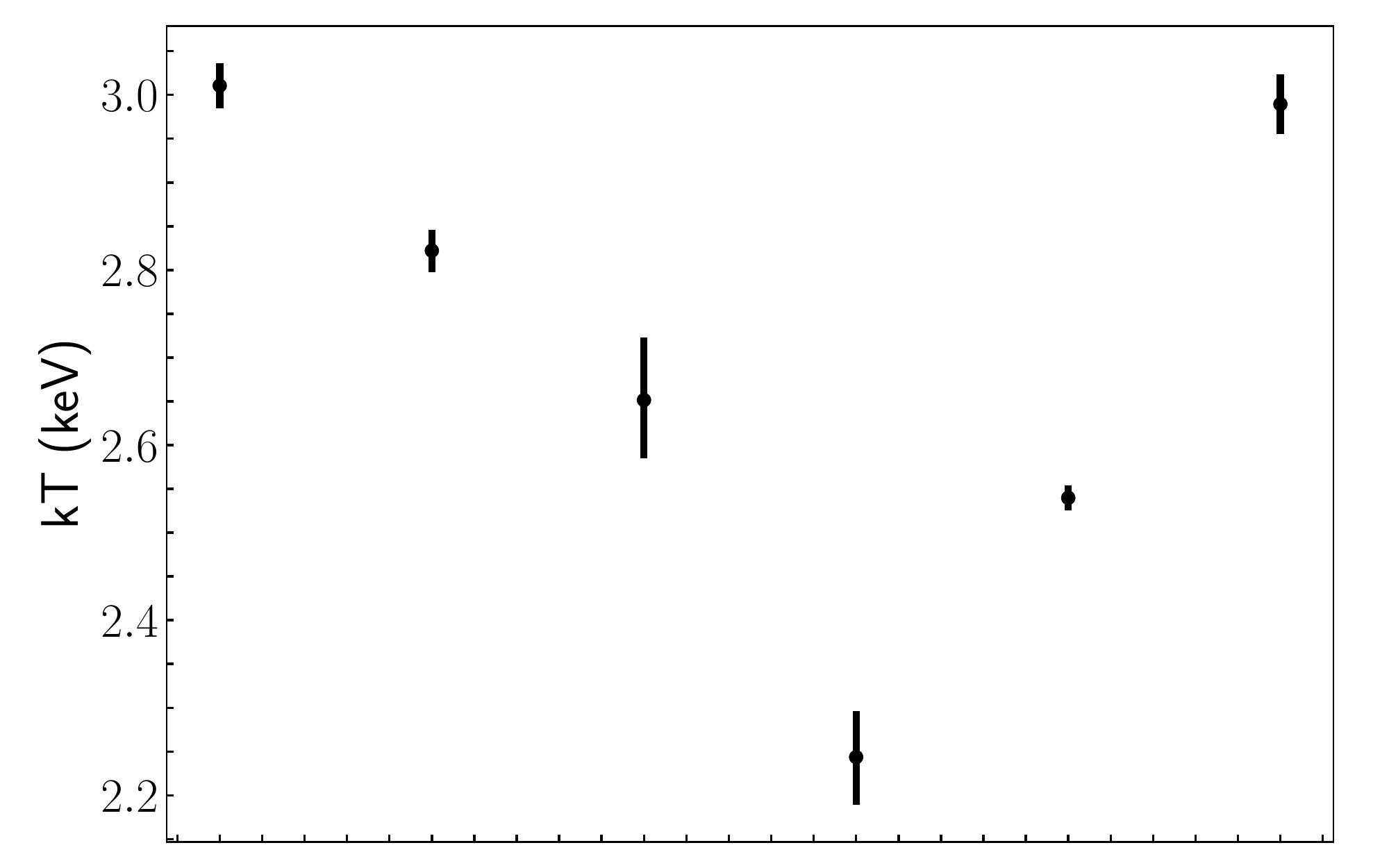}\\
\includegraphics[width=0.49\textwidth]{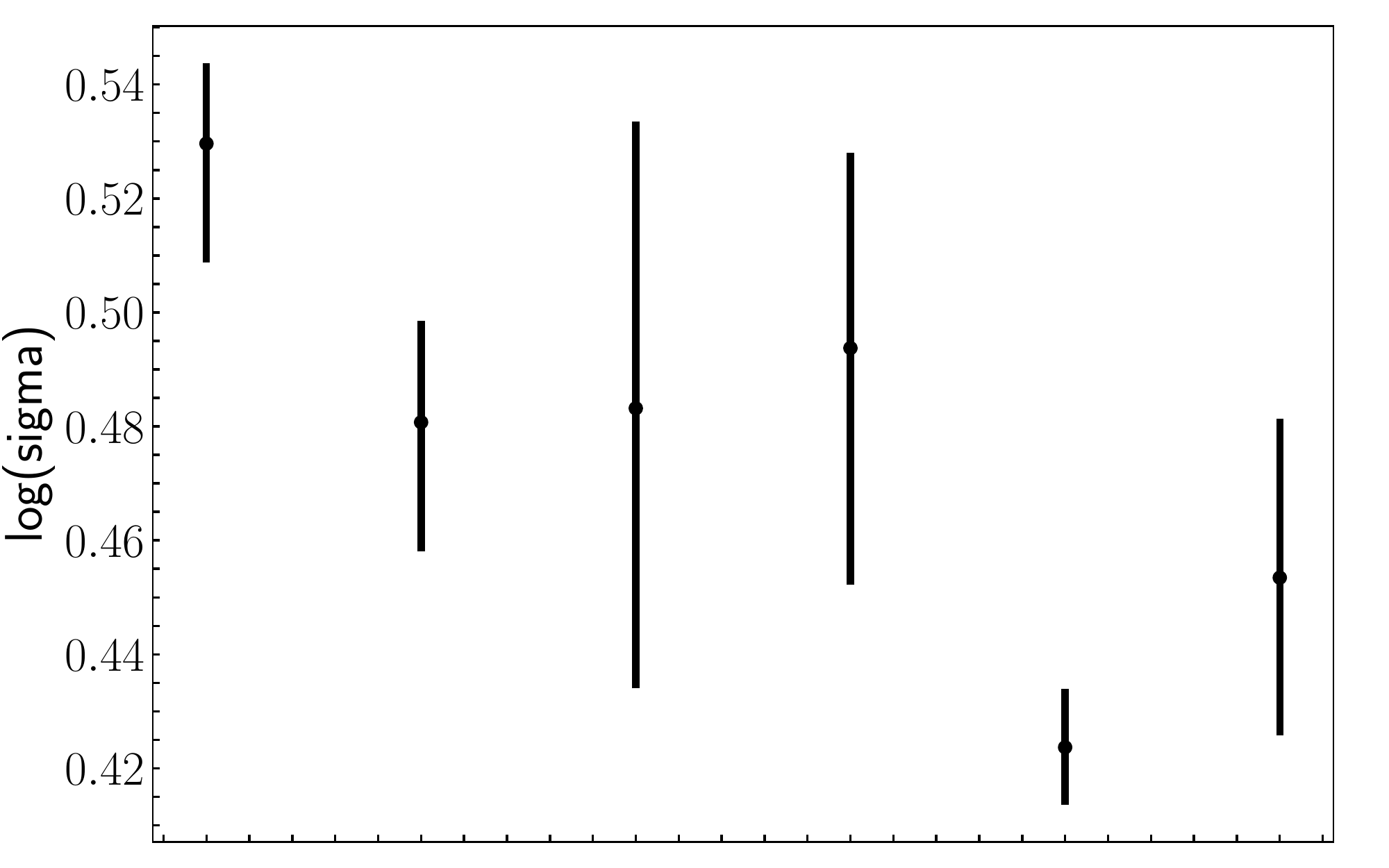}\\
\includegraphics[width=0.49\textwidth]{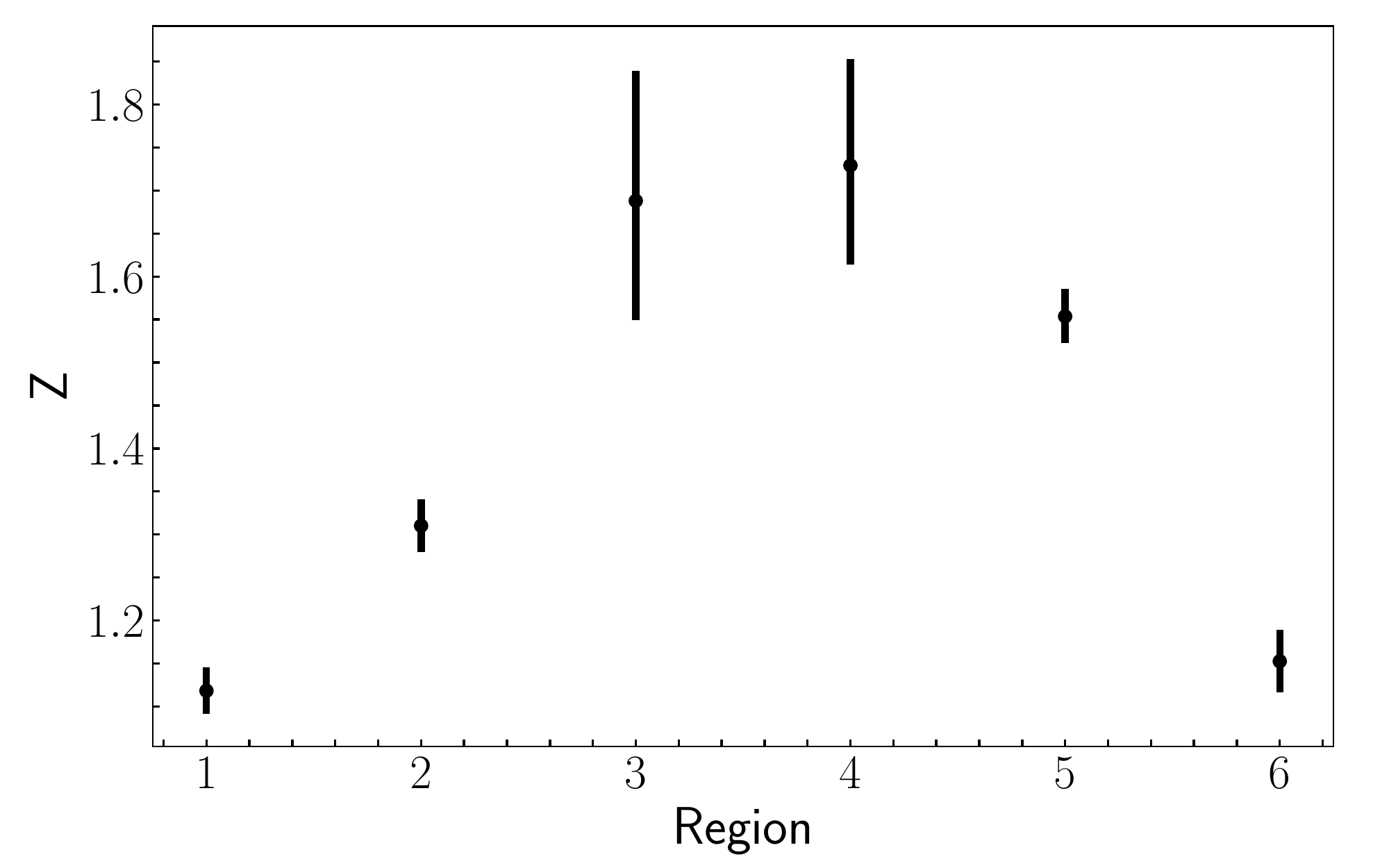}
        \caption{\emph{Top panel:} best-fit temperatures obtained for the inner region. \emph{Middle panel:} best-fit log(sigma) values obtained. \emph{Bottom panel:} best-fit metallicities obtained (See Section~\ref{regions_cold_fronts}). } \label{fig_cold_fronts2b} 
\end{figure}

\begin{table*}
\scriptsize 
\caption{\label{tab_coldfront2}Centaurus cluster best-fit parameters for regions following the fractional difference in surface brightness. }
\centering
\begin{tabular}{ccccccc}
\\
Region &\multicolumn{6}{c}{{\tt lognorm} model}  \\
\hline
 &$kT$&$\sigma$& Z& $z$   & $norm$   & cstat/dof\\ 
  & & & &  ($\times 10^{-3}$) &   ($\times 10^{-3}$)  \\ 
1&$3.01\pm 0.03$&$0.53\pm 0.02$&$1.12\pm 0.03$&$10.89\pm 0.35$&$55.88\pm 0.45$&$2163/1688$\\
2&$2.82\pm 0.02$&$0.48\pm 0.02$&$1.31\pm 0.03$&$10.76\pm 0.33$&$46.07\pm 0.40$&$1919/1687$\\
3&$2.65\pm 0.07$&$0.48\pm 0.05$&$1.69\pm 0.15$&$10.54\pm 1.13$&$11.36\pm 0.43$&$1279/1418$\\
4&$2.24\pm 0.05$&$0.49\pm 0.04$&$1.73\pm 0.12$&$10.27\pm 0.93$&$13.72\pm 0.49$&$1518/1504$\\
5&$2.54\pm 0.01$&$0.42\pm 0.01$&$1.55\pm 0.03$&$10.51\pm 0.27$&$41.43\pm 0.36$&$2398/1688$\\
6&$2.99\pm 0.03$&$0.45\pm 0.03$&$1.15\pm 0.04$&$10.54\pm 0.42$&$14.34\pm 0.15$&$2009/1685$\\
\\ 
 \hline
\end{tabular}
\end{table*}

  \begin{figure} 
\centering 
\includegraphics[width=0.40\textwidth]{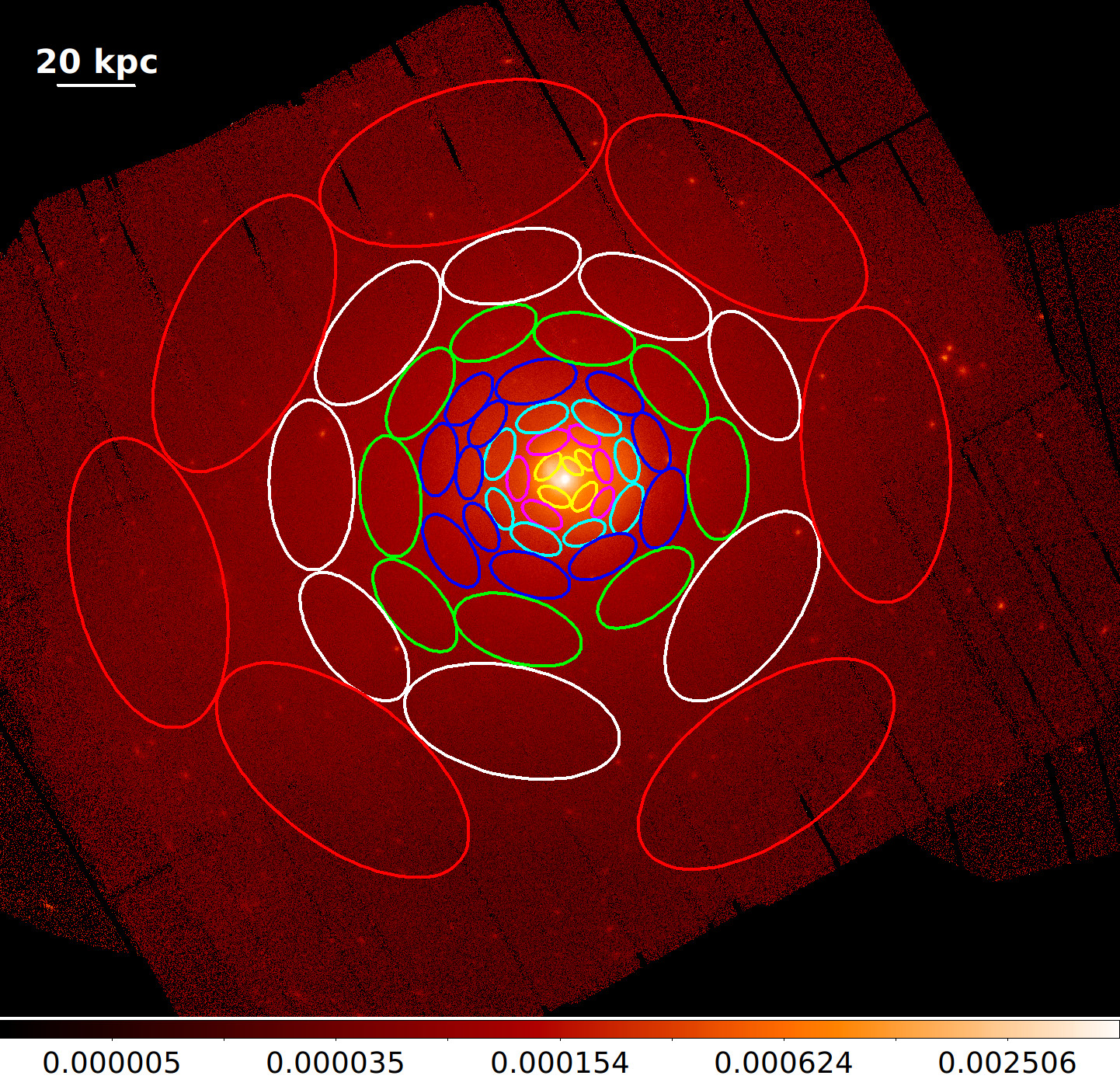}
\caption{Non-overlapping regions selected to analyze the energy profile for the spectral map obtained in Section~\ref{spec_maps}. Different colors indicate the different radius assigned.} \label{fig_energies1} 
\end{figure}

  \begin{figure} 
\centering 
\includegraphics[width=0.40\textwidth]{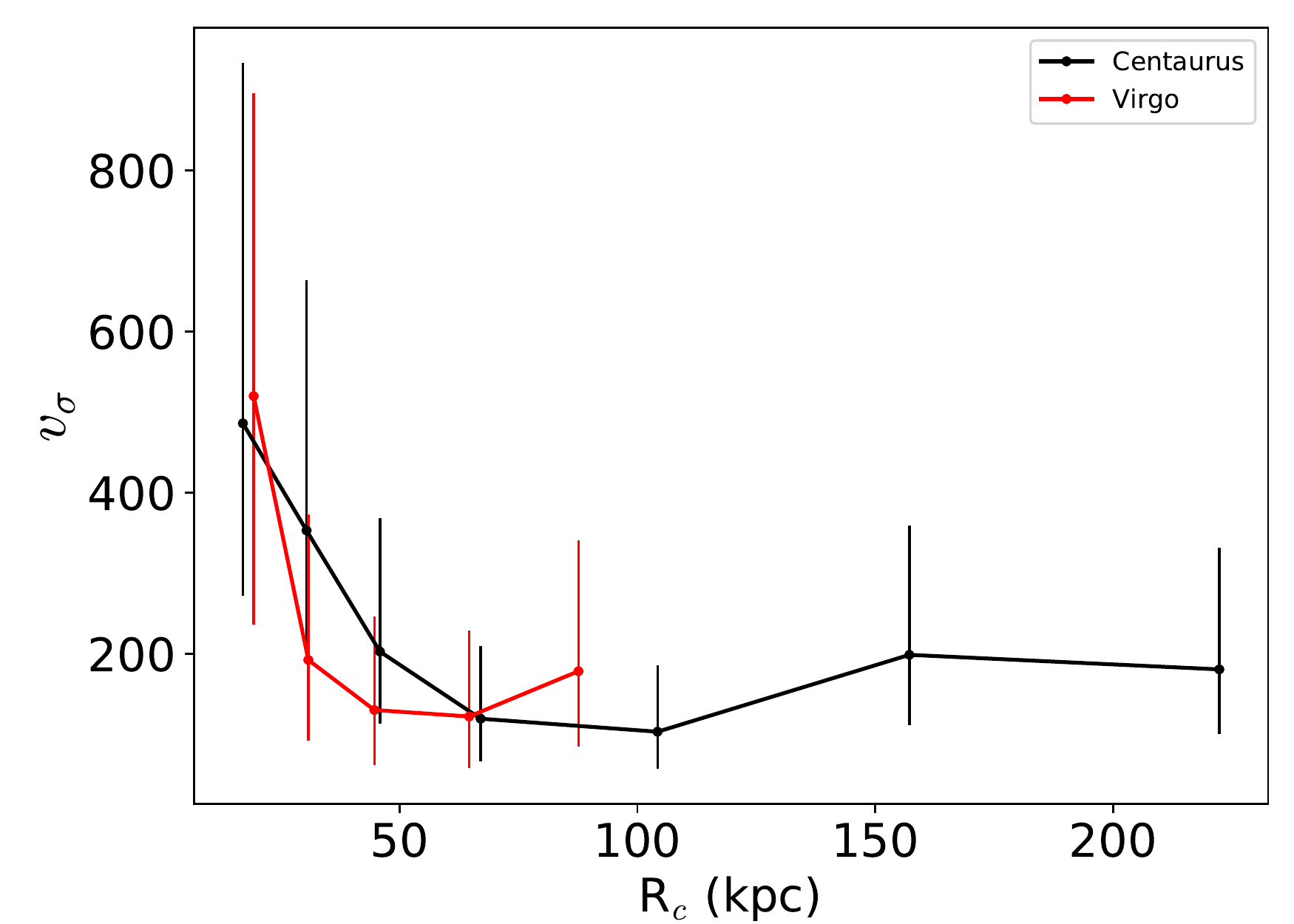}\\
\includegraphics[width=0.40\textwidth]{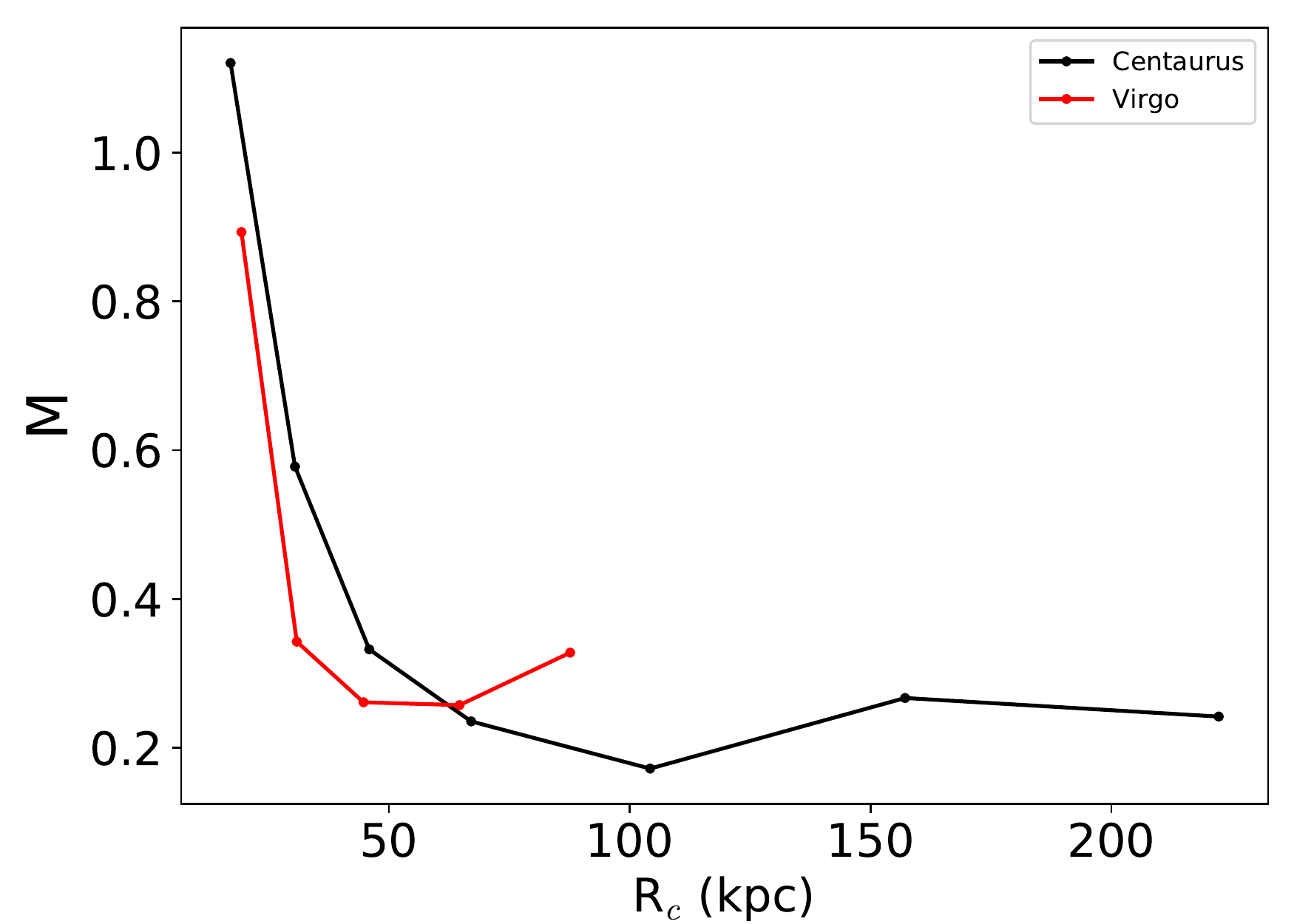} \\
\includegraphics[width=0.40\textwidth]{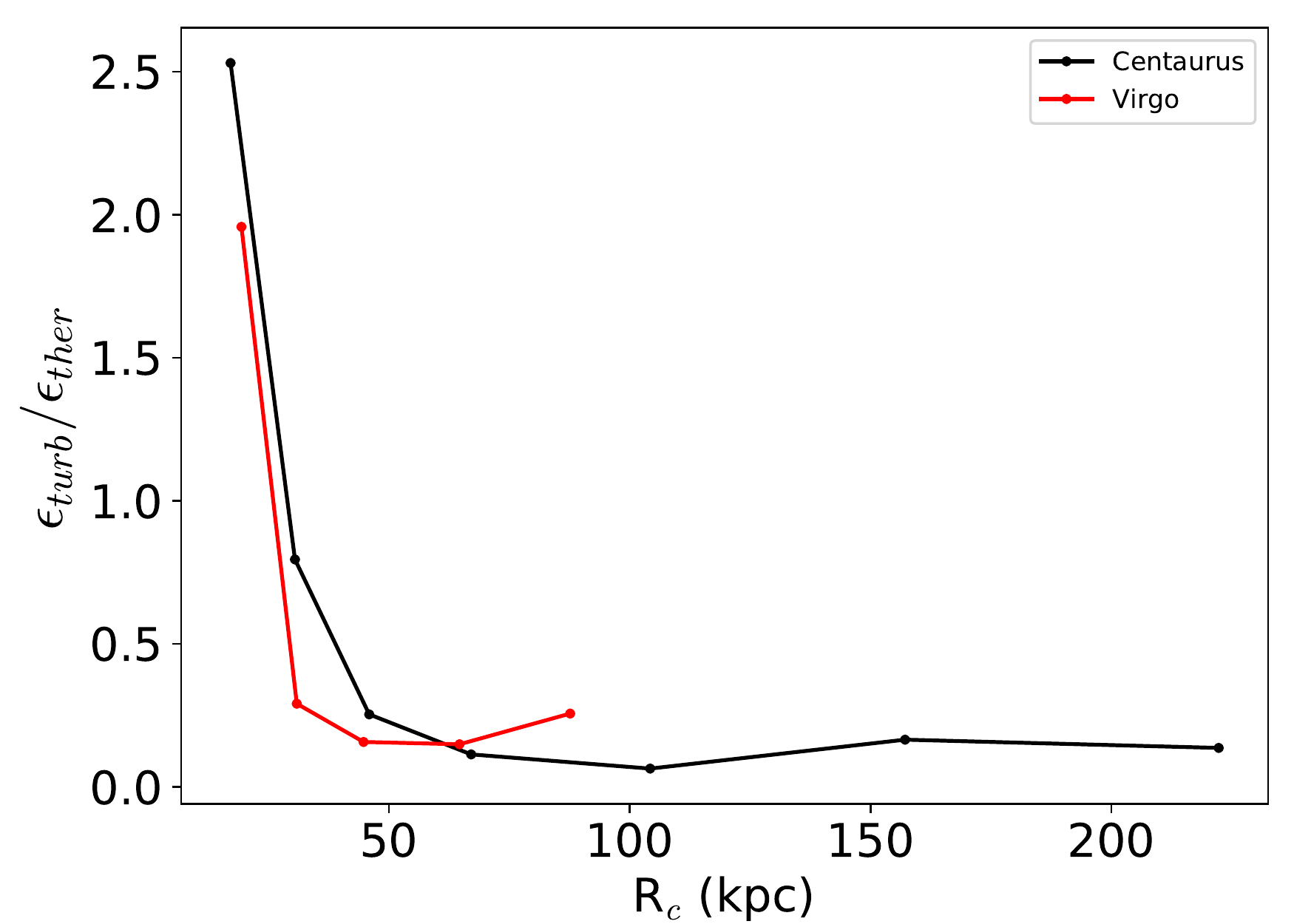} 
\caption{\emph{Top panel:} width of the velocity distribution as function of the distance from the cluster center. \emph{Middle panel:} mach number as function of the distance from the cluster center. \emph{Bottom panel:} upper limits for the ratio of turbulent to thermal energy ($\epsilon_{tur}/\epsilon_{ther}$). For comparison, values obtained for the Virgo cluster by \citet{gat21} are also included. } \label{fig_energies2} 
\end{figure}

\subsection{Cluster velocity substructures}\label{regions_xmm_proposal}
We extracted spectra for spatial regions following the X-ray surface brightness and having roughly equal number of counts in the Fe-K complex ($\sim$2000). Figure~\ref{fig_xmm_proposal_half1}  top panel shows the specific regions analyzed. The best-fit results are listed in Table~\ref{tab_xmm_proposal_half}. Figure~\ref{fig_xmm_proposal_half1} bottom panel shows the complex distribution of velocities obtained.  We have obtained velocities measurements with uncertainties down to $\Delta v\sim 120$ km/s (for region 12).  The largest blueshift/redshift, with respect to the Centaurus cluster, correspond to $-223\pm 250$ km/s and $1760^{+514}_{-564}$ km/s for regions 3 and 17, respectively. In the following, we compare the ICM X-ray velocities obtained with the optical spectroscopic redshifts measured for the individual galaxies NGC~4696, NGC~4709 and NGC~4696B.

We have found that, for the regions 1,2 and 3, close to the main system NGC~4696, there is an excellent agreement between the obtained redshift and the Centaurus cluster redshift. For these points we obtain an average redshift of $0.0102\pm 0.0007$. That is $\sim -61$ km/s with respect to the Centaurus cluster. Other regions surrounding the Cen~30 system display similar velocities to that of the system (indicated in blue color in Figure~\ref{fig_xmm_proposal_half1} top panel), except for region 7 ($261\pm 200$ km/s). Figure~\ref{fig_xmm_proposal_half2} shows the temperatures, $\log(\sigma)$ and metallicites profile obtained. The overall metallicity and temperature distribution is in good agreement with the spectral map shown in Figure~\ref{fig_velocity_ellipses2}. Regarding the metallicity and temperature, we have found that region 4 and 5, located south-west from the cluster core, has lower metallicities and larger temperatures than those located north-east from the cluster core (e.g. regions 6 and 7). The largest metallicity and lowest temperature correspond to the cluster core (i.e. region 1 and 2).

We note that region 22, containing NGC~4709 (i.e. the main galaxy of Cen~45), displays a velocity of $216\pm 258$ km/s (with respect to the Centaurus cluster), within the velocity of the Cen~30 system, but significatively smaller than the Cen~45 system velocity ($\sim 1500$ km/s above Cen 30). \citet{ota16} found an upper limit for the velocity around NGC~4709 $<750$~km/s. They suggested that such results indicate that the Centaurus cluster has experienced a subcluster merger along the line of sight and gas near NGC~4969. The temperature obtained for that region is in good agreement with the temperature obtained by \citet{wal13a} which could large be due to heating of gas near NGC~4696 by the strong shock along the merger process. On the other hand, for the region beyond NGC~4709 the velocity is $ 696\pm 310$ km/s (region 23). For region 17 we have found a velocity of $ 1759^{+565}_{-515} $ km/s, in good agreement with the NGC~4709 system velocity. As suggested by \citet{ota16}, such offset between the galaxy distribution and the mass centroids of the ICM along the sightline could be a fingerprint of a previous subcluster merger. For region 15, which contains the NGC~4696B, we have found a velocity of $642_{-249}^{+267}$ km/s, much larger than the system velocity which is close to the Centaurus cluster velocity. We found that the most external region beyond NGC 4709 (region 23) displays the largest temperature ($4.48^{+0.11}_{-0.17}$ K) and lowest metallicity ($0.47 \pm 0.03$). However, for regions with high kT and low Z the Fe~K lines become fainter leading to larger uncertainties.

 \subsection{The cold fronts}\label{regions_cold_fronts}
We have created regions to measure the velocity structure in the cold fronts identified in the {\it Chandra} observations by \citet{san16}. Figure~\ref{fig_cold_frontsa} shows the exact regions analyzed (top panel) while the best-fit results are listed in Table~\ref{tab_coldfront}. Figure~\ref{fig_cold_frontsa} shows the velocity obtained for each region (bottom panel). For two cold fronts, there is a hint of larger velocities in the hottest region (i.e. region 2 compared to regions 1), however the uncertainties are too large to provide a good constrain on the velocities. These results suggest that the cold front motion is difficult to observe because the gas is moving in a plane perpendicular to our line of sight. Figure~\ref{fig_cold_frontsb} shows the temperatures, the $\log(\sigma)$ and the metallicities obtained. As expected, there are clear differences in metallicity and temperatures in the interface between the cold fronts. The lack of significant metal mixing in cold fronts is likely due to its properties as a transient wave phenomena \citep{roe11}.

Following the fractional difference in 0.5 to 9.25 keV surface brightness, we have manually selected multiple regions in the innermost part of the cluster to study the velocity structure. Top panel in Figure~\ref{fig_cold_fronts2a} shows the extraction regions analyzed. Figures~\ref{fig_cold_fronts2a} and~\ref{fig_cold_fronts2b} shows the best-fit results while Table~\ref{tab_coldfront2} list the best-fit parameters. We have obtained velocities measurements with uncertainties down to $\Delta v\sim  79$ km/s (for region 5). The average velocity for all regions is $52\pm 171$. We note that regions in the east direction (i.e. 1 and 2) display lower velocity, larger temperature and larger metallicity than regions located in the west direction (i.e. regions 5 and 6).

\subsection{Energy budget}\label{sec_dis}  
We estimated the ratio of turbulent to thermal energy by measuring velocities for non-overlapping elliptical regions obtained in the spectral maps for different radius. Figure~\ref{fig_energies1} shows the ellipse extraction regions analyzed, where different radius are indicated by different colors. For each radius we measured the velocity mean and $\sigma$-width, assuming a gaussian distribution, as function of the distance to the cluster center. Figure~\ref{fig_energies2} top panel shows that the $\sigma$-width decreases as we move away from the inner radius. As a comparison, we included the values obtained for the Virgo cluster by \citet{gat21}. The $\sigma$-width decreases faster for the former, due to the strong influence of the AGN outflows near the cluster center. However, uncertainties are large in both cases.   

We compute the sound speed for each region as $c_{s}=\sqrt{\gamma kT/\mu m_{p}}$, where $kT$ is the best-fit temperature, $\gamma$ is the adiabatic index, $\mu$ is the mean particle mass and $m_{p}$ is the proton mass. Then, we computed the Mach number as a function of the radius, by dividing the bulk velocity by the sound speed and assuming that velocities are isotropic. We have found that for the innermost radius the Mach number is $\sim 1.12$, a value expected for AGN-driven outflow, while for radius $>30$~kpc the Mach number is $M\sim 0.17-0.33$, a range of values expected for gas sloshing (see middle panel in Figure~\ref{fig_energies2}). Bottom panel in Figure~\ref{fig_energies2} shows upper limits for the ratio of turbulent ($\epsilon_{turb}$) to thermal ($\epsilon_{ther}$) energy computed as $\epsilon_{turb}/\epsilon_{ther}=\frac{\gamma}{2}M^{2}$. For radius $>30$~kpc we found a contribution from the turbulent component $<25\%$. It is important to note that we are assuming that motions are isotropic. Simulations computed by \citet{zuh18} have shown that the kinetic energy may be underestimated near the cluster core by $\sim10-60\%$ depending on the line of sight, due to the conservative assumption of isotropy.

\section{Discussion}\label{sec_dis} 
\citet{san16} identified multiple substructures in the Centaurus cluster by analyzing {\it Chandra} observations, including a $1.9$~kpc radius shell around the core which
could be a shock generated by an AGN outburst, small cavities, filaments and a plume-like structure. \citet{wal15}, in their analysis of {\it Chandra} observations, found one-component velocities in the range 100-150 km/s on spatial scales of 4-10 kpc. Due to the lower spatial resolution in the {\it XMM-Newton} observations we are not able to study the velocity distribution in such structures. We have found for most of regions velocities similar to the Centaurus cluster velocity, including for regions near the NGC~4709 subsystem, in good agreement with the upper limits obtained in the {\it Suzaku} data analysis done by \citet{ota16}. The velocities measured near the cluster center are much lower than those found by \citet{ren06} in their analysis of {\it Chandra} observations.  Finally, the temperature and metallicity profiles obtained in our analysis in the $>20$~kpc scale (see Figures~\ref{fig_cas1_resultsb} and~\ref{fig_cas2_resultsb}) are in good agreement with previous {\it Chandra} and {\it XMM-Newton} results \citep{tak09,pan13,wal13a,lak19}.

Idealized merger simulations indicate that sloshing motions can produce line shifts in the order of $100-200$ km/s near the cluster core \citep{zuh16,zuh18}. Idealized AGN simulations are also capable of produce low line-of-sight velocity dispersions $150$ km/s consistent with those observed by {\it Hitomi} in the Perseus cluster \citep{bou17,ehl21}. Velocities obtained from cosmological simulations, on the other hand, vary between $100-400$ km/s \citep{lau17,ron18}. In this sense, the theoretical predictions are in good agreement with the velocity measurements.

\section{Conclusions and summary}\label{sec_con} 
We have analyzed the velocity structure in the Centaurus cluster using the technique developed by \citet{san20} to calibrate the absolute energy scale of the {\it XMM-Newton} EPIC-pn detector. Our results indicates that the ICM is dominated by the low velocity motions and, possibly, by the impact of the AGN in the NGC~4696 system. In this Section we briefly summarize our findings.

\begin{enumerate}
\item  We have obtained accurate velocity measurements with uncertainties down to $\Delta v \sim 79$ km/s.

\item  We have created 2D projected maps for velocity, temperature, $\log(\sigma)$, metallicity, density, entropy and pressure distribution for the Centaurus cluster. There is no clear indication of the spiral pattern in the velocity map associated to gas sloshing. While there is a hint for blueshifted motion near the cluster center, most likely due to the influence of AGN ouflows, the velocity uncertainties are large.

\item  We have studied the velocity distribution by creating non-overlapping circular regions. We found that the gas located at $<20$~kpc from the cluster core displays low blueshifted velocities (with respect to the Centaurus cluster velocity) while gas located at $20-150$~kpc displays a velocity close to the velocity of the cluster. Finally, gas located at large distance (i.e. $>150$~kpc) displays a redshift behavior with low velocities ($<500$~km/s).

\item We have analyzed the velocity distribution along E and W directions, by creating non-overlapping circular regions. We have found that, while the metallicity distribution shows larger values along the E direction for large distances, together with the lower temperatures in the same direction, the velocity distribution is similar along both directions (including the uncertainties).  The velocities are $<600$~km/s.

\item We have analyzed spectra for spatial regions with similar number of counts to study the substructures within the cluster. We have found that the gas located near the cluster center display velocities similar to the velocity from the main system NGC~4696. We have found a region that displays a large redshift similar to the velocity of the system NGC~4709, suggesting an offset between the galaxy and the ICM velocity distribution.

\item We have analyzed the velocity structure following the surface brightness near the cluster core. We have found that the velocities are similar to the velocity of the Centaurus cluster. 

\item We have analyzed the velocity structure following the 3 cold fronts located towards the E and W direction from the cluster core. Our measurements indicate that despite lack of significant metal mixing, the velocities are close to the velocity from the NGC~4696 system. In that sense, the cold front are likely moving tangentially (i.e. in a plane perpendicular to our line of sight with low velocity).

\item By fitting multiple non-overlapping regions, we have found the the width of the velocity distribution decreases as we move away from the cluster center. We have found a Mach number of $M\sim 0.17-0.33$ for radius $>30$~kpc from the cluster core, a range of values expected for gas sloshing. We also have found a contribution from the turbulent component of $<25\%$ to the total energetic budget for radius $>30$ kpc.

\end{enumerate}

Future work will include the analysis of the soft band ($<1.5$ keV).

\section{Acknowledgements} 
This work was supported by the Deutsche Zentrum f\"ur Luft- und Raumfahrt (DLR) under the Verbundforschung programme (Kartierung der Baryongeschwindigkeit in Galaxienhaufen). This work is based on observations obtained with XMM-Newton, an ESA science mission with instruments and contributions directly funded by ESA Member States and NASA.

\subsection*{Data availability}
The observations analyzed in this article are available in {\it XMM-Newton} Science Archive (XSA\footnote{\url{http://xmm.esac.esa.int/xsa/}}).

\bibliographystyle{mnras}

\newpage

\appendix\label{sec_apx}

\section{XMM-Newton spectra and fitting procedure}\label{sec_xmm_spec}
\begin{figure} 
\centering 
\includegraphics[width=0.48\textwidth]{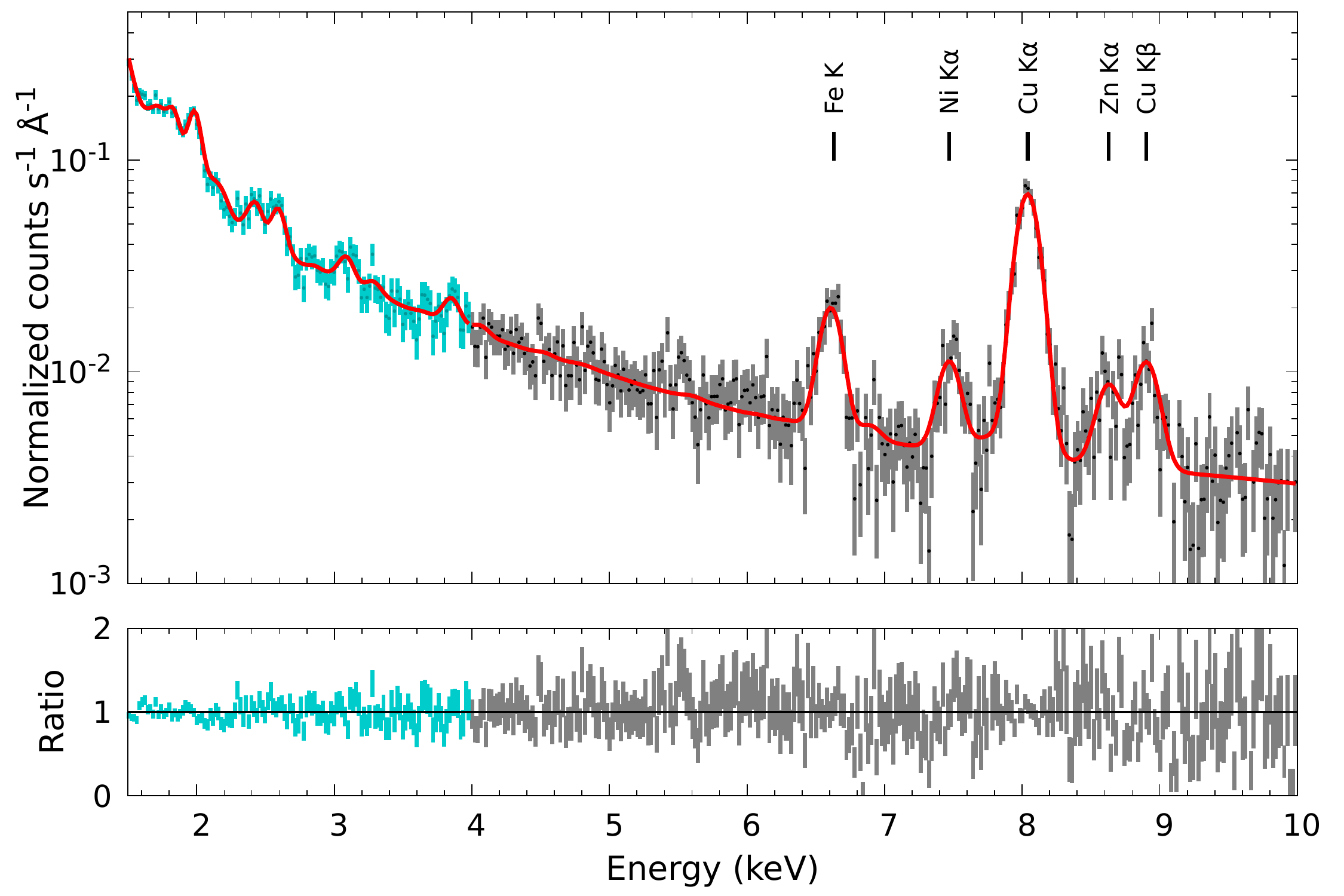} 
\caption{Best-fit spectra obtained for the innermost region in the Case 1 analysis. The spectra have been rebinned for illustrative purposes. The background lines as well as the astrophysical Fe K line are indicated. } \label{fig_cas1_spectra} 
\end{figure}

Figure~\ref{fig_cas1_spectra} shows best-fits spectra obtained for the inner region analyzed in Case 1 in Section~\ref{circle_rings}, as an example of the EPIC-pn spectra analyzed in this work. As described in Section~\ref{sec_fits}, we load the data twice in order to fit separately, but simultaneously, the 1.5-4.0 keV and hard 4.0-10 keV energy bands. The instrumental emission lines used as part of the background emission for the energy scale calibration are Ni K$\alpha$ ($7.47$ keV), Cu K$\alpha$ ($8.04$ keV),  Zn K$\alpha$ ($8.63$ keV) and Cu K$\beta$ ($8.90$ keV).

\begin{figure} 
\centering 
\includegraphics[width=0.48\textwidth]{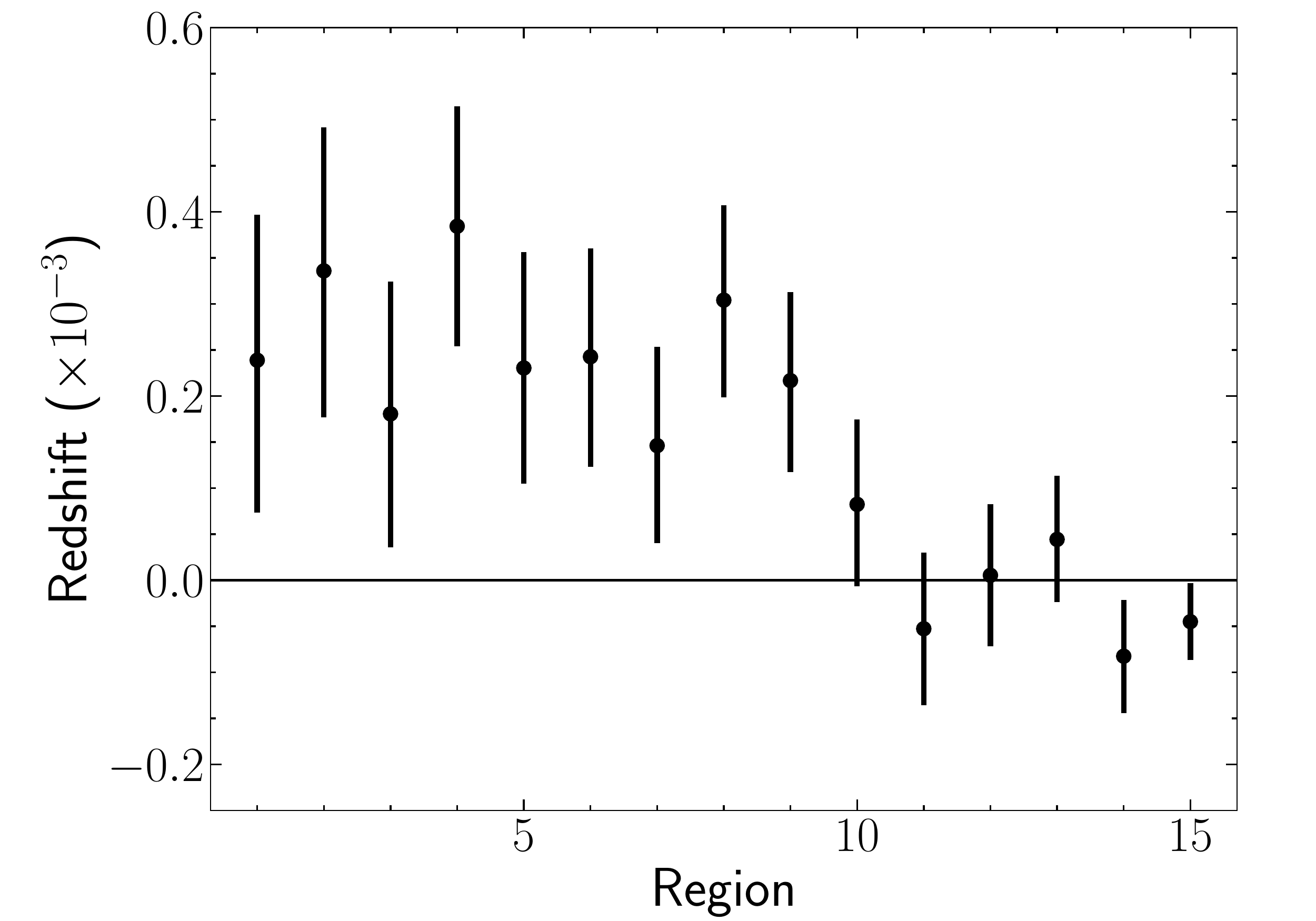} 
\caption{Instrumental Cu-K$\alpha$ redshift variation for each region in the Case 1 analysis, obtained from the background best-fit resuts.} \label{fig_instrument_redshift} 
\end{figure}

Figure~\ref{fig_instrument_redshift} shows the instrumental Cu-K$\alpha$ redshift variation for each region in the Case 1 analysis, after applying the energy calibration scale correction. This plot shows a systematic uncertainty $\sim 75$ km/s  coming from the calibration uncertainty. This could be compared with the residual calibration uncertainties at $\sim 150$ km/s level found by \citet{san20}, which depends on the CCD spatial location. We also note that the Cu K$\alpha$ redshift is obtained from the background fit which also includes the Ni K$\alpha$, Zn K$\alpha$ and Cu K$\beta$ lines, thus increasing the uncertainty.

\begin{figure} 
\centering 
\includegraphics[width=0.48\textwidth]{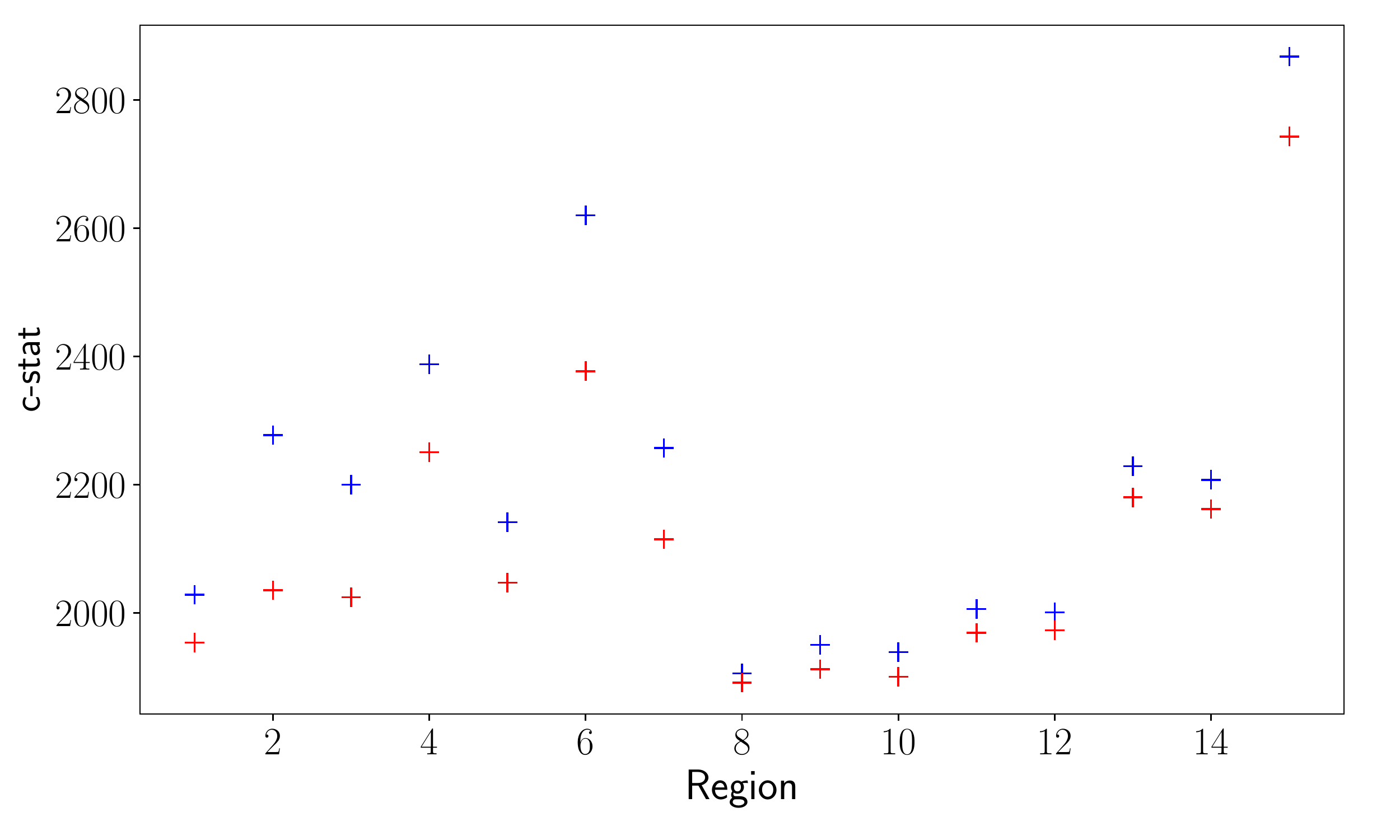} 
\caption{Statistical comparison between the {\tt lognorm} (red) and the {\tt apec} (blue) models for each region in the Case 1 analysis.} \label{fig_cstat_change} 
\end{figure}

Previous analysis of the Centaurus cluster have shown the presence of a multi-temperature component near the Centaurus cluster center \citet{san08,pan13,lak19}. In that sense, we have tested a two-temperature model ({\tt tbabs*(apec+apec)}) in our fitting procedure, however we note that for multiple extraction regions the second temperature component was not well constrained. In order to avoid switching between 1T and 2T models in some arbitrary way, we decided to use the {\tt lognorm} model. 
Figure~\ref{fig_cstat_change} shows a comparison of the best fit statistic obtained for all regions in Case 1 analysis when fitting the spectra with the {\tt lognorm} model and with 1-{\tt apec} mode. In all cases the {\tt lognorm} leads to a better fit, from the statistical point of view.

\end{document}